\documentclass[amssymb,nobibnotes,nofootinbib,aps,prd]{revtex4}

 \usepackage{here}

\usepackage[dvipdfmx]{graphicx}
\usepackage{amssymb,amsfonts,amsmath,bm,color,multirow,cases,empheq,hyperref}
\usepackage{mathrsfs}
\definecolor{darkergreen}{rgb}{0,0.5,0}

\usepackage{enumerate}
\usepackage{ulem}
\usepackage{afterpage}

\hypersetup{%
 setpagesize=false,
 bookmarksnumbered=true,%
 bookmarksopen=true,%
 colorlinks=true,%
 linkcolor=blue,%
 citecolor=red}

\usepackage{tikz}
\usetikzlibrary{positioning,fit}
\usetikzlibrary{calc,intersections,through,backgrounds}
\usetikzlibrary{calc,arrows,decorations.markings,shapes.arrows,patterns, decorations.pathmorphing}
\usetikzlibrary{positioning,matrix,arrows,decorations.pathmorphing,decorations.pathreplacing}
\usetikzlibrary{arrows,matrix,positioning,shapes,arrows}
\usetikzlibrary{shapes.geometric}
\usetikzlibrary{decorations.text}
\usetikzlibrary{arrows.meta}

\tikzset{
  ->-/.style={decoration={markings, mark=at position 0.5 with {\arrow{to}}},
              postaction={decorate}},}
\tikzset{
  -<-/.style={decoration={markings, mark=at position 0.5 with {\arrow{to reversed}}},
              postaction={decorate}},}
              
\tikzset{
  pics/torus/.style n args={3}{
    code = {
      \providecolor{pgffillcolor}{rgb}{1,1,1}
      \begin{scope}[
          yscale=cos(#3),
          outer torus/.style = {draw,line width/.expanded={\the\dimexpr2\pgflinewidth+#2*2},line join=round},
          inner torus/.style = {draw=pgffillcolor,line width={#2*2}}
        ]
        \draw[outer torus] circle(#1);\draw[inner torus] circle(#1);
        \draw[outer torus] (180:#1) arc (180:360:#1);\draw[inner torus,line cap=round] (180:#1) arc (180:360:#1);
      \end{scope}
    }
  }
}

\newcommand{\tikznode}[2]{\relax
    \ifmmode%
    \tikz[remember picture,baseline=(#1.base),inner sep=0pt] \node (#1) {$#2$};
    \else
    \tikz[remember picture,baseline=(#1.base),inner sep=0pt] \node (#1) {#2};%
    \fi
}

\newcommand{\no}{\nonumber}

\newcommand{\cF}{\mathcal F}

\newcommand{\cM}{\mathcal M}
\newcommand{\cO}{\mathcal O}

\newcommand{\la}{\lambda}
\newcommand{\Tr}{{\rm Tr}}

\allowdisplaybreaks

\begin{document}

\begin{flushright}
TTI-MATHPHYS-37
\end{flushright}
\vspace*{0.5cm}

\title{
Monodromy-Matrix Description of Doubly Rotating Black Rings
}

\author{Jun-ichi Sakamoto}
\email{jsakamoto@toyota-ti.ac.jp}
\author{Shinya Tomizawa}
\email{tomizawa@toyota-ti.ac.jp}
\affiliation{\vspace{3mm}Mathematical Physics Laboratory, Toyota Technological Institute\vspace{2mm}\\Hisakata 2-12-1, Tempaku-ku, Nagoya, Japan 468-8511\vspace{3mm}}

\begin{abstract}
Extending the single-angular-momentum case analyzed in our previous work, we investigate the solution-generating technique based on the Breitenlohner-Maison (BM) linear system for asymptotically flat, stationary, bi-axisymmetric black hole solutions with two angular momenta in five-dimensional vacuum Einstein theory.
In particular, we construct the monodromy matrix associated with the BM linear system for the doubly rotating Myers-Perry black holes and the Pomeransky-Sen'kov black rings.
Conversely, by solving the corresponding Riemann-Hilbert problem using the procedure developed by Katsimpouri et al., we demonstrate that the factorization of the monodromy matrix precisely reproduces these vacuum solutions, thereby reconstructing both geometries.
\end{abstract}

\date{\today}

\maketitle

\section{Introduction}

\medskip
The study of higher-dimensional black holes has evolved into a central topic in modern gravitational physics. Motivated by string theory and brane-world scenarios, these solutions reveal a remarkable diversity of horizon topologies and rotation structures absent in four dimensions. 
This breakdown of uniqueness has prompted new approaches to classification based on symmetry, topology, and conserved charges. 
Recent advances employ powerful mathematical tools, including the inverse scattering method, and the solution-generating based on hidden symmetry, to construct exact solutions.
Emparan and Reall~\cite{Emparan:2001wn} first obtained the exact solution for an $S^1$-rotating black ring, thereby demonstrating that five-dimensional vacuum Einstein theory admits not only the $S^1$-rotating Myers-Perry black hole~\cite{Myers:1986un} but also two distinct black ring solutions with identical conserved charges, providing a clear manifestation of non-uniqueness in higher dimensions. 
Given that five-dimensional black holes can in general rotate along two independent axes, it is natural to inquire whether more general black ring solutions exist that carry two independent angular momenta, associated with rotations along both the $S^1$ and $S^2$ directions. Using the B\"acklund transformation, Mishima and Iguchi first derived the $S^2$-rotating black ring solution~\cite{Mishima:2005id}, and Figueras~\cite{Figueras:2005zp} independently obtained the same solution via a different approach. This solution, however, inevitably exhibits conical singularities inside the ring. The inverse scattering method (ISM), originally developed by Belinski and Zakharov~\cite{Belinsky:1979mh}, has since played a crucial role in generating more general black ring solutions carrying both $S^1$ and $S^2$ rotations. However, direct application of ISM in higher dimensions often produces singular spacetimes. Pomeransky overcame this difficulty by refining the method and successfully applying it to five-dimensional black holes, deriving the Myers-Perry solution from the five-dimensional Schwarzschild seed~\cite{Pomeransky:2005sj}. 
The ISM also enabled the construction of $S^2$-rotating black rings~\cite{Tomizawa:2005wv}. Nevertheless, generating the $S^1$-rotating black ring via ISM proved significantly more challenging, as attempts from regular seeds invariably led to naked curvature singularities. A major breakthrough was achieved in Refs.~\cite{Iguchi:2006rd,Tomizawa:2006vp}, where the appropriate seed---a certain singular configuration---was identified to produce the $S^1$-rotating black ring. This development paved the way for constructing the doubly rotating black ring via ISM, and ultimately Pomeransky and Sen’kov succeeded in obtaining the balanced doubly rotating black ring solution~\cite{Pomeransky:2006bd}. Although only the balanced case was presented in their work~\cite{Pomeransky:2006bd}, an explicit unbalanced generalization was later constructed in Ref.~\cite{Morisawa:2007di}, followed by a more compact representation in Ref.~\cite{Chen:2011jb}. 

\medskip
In gravity theories coupled to a Maxwell field, a rotating black ring can carry a dipole charge, which - although not a conserved quantity - serves as an additional parameter characterizing the ring.
Such a dipole black ring solution was first discovered by Emparan~\cite{Emparan:2004wy}.
Subsequently, Elvang et al.~\cite{Elvang:2004xi} constructed further examples of dipole rings in five-dimensional minimal supergravity, based on a seven-parameter family of non-supersymmetric black ring solutions.
However, this dipole black ring does not admit a supersymmetric limit~\cite{Elvang:2004rt}, because the dipole is related to the other conserved charges: among the four conserved and one dipole charges, only three are truly independent.
As suggested in Ref.~\cite{Elvang:2004xi}, a most general non-Bogomol'nyi-Prasad-Sommerfield (non-BPS) black ring solution is expected to exist, characterized by its mass, two independent angular momenta, electric charge, and a dipole charge that remains distinct from the asymptotic conserved quantities.
Furthermore, the uniqueness theorem for black rings in minimal supergravity established in Ref.~\cite{Tomizawa:2009tb} states that, under the assumption of a topologically trivial domain of outer communication, an asymptotically flat, stationary, and bi-axisymmetric black ring with a non-degenerate, connected event horizon of topology $S^1 \times S^2$ - if it exists - is uniquely specified by five physical parameters: its mass, electric charge, two independent angular momenta, dipole charge, and additional geometric data from the rod structure, such as the ratio of the $S^2$ radius to the $S^1$ radius.
Recently, another non-BPS solutions~\cite{Bouchareb:2007ax} describing charged rotating black rings with three independent parameters in five-dimensional minimal supergravity - distinct from the three-parameter black ring of Ref.~\cite{Elvang:2004xi} - have been constructed by means of the electric Harrison transformation.
Moreover, a more general non-BPS black ring solution with four independent parameters was constructed in Ref.~\cite{Suzuki:2024vzq}, where the four conserved charges are independent but the dipole charge is related to other conserved quantities.
These works demonstrated that in contrast to the black ring in Ref.~\cite{Elvang:2004xi}, the resulting black ring solutions include certain special cases of the supersymmetric solutions of Ref.~\cite{Elvang:2004rt}, though not the complete supersymmetric black ring.
The most general black ring encompassing the full supersymmetric configuration is expected to possess five independent parameters, but such a fully general solution has yet to be constructed. 
Since these methods rely on a specific choice of singular seed solutions, the construction of such a solution is not so easy.

\medskip
It can be expected that the solution-generating technique based on the Breitenlohner-Maison (BM) linear system, as developed in \cite{Breitenlohner:1986um,Chakrabarty:2014ora,Katsimpouri:2012ky,Katsimpouri:2013wka,Katsimpouri:2014ara}, provides a powerful and systematic framework for constructing exact solutions in higher-dimensional gravity theories coupled to matter fields.
In this formulation, the monodromy matrix ${\cal M}(w)$ plays a central role: it is a meromorphic function of an auxiliary complex variable $w$--the spectral parameter--and takes values in the Geroch group.
This approach offers two possibilities:
\begin{itemize}
\item[(1)] given a prescribed rod structure, one can construct a corresponding monodromy matrix ${\cal M}^{\rm new}(w)$  directly and read off the associated gravitational solution; and
\item[(2)] given a monodromy matrix ${\cal M}(w)$ corresponding to a known spacetime, one can generate new solutions by applying suitable global transformation denoted by $g(w)\in G$ ($G$: a global symmetry group depending on theories with matter fields) to it and then reconstructing the geometry from the transformed matrix ${\cal M}^{\rm new}(w)=g^\sharp(w){\cal M}(w)g(w)$. 
\end{itemize}
For both cases, as a final step, when reading off the gravitational solution from the monodromy matrix, one needs to factorize it as ${\cal M}^{\rm new}(w)={\cal V}(\lambda,x)^\sharp {\cal V}(\lambda,x)$, where $\lambda$ is another spectral parameter and $x$ denotes the two-dimensional coordinates.
This is the so-called Riemann-Hilbert problem, which is, in general, highly nontrivial to solve.
This method possesses several notable advantages:
\begin{itemize}
\item[(i)] Since the BM system forms the Geroch group, it provides a unified algebraic framework that encompasses a wide range of established techniques - such as the inverse scattering method, Ehlers transformations, and Harrison transformations - within a single formalism.
\item[(ii)]  It can generate both extremal and non-extremal black hole solutions within the same framework, in contrast to the ISM and the Ehlers or Harrison transformation methods, which are typically restricted to the latter.
\item[(iii)]  In case (1), the construction does not depend on a particular choice of seed solution, provided that the monodromy matrix corresponding to a given rod structure can be explicitly determined.
\end{itemize}
In case (1), as shown in Ref.~\cite{Sakamoto:2025jtn}, we explicitly constructed the monodromy matrix for multi-black-string configurations, by analogy with the monodromy matrix of the single black-string solution.
However, for case (1), the precise manner in which the monodromy matrix encodes the rod structure remains insufficiently understood.
In our previous work~\cite{Sakamoto:2025xbq}, we investigated three distinct asymptotically flat, stationary, and bi-axisymmetric black hole solutions with a single angular momentum in five-dimensional vacuum Einstein gravity, each characterized by a different horizon topology: the singly rotating Myers-Perry black hole, the Emparan-Reall black ring, and the Chen-Teo rotating black lens.
We constructed the monodromy matrices corresponding to these solutions and, conversely, by solving the associated Riemann-Hilbert problem, demonstrated that the factorization of the monodromy matrix exactly reproduces the original vacuum geometries.
These results confirm that the BM method can be successfully applied to black holes with non-spherical horizon topologies.

\medskip
As future directions, we aim to construct black ring solutions and multi-black-ring configurations---such as di-ring and black Saturn systems---in various gravitational theories that include matter fields. 
Therefore, as the first step toward this goal, the present paper introduces a new construction of a vacuum solution describing a doubly rotating black ring in five-dimensional Einstein gravity. 
Our approach is based on the BM linear system, formulated for asymptotically flat, stationary, bi-axisymmetric black hole spacetimes with two independent angular momenta. 
In particular, extending our previous work on five-dimensional black hole solutions with a single angular momentum, we construct the monodromy matrices associated with the BM linear system for the doubly rotating Myers-Perry black hole and the Pomeransky-Sen’kov black ring.
Conversely, by solving the corresponding Riemann-Hilbert problem following the procedure developed by Katsimpouri et al., we demonstrate that the factorization of the monodromy matrix precisely reproduces these vacuum solutions, thereby reconstructing both geometries in full detail.

\medskip
In Sec.~\ref{5dsugra}, we begin by reviewing the fundamental aspects of the Breitenlohner-Maison (BM) linear system for five-dimensional vacuum Einstein gravity. We explain how the $SO(4,4)$-valued monodromy matrix arises in this framework, outline the procedure for constructing such matrices from known solutions, and describe how the corresponding gravitational fields can be recovered through the factorization of the monodromy matrix---an algebraic procedure equivalent to solving the so-called Riemann-Hilbert problem. 
In Sec.~\ref{sec:mp}, as a concrete application of this formalism, we construct the explicit $SO(4,4)$-valued monodromy matrix associated with the doubly rotating Myers-Perry black hole and demonstrate in detail how its factorization reproduces the full spacetime metric and other physical quantities. We further analyze the structure of the charge matrix obtained from the residues of the monodromy matrix and clarify its algebraic properties, including the emergence of nilpotency in the extremal limit.
In Sec.~\ref{sec:ups}, we turn to the case of the Pomeransky-Sen'kov black ring, deriving its corresponding monodromy matrix and carrying out its factorization within the same $SO(4,4)$ framework. 
We also discuss the structure of the associated charge matrix, showing that, unlike the Myers-Perry case, the extremal limit of the black ring does not lead to a nilpotent degeneration of the charge matrix.
Finally, Sec.~\ref{sec:dis} summarizes the results and discusses their broader implications, including possible extensions of the present analysis to more general configurations such as multi-center or charged solutions in higher-dimensional supergravity theories.

\section{Summary of our set up for constructing five-dimensional asymptotically flat black hole solutions}\label{5dsugra}

In this work, we consider the construction of asymptotically flat black hole solutions in five-dimensional pure Einstein gravity, governed by the Einstein-Hilbert action,
\begin{align}\label{5d_sugra_action}
S_5 &= \int d^5x\,\sqrt{-g_5}\,R_5\,.
\end{align}
We focus on a stationary, bi-axisymmetric, and asymptotically flat class of five-dimensional black hole solutions that admit three mutually commuting Killing vectors, $(\partial_t,\partial_{\tilde{\phi}},\partial_{\tilde{\psi}})$.
The metric of these solutions asymptotically approaches the flat spacetime form at spatial infinity ($r \to \infty$)~\cite{Harmark:2004rm}:
\begin{align}\label{asym-met}
    ds^2_5&\simeq \left(-1+\frac{8M}{3\pi}\frac{1}{r^2}+\cO\left(\frac{1}{r^3}\right)\right)dt^2-2\left(\frac{4J_{1}}{\pi }\frac{\sin^2\theta}{r^2}+\cO\left(\frac{1}{r^3}\right)\right)dtd\tilde{\phi}\no\\
    &\quad-2\left(\frac{4J_{2}}{\pi}\frac{\cos^2\theta}{r^2}+\cO\left(\frac{1}{r^3}\right)\right)dtd\tilde{\psi}\no\\
    &\quad+\left(1+\cO\left(\frac{1}{r}\right)\right)\left(dr^2+r^2(d\theta^2+\sin^2\theta d\tilde{\phi}^2+\cos^2\theta d\tilde{\psi}^2)\right)\,,
\end{align}
where $M$ and $J_{1,2}$ denote the ADM mass and  ADM angular momenta, respectively. 
The coordinates $\tilde \phi$ and $\tilde \psi$ have the identifications $\tilde \phi\sim \tilde \phi+2\pi$ and $\tilde \psi\sim \tilde \psi+2\pi$, respectively, and  $\theta$ takes the range $0\leq \theta\le \frac{\pi}{2}$.

\medskip
As explained in Ref.~\cite{Sakamoto:2025xbq}, in order to apply the solution-generating technique described below, the dimensional reduction is performed not with respect to the standard angular coordinates $(\tilde{\phi},\tilde{\psi})$, but instead using the Euler angles $(\phi,\psi)$ defined by
\begin{align}\label{new-angle}
\tilde{\phi} = \frac{\phi + \psi}{2}, \qquad \tilde{\psi} = \frac{\phi - \psi}{2}\,,
\end{align}
with the identifications of $\psi\sim \psi+4\pi,\phi\sim \phi+2\pi$.
By introducing the Weyl-Papapetrou coordinates $z\in \mathbb{R}$ and $\rho\in [0,\infty)$, the five-dimensional spacetime metric can be written as
\begin{align}
\begin{split}
    ds_5^2&=-f^2(dt+\check{A}^0)^2+f^{-1}e^{2U}(d\psi+\omega_3)^2+f^{-1}e^{-2U}(e^{2\nu}(d\rho^2+dz^2)+\rho^2d\phi^2)\,,\\
    \check{A}^0&=\zeta^{0}(d\psi+\omega_3)+\hat{A}^{0}\,,\qquad \omega_3=\omega_{3,\phi}d\phi\,,\qquad \hat{A}^{0}=\hat{A}^{0}_{\phi}d\phi\,,
\end{split}
\end{align}
where the functions $f$, $U$, $\nu$, $\zeta^0$, $\omega_{3,\phi}$ and $\hat A^{0}_\phi$ depend only on $\rho$ and $z$.
In this setting, by performing the dimensional reduction along the Killing directions and dualizing the resulting one-form fields $\check{A}^0$ and $\omega_3$ into scalars, the Einstein-Hilbert action (\ref{5d_sugra_action}) reduces to a two-dimensional dilaton gravity theory coupled to a classically integrable two-dimensional coset sigma model,  whose action is given by
\begin{align}\label{2d-gravity-sigma}
    S_{\rm 2}=\int d\rho dz\,\sqrt{g_2}\,\rho\Bigl[R_2-2g_2^{mn}\Tr(\partial_{m}MM^{-1}\,\partial_{n}MM^{-1})\Bigr]
\end{align}
defined on the conformal flat space
\begin{align}
    ds^2_2=e^{2\nu(z,\rho)}(d\rho^2+dz^2)\,.
\end{align}
As in our previous paper, we formulate the coset sigma model on the symmetric coset space
\begin{align}\label{sym-coset}
\frac{G}{H}=\frac{SO(4,4)}{SO(2,2)\times SO(2,2)},
\end{align}
rather than $SL(3,{\bm R})/SO(3)$ discussed in Ref.~\cite{Maison:1979kx}.
Each Lie group has a $8\times 8$ matrix realization defined by
\begin{align}
    G&=SO(4,4)=\{~g \in GL(8,\mathbb{R}) ~\lvert~ g^{T}\eta g=\eta\,,~ {\rm det}g=1~\}\,,\\
    H&=SO(2,2)\times SO(2,2)=\{~g \in SO(4,4)~ \lvert ~g^{T}\eta' g=\eta'~\}\,,
\end{align}
where the invariant metrics $\eta$ and $\eta'$ for $G$ and $H$ are
\begin{align}
    \eta=
    \begin{pmatrix}
        0_4&1_4\\
        1_4&0_4
    \end{pmatrix}\,,\qquad
    \eta'=\text{diag}(-1,-1,1,1,-1,-1,1,1)\,.
\end{align}
For the explicit parametrization of $M(z,\rho)$, see Eqs.~(11) and (13) in Ref.~\cite{Sakamoto:2025xbq}.
It is known that the 2D sigma model is classically integrable \cite{Maison:1979kx}, and this fact allows us to exploit powerful solution-generating techniques to construct a wide class of exact higher-dimensional black hole solutions.
Two approaches are commonly employed, each associated with a distinct linear system that encodes the classical integrability of the underlying sigma model.
The first is the inverse-scattering method (ISM), based on the Belinski-Zakharov linear system~\cite{Belinsky:1979mh}, while the second involves the factorization of the monodromy matrix associated with the Breitenlohner-Maison (BM) linear system~\cite{Breitenlohner:1986um}.

\medskip
In this work, we employ the latter solution-generating technique developed in Refs.~\cite{Breitenlohner:1986um,Chakrabarty:2014ora,Katsimpouri:2012ky,Katsimpouri:2013wka,Katsimpouri:2014ara}, and we summarize only the essential elements of the procedure.
For detailed definitions and conventions, we refer the reader to our previous work~\cite{Sakamoto:2025xbq} (see also~\cite{Sakamoto:2025jtn}).
A key object in this framework is the monodromy matrix $\cM(w)$, a $G$-valued function that depends meromorphically on an auxiliary complex variable $w \in \mathbb{C}$, known as the spectral parameter.
It satisfies the algebraic constraints
\begin{align}\label{m-con}
    \cM^{-1}=\eta \cM^{T}\eta\,,\qquad   \cM^{\natural}=\cM\,,
\end{align}
where $\natural:G\to G$ is an anti-involutive automorphism 
\begin{align}
    x^{\natural}=\eta' x^{T}\eta'\qquad \text{for}\quad x\in G\,.
\end{align}
So far, there exists no general algorithm for systematically constructing the monodromy matrices associated with physically meaningful gravitational solutions.
However, for known exact solutions, the monodromy matrix $\cM(w)$ can be obtained from the coset representative $M(z,\rho)$.
For a sufficiently large positive real constant $R$, taking the limit $\rho \to 0^{+}$ of the coset matrix $M(z,\rho)$ in the region $z < -R$ yields the corresponding monodromy matrix~\cite{Breitenlohner:1986um}:
\begin{align}\label{sub-rule}
    \cM(w)=\lim_{\rho \to 0^+}M(z=w,\rho)\,.
\end{align}
Furthermore, we demonstrate that factorizing this monodromy matrix---as shown below Eq.~(\ref{fac-m})---precisely reproduces the coset  matrix $M(\rho,z)$ corresponding to  the original gravitational solutions.
Empirically, from many examples, it has been observed that the monodromy matrices corresponding to asymptotically flat, five-dimensional non-extremal black hole solutions can be expressed as meromorphic functions with only simple poles in $w$:
\begin{align}
    \cM(w)=Y_{\rm flat}+\sum_{i=1}^{N}\frac{A_i}{w-w_i}\,,
\end{align}
where the constant matrix $Y_{\rm flat} = Y_{\rm flat}^{\natural}$ characterizes the asymptotic flatness of the gravitational solution, and each residue matrix $A_i$ is a rank-2 constant matrix determined by the physical charges of the spacetime and the rod data.
Because certain scalar fields that parameterize $M(z,\rho)$ possess constant-shift ambiguities, the normalization of $Y_{\text{flat}}$ is not unique.
Following Ref.~\cite{Sakamoto:2025xbq}, we fix the gauge of the scalar fields such that the asymptotic constant matrix $Y_{\text{flat}}$ takes the form
\begin{align} \label{eq:yflat}
    Y_{\text{flat}}=
    \begin{pmatrix}
        1&0&0&0&0&0&0&0\\
        0&1&0&0&0&0&0&0\\
        0&0&0&0&0&0&0&1\\
        0&0&0&0&0&0&-1&0\\
        0&0&0&0&1&0&0&0\\
        0&0&0&0&0&1&0&0\\
        0&0&0&1&0&0&0&0\\
        0&0&-1&0&0&0&0&0
    \end{pmatrix}
    \,.
\end{align}
The positions $w_j$ of the simple poles is identified with the positions $z_i$ the intersection points of neighboring rods, and the total number $N$ of simple poles corresponds to the number of such intersection points.
In terms of the new spectral parameter $\lambda$,  defined  by 
\begin{align}\label{r-alg}
    \frac{1}{\la}-\la=\frac{2}{\rho}(w-z)\,,
\end{align}
we can find the factorization form of the monodromy matrix
\begin{align}\label{fac-m}
    \cM(w(\la,z,\rho))=X_-^{}(\la,z,\rho)M(z,\rho)X_+(\la,z,\rho)\,,
\end{align}
where the matrix-valued functions $X_+(\la,z,\rho)$ and $X_-(\la,z,\rho)=X_+^{\natural}(-1/\la,z,\rho)$ are required to satisfy the boundary conditions
\begin{align}\label{Xpm-bc}
    X_+(0,z,\rho)=1_{8\times 8}=X_-(\infty,z,\rho)\,.
\end{align}

\section{5D Myers-Perry solution}\label{sec:mp}

We begin by presenting the monodromy matrix corresponding to the five-dimensional Myers-Perry black hole with two independent angular momenta and subsequently perform its factorization.
The $SL(3,\mathbb{R})$ monodromy matrix for this solution was first obtained in Ref.~\cite{Chakrabarty:2014ora}.
In the present work, we extend that analysis to the $SO(4,4)$ case, which provides a broader framework applicable to the construction of a variety of charged black hole solutions through suitable transformations ${\cal M}\to {\cal M}'=g^\natural{\cal M}g$ for $g\in G$.

\begin{figure}
\begin{center}
\begin{tikzpicture}[scale=0.65]
\node[font=\small ] at (-10,3) {$t$};
\node[font=\small ] at (-10,2) {$\tilde{\phi}$};
\node[font=\small ] at (-10,1) {$\tilde{\psi}$};
\node[font=\small ] at (-10,0) {$z$};
\node[font=\small ] at (-6.5,1.5) {$(0,0,1)$};
\node[font=\small ] at (6.5,2.5) {$(0,1,0)$};
\node[font=\small ] at (0,3.5) {$(1,\omega_{\tilde{\phi}},\omega_{\tilde{\psi}})$};
\node[font=\small ] at (-4,-0.5) {$w_1$};
\node[font=\small ] at (4,-0.5) {$w_2$};
\draw[gray,line width = 0.8] (-9,3) -- (9,3);
\draw[black,line width = 5] (-4,3) -- (4,3);
\draw[gray,line width = 0.8] (-9,3) -- (9,3);
\draw[gray,line width = 0.8] (-9,1) -- (9,1);
\draw[black,line width = 5] (4,2) -- (8.4,2);
\draw[black,line width = 5,dashed ] (8.5,2) -- (9,2);
\draw[gray,line width = 0.8] (-9,2) -- (9,2);
\draw[black,line width = 5,dashed ] (-9,1) -- (-8.5,1);
\draw[black,line width = 5] (-8.4,1) -- (-4,1);
\draw[black,dashed ] (-4,0) -- (-4,3);
\draw[black,dashed ] (4,0) -- (4,3);
\draw[->,black,line width = 1] (-9,0) -- (9,0);
\end{tikzpicture}
\caption{Rod diagram for 5D Myers-Perry solution. The positions are $w_1=-\frac{1}{2}\alpha\,, w_2=\frac{1}{2}\alpha$ with $\alpha>0$. }\label{mp-rod}
\end{center}
\end{figure}
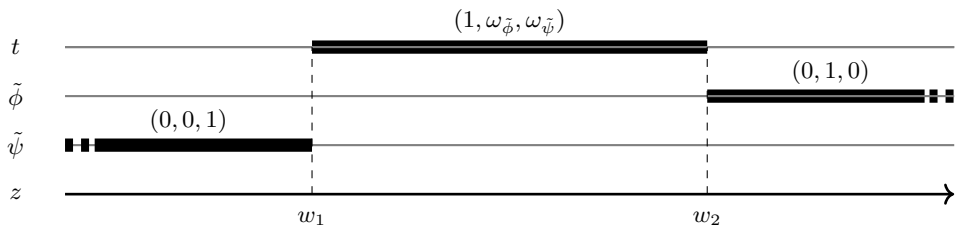

\subsection{5D solution}

The 5D Myers-Perry black hole solution with two angular momenta~\cite{Myers:1986un} is described by the metric
\begin{align}\label{mr-metric}
ds^2_{\rm MP}&=-dt^2+\frac{r_0^2}{\Sigma}\left[dt-a_1\,\sin^2\theta d\tilde{\phi}-a_2\cos^2\theta d\tilde{\psi}\right]^2\no\\
&\quad+(r^2+a_1^2)\sin^2\theta\,d\tilde{\phi}^2+(r^2+a_2^2)\,\cos^2\theta\, d\tilde{\psi}^2
+\frac{\Sigma}{\Delta}dr^2+\Sigma d\theta^2\,,    
\end{align}
where the parameters $r_0$ and $a_1,a_2$ denote the mass and rotational parameters, respectively, and the functions $\Delta$ and $\Sigma$ are defined as
\begin{align}\label{def-delta-sigma}
    \Delta&=r^2\left(1+\frac{a_1^2}{r^2}\right)\left(1+\frac{a_2^2}{r^2}\right)-r_0^2\,,\qquad
    \Sigma=r^2+a_1^2\cos^2\theta+a_2^2\sin^2\theta\,.
\end{align}
The angle variables $\theta\,, \tilde{\phi}$ and $\tilde{\psi}$ take values in the range
\begin{align}
   0\leq \theta \leq \frac{\pi}{2}\,,\qquad  0\leq \tilde{\phi}<2\pi\,,\qquad 0\leq \tilde{\psi}<2\pi\,.
\end{align}
The ADM mass and  two ADM angular momenta are written as
\begin{align}
\begin{split}\label{mp-asy}
    M&=\frac{3\pi}{8}r_0^2\,,\qquad J_1=\frac{\pi}{4}a_1 r_0^2\,,\qquad J_2=\frac{\pi}{4}a_2 r_0^2\,.
\end{split}
\end{align}

\medskip
To write down the corresponding monodromy matrix, we need to introduce the Weyl-Papapetrou coordinates $(\rho,z)$ defined as
\begin{align}
    \rho=\frac{1}{4}r \sqrt{\Delta} \sin2\theta\,,\qquad z=\frac{1}{4}r^2\left(1-\frac{r_0^2-a_1^2-a_2^2}{2r^2}\right)\cos2\theta\,.
\end{align}
The rod structure of this solution is shown in Fig.\,\ref{mp-rod}. It is characterized by the two intersection points
\begin{align}
     w_1=-\frac{1}{2}\alpha\,,\qquad w_2=\frac{1}{2}\alpha\,,
\end{align}
which divide the $z$-axis into three intervals (rods) : (i) the $\tilde{\psi}$-rotational axis: $I_1=\{(\rho,z) \lvert \rho=0,-\infty<z<w_1\}$, (ii) the horizon cross
section: $I_2=\{(\rho,z) \lvert \rho=0, w_1<z<w_2\}$, (iii) the $\tilde{\phi}$-rotational axis: $I_3=\{(\rho,z) \lvert \rho=0, w_2<z<\infty\}$. The rod
vector on the finite interval $I_2$ takes the form $v_2=(1,\omega_{\tilde{\phi}},\omega_{\tilde{\psi}})$ with the two angular velocities of the horizon, 
\begin{align}
    \omega_{\tilde{\phi}}=\frac{a_1^2-a_2^2-4\alpha+r_0^2}{2a_1r_0^2}\,,\qquad \omega_{\tilde{\psi}}=\frac{-a_1^2+a_2^2-4\alpha+r_0^2}{2a_2r_0^2}\,.
\end{align}
The rods vectors on the semi-infinite intervals $I_1$ and $I_3$ are given by $v_1= (0, 0, 1)$ and $v_3=(0,1,0)$, respectively,  corresponding to 
 the fixed-point sets of the $U(1)$ isometries generated by the Killing vectors $\partial_{\tilde{\psi}}$ and $\partial_{\tilde{\phi}}$, respectively.

\medskip
Furthermore, to express the corresponding coset matrix in a compact form, we define the prolate spherical coordinates $(x,y)$ as
\begin{align}
    x=\cos2\theta\,,\qquad y=\frac{2r^2+a_1^2+a_2^2-r_0^2}{4\alpha}-1\,,
\end{align}
where the real parameter $\alpha$ is
\begin{align}\label{eq:alpha}
    \alpha=\frac{1}{4}\sqrt{(r_0^2-a_1^2-a_2^2)^2-4a_1^2a_2^2}\,.
\end{align}
Here, $x$ and $y$ take values in the range 
\begin{align}
    -1\leq x \leq 1\,,\qquad  y\geq 1\,.
\end{align}
The relation between the Weyl-Papapetrou coordinates and the prolate spherical coordinates is given by
\begin{align}
    \rho=\frac{\alpha}{2}\sqrt{(1-x^2)(y^2-1)}\,,\qquad z=\frac{\alpha}{2}x y\,.
\end{align}
In terms of the prolate spherical coordinates, the metric~(\ref{mr-metric}) can be rewritten as
\begin{align}\label{mp-two}
    ds^2_{\rm MP}&=-\frac{H_-(x,y)}{H_+(x,y)}\left(dt-\Omega_{\tilde{\phi}}d\tilde{\phi}-\Omega_{\tilde{\psi}}d\tilde{\psi}\right)^2
    +\frac{H_+(x,y)}{8}
    \left(\frac{dx^2}{1-x^2}-\frac{dy^2}{y^2-1}\right)\no\\
   &\qquad +\frac{F_+(x,y)}{H_-(x,y)}d\tilde{\psi}^2
   -\frac{F_-(x,y)}{H_-(x,y)}d\tilde{\phi}^2+2\frac{J(x,y)}{H_-(x,y)}d\tilde{\phi}d\tilde{\psi}\,,
\end{align}
where $\Omega_{\tilde{\phi}}$ and $\Omega_{\tilde{\psi}}$ are given by
\begin{align}
    \Omega_{\tilde{\phi}}(x,y)&=-\frac{a_1r_0^2(1-x)}{H_-(x,y)}\,,\qquad
    \Omega_{\tilde{\psi}}(x,y)=-\frac{a_2r_0^2(1+x)}{H_-(x,y)}\,,
\end{align}
and the scalar functions $H_{\pm}(x,y), F_{\pm}(x,y)$ and $J(x,y)$ are defined as
\begin{align}
    H_{\pm}(x,y)&=4\alpha y+(a_1^2-a_2^2)x\pm r_0^2\,,\\
    F_{\pm}(x,y)&=\pm \frac{1\pm x}{4}\bigl(16\alpha^2 y^2\mp(a_1^2-a_2^2)^2 x+(a_1^2+a_2^2)r_0^2(1\pm x)\no\\
    &\qquad\mp 4\alpha(a_1^2-a_2^2)(1\mp x)y -r_0^4\bigr)\,,\\
    J(x,y)&=\frac{a_1a_2 r_0^2(1-x^2)}{2H_-(x,y)}\,.
\end{align}

\subsection{Coset space description}

To obtain the corresponding monodromy matrix, we first perform a dimensional reduction to three dimensions along the Killing directions $(t,\psi)$ and extract the 16 scalar fields $\{e^{2U},x^I,y^I,\zeta^{\Lambda},\tilde{\zeta}_{\Lambda},\sigma\}\,(I=1,2,3, \Lambda=0,1,2,3)$ that parametrize the coset matrix $M_{\rm MP}(z,\rho)\in SO(4,4)$, as described in the previous paper \cite{Sakamoto:2025xbq}.
The resulting 16 scalar fields are given by
\begin{align}\label{mr-scalar}
\begin{split}
 e^{2U}&=\frac{F_+(x,y)-F_-(x,y)-a_1a_2r_0^2(1-x^2)}{4\sqrt{H_+(x,y)H_-(x,y)}}\,,\quad
    x^I=0\,,\quad y^I=\sqrt{\frac{H_-(x,y)}{H_+(x,y)}}\,,\\
   \zeta^0&=\frac{\Omega_{\tilde{\psi}}(x,y)-\Omega_{\tilde{\phi}}(x,y)}{2}\,,\quad
   \tilde{\zeta}_{0}=-\frac{\Omega_{\tilde{\psi}}(-x,-y)-\Omega_{\tilde{\phi}}(-x,-y)}{2}\,,\quad   \zeta^I=0\,,\quad \tilde{\zeta}_{I}=0\,,\\
\sigma&=\frac{1}{8 (a_1-a_2)^2}\biggl[\frac{r_0^2 \left(\left(4 \alpha y+r_0^2\right)^2-(a_1-a_2)^4\right)}{H_+(x,y)}+\frac{r_0^2 \left(\left(r_0^2-4 \alpha y\right)^2-(a_1-a_2)^4\right)}{H_-(x,y)}\\
&\quad+2 x \left(a_1^2-a_2^2\right) \left(r_0^2-(a_1-a_2)^2\right)\biggr]+\alpha  \left(1-\frac{r_0^2}{(a_1-a_2)^2}\right)y\,.
\end{split}
\end{align}
The conformal factor $e^{2\nu}$ is given by
\begin{align}\label{conf-mp}
    e^{2\nu}=\frac{r_0^2(1-y^2)-(a_1+a_2)^2(x^2-y^2)}{(r_0^2-(a_1+a_2)^2)\,(x^2-y^2)}\,.
\end{align}
From the scalar fields (\ref{mr-scalar}), we can obtain the coset matrix $M_{\rm MP}(z,\rho)$, which approaches the following constant matrix at the spacial infinity $r\to\infty$:
\begin{align}\label{M-mp-y}
    \lim_{r\to \infty}M_{\rm MP}(z,\rho)= Y_{\text{flat}}\,.
\end{align}
By definitions, the twist potentials $\tilde{\zeta}_{\Lambda}$ and $\sigma$ have ambiguity under constant shifts but we fix the gauge such that $M_{\rm MP}(z,\rho)$ satisfies the asymptotic condition (\ref{M-mp-y}) with Eq.~(\ref{eq:yflat}).

\subsection{Monodromy matrix}

We now compute the monodromy matrix $\cM_{\rm MP}(w)$ corresponding to the five-dimensional Myers-Perry black hole.
According to the relation~(\ref{sub-rule}) between the monodromy matrix $\cM_{\rm MP}(w)$ and the coset matrix $M_{\rm MP}(z,\rho)$, the former can be obtained by taking the limit $\rho \to 0$ in the region where $z$ is sufficiently negative.
The resulting monodromy matrix $\cM_{\rm MP}(w)$ takes the form
\begin{align}\label{mr-mono}
    \cM_{\rm MP}(w)=Y_{\rm flat}+\sum_{i=1}^{2}\frac{A_i}{w-w_i}\,.
\end{align}
The explicit expressions of the residue matrices $A_j$ are given by 
\begin{align}
\begin{split}\label{mp-A}
     \eta'A_1&=
\left(
\begin{array}{cccccccc}
 \frac{r_0^2 \Delta_{++}}{32 \alpha} & 0 & 0 & \frac{a_1 r_0^2}{8 \alpha} & 0 & 0 &A_{71}^{+} & 0 \\
 0 & 0 & 0 & 0 & 0 & 0 & 0 & 0 \\
 0 & 0 & \frac{\Delta_{--}}{8 \alpha} & 0 & -\frac{a_2 r_0^2}{8 \alpha} & 0 & 0 & A_{83}^{+} \\
 \frac{a_1 r_0^2}{8 \alpha} & 0 & 0 & -\frac{\Delta_{-+}}{8 \alpha} & 0 & 0 & A_{74}^{+} & 0 \\
 0 & 0 & -\frac{a_2 r_0^2}{8 \alpha} & 0 & -\frac{r_0^2 \Delta_{+-}}{32 \alpha} & 0 & 0 & A_{85}^{+} \\
 0 & 0 & 0 & 0 & 0 & 0 & 0 & 0 \\
A_{71}^{+} & 0 & 0 & A_{74}^{+} & 0 & 0 &A_{77}^{+} & 0 \\
 0 & 0 & A_{83}^{+} & 0 & A_{85}^{+} & 0 & 0 & A_{88}^{+} \\
\end{array}
\right)\,,\\
     \eta'A_2&=\left(
\begin{array}{cccccccc}
 \frac{r_0^2 \Delta_{-+}}{32 \alpha} & 0 & 0 & -\frac{a_1 r_0^2}{8 \alpha} & 0 & 0 &A_{71}^{-} & 0 \\
 0 & 0 & 0 & 0 & 0 & 0 & 0 & 0 \\
 0 & 0 & \frac{\Delta_{+-}}{8 \alpha} & 0 & \frac{a_2 r_0^2}{8 \alpha} & 0 & 0 & A_{83}^{-} \\
 -\frac{a_1 r_0^2}{8 \alpha} & 0 & 0 & -\frac{\Delta_{++}}{8 \alpha} & 0 & 0 & A_{74}^{-} & 0 \\
 0 & 0 & \frac{a_2 r_0^2}{8 \alpha} & 0 & -\frac{r_0^2 \Delta_{--}}{32 \alpha} & 0 & 0 & A_{85}^{-} \\
 0 & 0 & 0 & 0 & 0 & 0 & 0 & 0 \\
A_{71}^{-} & 0 & 0 & A_{74}^{-} & 0 & 0 &A_{77}^{-} & 0 \\
 0 & 0 & A_{83}^{-} & 0 & A_{85}^{-} & 0 & 0 & A_{88}^{-} \\
\end{array}
\right)\,,
\end{split}
\end{align}
where we have introduced the functions: 
\begin{align}
\begin{split}
    \Delta_{\pm\pm}&=4\alpha\pm(a_1^2-a_2^2\pm r_0^2)\,,\qquad \Delta_{\pm\mp}=4\alpha\pm(a_1^2-a_2^2\mp r_0^2)\,,\\
    A_{71}^{\pm}&=-\frac{r_0^2 \left((a_1-a_2)\Delta_{\pm+} \mp a_1 r_0^2\right)}{64 \alpha}\,,\\
    A_{74}^{\pm}&=-\frac{r_0^2 \left(\Delta_{\mp-}\pm 2 (a_1-a_2)^2\right)}{64 \alpha}\,,\\
    A_{77}^{\pm}&=\frac{r_0^2 \left(\left(4 (a_1-a_2)^2-r_0^2\right)\Delta_{\pm-} \pm 2 r_0^2 (a_1-a_2) (a_1-3 a_2)\right)}{512 \alpha}\,,\\
    A_{83}^{\pm}&=\frac{r_0^2 \left(\Delta_{\pm +}\mp 2 (a_1-a_2)^2\right)}{64 \alpha}\,,\\
    A_{85}^{\pm}&=\frac{r_0^2 \left((a_1-a_2)\Delta_{\pm -} \mp a_2 r_0^2\right)}{64 \alpha}\,,\\
    A_{88}^{\pm}&=-\frac{r_0^2 \left(\left(4 (a_1-a_2)^2-r_0^2\right)\Delta_{\pm+} \mp2 r_0^2 (a_1-a_2) (3 a_1-a_2)\right)}{512 \alpha}\,.
\end{split}
\end{align}
While the residue matrices in the $SL(3,\mathbb{R})$ case have rank one~\cite{Chakrabarty:2014ora}, both residue matrices $A_j$ in the $SO(4,4)$ case are of rank two.
As observed in many examples, the pole positions of the monodromy matrix correspond to the corner points of the rod structure, as illustrated in Fig.~\ref{mp-rod}.

\subsubsection*{Charge matrix}

To examine how the geometric structure of the 5D Myers-Perry black hole is encoded in the monodromy matrix, we compute the charge matrix $Q\in \mathfrak{so}(4,4)$, which characterizes the asymptotic behavior of the monodromy matrix in the large spectral parameter region defined as \cite{Chakrabarty:2014ora}\footnote{The charge matrix was originally introduced to characterize the asymptotic behavior of the coset matrix $M(z,\rho)$ at spatial infinity (e.g., for asymptotically flat four-dimensional black holes \cite{Bossard:2009at}).}
\begin{align}
    \cM(w)= Y_{\rm flat}\left(1+\frac{Q}{w}+\cO(w^{-2})\right)\,.
\end{align}
In this case, the charge matrix $Q$ is given by
\begin{align}
    Q&=Y_{\rm flat}^{-1}\left(\sum_{j=1}^{2}A_j\right)\in \mathfrak{so}(4,4)\,.
\end{align}
From the expressions (\ref{mp-A}) of $A_j$, $Q$ can be expanded as
\begin{align}\label{mp-charge-expansion}
    Q&=-\frac{1}{8}r_0^2 \sum_{j=1}^{3}H_j
    +\frac{1}{64}(r_0^2-4(a_1-a_2)^2)r_0^2E_0+F_0+\frac{1}{8}(a_1-a_2)r_0^2(E_{p_0}+E_{q^0})\,,
\end{align}
where $\{H_j,E_{\Lambda},E_{p_{\Lambda}},E_{q^{\Lambda}},F_{\Lambda},F_{p_{\Lambda}},F_{q^{\Lambda}}\}\,(\Lambda=0,1,2,3)$ is the basis of $\mathfrak{so}(4,4)$ and we employ the matrix representation given in \cite{Sakamoto:2025jtn}.
A part of this charge matrix is uniquely determined by the asymptotic quantities of the black hole solution, and we can verify that it satisfies the following general form \cite{Sakamoto:2025xbq}:
\begin{align}\label{gQ-ex}
    Q&=-\frac{M}{3 \pi}\sum_{j=1}^{3}H_j+Q_{E_0} E_0+F_0 +\frac{J_1-J_2}{2 \pi} (E_{p_0}+E_{q^0})\,.
\end{align}
Here, the constant $Q_{E_0}$ depends on the parameters of the gravitational solution, although its explicit relation to the asymptotic conserved quantities has not yet been fully clarified~\footnote{If we require the general expression (\ref{gQ-ex}) of the charge matrix satisfies the cubic relation (\ref{mp-q-rel}), the constant $Q_{E_0}$ is expressed in terms of the asymptotic quantities: $Q_{E_0}=\frac{M^2}{9\pi^2}-\frac{3(J_1-J_2)^2}{8M\pi}$. However, as we shall see in the next section, this cubic relation is modified for the black ring solution, and therefore it does not hold in general for asymptotically flat vacuum solutions of five-dimensional black holes. }.
We can also show that the charge matrix $Q$ satisfies the cubic relation
\begin{align}\label{mp-q-rel}
    Q^3-\frac{1}{4}\Tr(Q^2)Q=0\,,
\end{align}
where
\begin{align}
    \Tr(Q^2)=\frac{1}{4}\left(r_0^2-(a_1-a_2)^2\right)r_0^2\,.
\end{align}
In the $SL(3,\mathbb{R})$ case~\cite{Chakrabarty:2014ora}, the charge matrix also satisfies a similar cubic relation.
However, an essential difference arises in the $SO(4,4)$ case: unlike the $SL(3,\mathbb{R})$ charge matrix, the $SO(4,4)$ charge matrix~(\ref{mp-charge-expansion}) explicitly contains the angular momentum parameters $a_1$ and $a_2$.
This allows for a more direct investigation of the correspondence between the nilpotency condition of the charge matrix and the extremality condition of the doubly rotating Myers-Perry black hole.
When the parameter $r_0$ takes one of the following values,
\begin{align}
    r_0=0\,,\qquad \text{or}\qquad r_0^2=(a_1-a_2)^2\,,
\end{align}
the charge matrix $Q$ becomes nilpotent $Q^3=0$.
On the other hand, the horizons are located at the roots of $\Delta=0$, defined in~(\ref{def-delta-sigma}), and are given by
\begin{align}
    r_{\pm}^{2}=\frac{1}{2}\Bigl[r_0^2-a_1^2-a_2^2\pm \sqrt{(r_0^2-a_1^2-a_2^2)^2-4a_1^2a_2^2}\Bigr]=\frac{1}{2}\Bigl[r_0^2-a_1^2-a_2^2\pm\alpha\Bigr]\,.
\end{align}
Taking the second condition $r_0^2=(a_1-a_2)^2$ corresponds to an extremal limit $r_+\to r_-$. The metric is regular only when both angular momenta are non-zero.
It should be noted that the nilpotent condition of the charge matrix (\ref{mp-charge-expansion}) covers only one branch of the two extremal conditions for the Myers-Perry black hole, while in the other branch $r_0^2=(a_1+a_2)^2$ the charge matrix is not nilpotent.

\subsection{Factorization of monodromy matrix}

We explicitly perform the factorization of the monodromy matrix (\ref{mr-mono}) by following the procedure developed in \cite{Katsimpouri:2013wka}.
To this end, we express the residue matrices $A_j$ in terms of the eight-component vectors $a_j$ and $b_j$
\begin{align}\label{rank2-mat}
    A_j=\alpha_j(a_j\otimes a_j)\eta'-\beta_j((\eta b_j)\otimes (\eta b_j))\eta'\,,
\end{align}
where $\alpha_j$ and $\beta_{j}$ are constants.
The constant vectors $a_j$ and $b_j$ satisfy
\begin{align}\label{ab-cond}
    a_j^{T}\eta a_j=0\,,\qquad  b_j^{T}\eta b_j=0\,,\qquad a_j^{T}b_j=0\,.
\end{align}
The vectors $a_j$ and $b_j$ can be constructed from the eigenvectors of matrix $A_j$ with the non-zero eigenvalues, and they are taken as
\begin{align}
\begin{split}\label{mp-abvec}
    a_1^{T}&=\left(-\frac{8 a_1}{\Delta_{+-}+2 (a_1-a_2)^2},0,0,\frac{8 a_1}{\Delta_{++} (a_1-a_2)-a_1 r_0^2},0,0,1,0\right)\,,\\
    b_1^{T}&=\left(0,0,\frac{8 a_2}{ (a_1-a_2)\Delta_{+-}-a_2 r_0^2},0,\frac{8 a_2}{\Delta_{++}-2 (a_1-a_2)^2},0,0,1\right)\eta\,,\\
    a_2^{T}&=\left(\frac{8 a_1}{\Delta_{--}-2 (a_1-a_2)^2},0,0,-\frac{8 a_1}{(a_1-a_2)\Delta_{-+} +a_1 r_0^2},0,0,1,0\right)\,,\\
    b_2^{T}&=\left(0,0,-\frac{8 a_2}{ (a_1-a_2)\Delta_{--}+a_2 r_0^2},0,-\frac{8 a_2}{\Delta_{-+}+2 (a_1-a_2)^2},0,0,1\right)\eta\,,
\end{split}
\end{align}
and the constants $\alpha_j$ and $\beta_j$ are given by
\begin{align}
    \alpha_1&=A_{77}^{+}\,,\qquad 
    \beta_1=-A_{88}^{+}\,,\qquad
    \alpha_2=A_{77}^{-}\,,\qquad \beta_2=-A_{88}^{-}\,.
\end{align}
By following \cite{Katsimpouri:2013wka}, we take ansatz of the matrix-valued function $X_+(\la,z,\rho)$ in the factorized monodromy matrix (\ref{fac-m}) as
\begin{align}\label{xp-ex}
    X_+(\la,z,\rho)=1-\sum_{j=1}^{2}\frac{\la C_j}{1+\la \la_j}\,,
\end{align}
where each residue $C_j$ is defined as
\begin{align}\label{c-def}
    C_j=(c_j\otimes a_j)\eta'-\left((\eta d_j)\otimes (\eta b_j)\right)\eta'\,.
\end{align}
The vectors $c_j$ and $d_j$ are constructed by solving the equations \cite{Katsimpouri:2013wka,Sakamoto:2025jtn}
\begin{align}\label{abcd-eq2}
    \eta'a&=d\,\Gamma^{(0)T}-(\eta c) \Gamma^{(a)T}\,,\qquad \eta'b=c\Gamma^{(0)}+(\eta d)\,\Gamma^{(b)T}\,,
\end{align}
where the $8\times 2$ matrices $a,b,c,d$ are
\begin{align}
    a&=(a_1,a_2)\,,\quad  b=(b_1,b_2)\,,\quad c=(c_1,c_2)\,,\quad  d=(d_1,d_2)\,.
\end{align}
The $2\times 2$ matrices $\Gamma^{(0)}$ and $\Gamma^{(a)}\,, \Gamma^{(b)}$ are expressed in terms of the vectors $a_j,b_j$, and their definitions can be found in Sec.\,3 in \cite{Sakamoto:2025jtn}. These matrices corresponding to the vectors (\ref{mp-abvec}) are given by
\begin{align}
\begin{split}\label{gamma-mp}
    \Gamma_{11}^{(0)}&=\frac{32 \left((a_1+a_2)^2 \left(r_0^4-5 r_0^2 a_{12}^2+4 a_{12}^4\right)-4 \alpha \left(a_1^2-a_2^2\right) \left(4 a_{12}^2-3 r_0^2\right)\right)}{r_0^4 \left(r_0^4 (2 a_1-a_2) (a_1-2 a_2)-6 r_0^2 a_{12}^4+4 a_{12}^6\right)}\frac{1}{\la_1\nu_1}\,,\\
    \Gamma_{22}^{(0)}&=\frac{32 \left((a_1+a_2)^2 \left(r_0^4-5 r_0^2 a_{12}^2+4 a_{12}^4\right)+4 \alpha \left(a_1^2-a_2^2\right) \left(4 a_{12}^2-3 r_0^2\right) \right)}{r_0^4 \left(r_0^4 (2 a_1-a_2) (a_1-2 a_2)-6 r_0^2 a_{12}^4+4 a_{12}^6\right)}\frac{1}{\la_2\nu_2}\,,\\
    \Gamma_{12}^{(0)}&=-\frac{32 \left(4r_0^2\alpha^2 + (a_{12}^2-r_0^2)(4a_{12}^2-r_0^2 ) \alpha\right)}{r_0^2 \left(r_0^4 (2 a_1-a_2) (a_1-2 a_2)-6 r_0^2 a_{12}^4+4 a_{12}^6\right)}\frac{1}{\la_{12}}\,,\\
    \Gamma_{21}^{(0)}&=-\frac{32 \left(-4r_0^2\alpha^2 + (a_{12}^2-r_0^2)(4a_{12}^2-r_0^2 ) \alpha\right)}{r_0^2 \left(r_0^4 (2 a_1-a_2) (a_1-2 a_2)-6 r_0^2 a_{12}^4+4 a_{12}^6\right)}\frac{1}{\la_{12}}\,,
\end{split}
\end{align}
and
\begin{align}
    \Gamma^{(a)}=\Gamma^{(a)}=0_{2\times 2}\,.
\end{align}
In contrast to the static case, the matrix $\Gamma^{(0)}$ in this setting possesses non-zero diagonal components. 
Then, by computing the matrix $X_+$ from the above matrices $C_j$, we can show that the monodromy matrix $\cM_{\rm MP}(w)$ can be factorized
\begin{align}\label{mon-fac-mr}
    \cM_{\rm MP}(w)=X_-(\la,z,\rho)M_{\rm MP}(z,\rho)X_+(\la,z,\rho)\,.
\end{align}

\subsubsection*{Conformal factor}

Finally, we compute the conformal factor $e^{2\nu}$.
Since $\Gamma^{(a,b)}=0_{2\times 2}$, the conformal factor $e^{2\nu}$ can be obtained by using the simplified formula \cite{Katsimpouri:2013wka} (see also \cite{Katsimpouri:2012ky})
\begin{align}\label{conf}
     e^{2\nu}&=k_{\rm BM}\prod_{j=1}^{2}(\la_j\nu_j)\,{\rm det}(\Gamma^{(0)})\,,
\end{align}
where $k_{\rm BM}$ is the integration constant and $\nu_j$ is defined as
\begin{align}
      \nu_j=-\frac{2}{\rho\left(\la_j+\la_j^{-1}\right)}\,.
\end{align}
By evaluating the determinant of $\Gamma^{(0)}$ given in (\ref{gamma-mp}), the formula (\ref{conf}) precisely leads to the conformal factor (\ref{conf-mp}) for the 5D Myers-Perry black hole by taking the overall constant $k_{\rm BM}$ as
\begin{align}
    &k_{\rm BM}=\frac{r_0^6 \left(r_0^2 (a_1-2 a_2)-2 (a_1-a_2)^3\right) \left(r_0^2 (2 a_1-a_2)-2 (a_1-a_2)^3\right)}{4096 \left(r_0^2-(a_1+a_2)^2\right) \left(r_0^2-(a_1-a_2)^2\right)}\,.
\end{align}

\subsection{Monodromy matrix for extremal limit and its factorization}

Before closing this section, let us consider to take the extremal limit $r_0^2 \to (a_1-a_2)^2 $ for the monodromy matrix~(\ref{mr-mono})  of the five-dimensional Myers-Perry black hole and its factorization .
In this limit, the two simple poles $z = \pm \alpha$ of the monodromy matrix~(\ref{mr-mono}) collapse into a single pole at $w = 0$, since the extremal condition $r_0^2=(a_1-a_2)^2$ is one of the roots of the quadratic equation $\alpha^2=(r_0^2-a_1^2-a_2^2)^2-4a_1^2a_2^2=0$ with respect to $r_0^2$.
Therefore, the mondoromy matrix (\ref{mr-mono}) becomes 
\begin{align}\label{exmp-mono}
    \cM_{\rm exMP}(w)=Y_{\rm flat}+\frac{A^{(1)}}{w}+\frac{A^{(2)}}{w^2}\,,
\end{align}
where the residue matrices are
\begin{align}\label{mpex-res1}
A^{(1)}&=\left(
\begin{array}{cccccccc}
 -\frac{1}{4} a_{12}^2 & 0 & 0 & 0 & 0 & 0 & \frac{1}{8} a_{12}^3 & 0 \\
 0 & 0 & 0 & 0 & 0 & 0 & 0 & 0 \\
 0 & 0 & -1 & 0 & 0 & 0 & 0 & -\frac{1}{8} a_{12}^2 \\
 0 & 0 & 0 & -1 & 0 & 0 & -\frac{1}{8} a_{12}^2 & 0 \\
 0 & 0 & 0 & 0 & \frac{1}{4} a_{12}^2 & 0 & 0 & -\frac{1}{8} a_{12}^3 \\
 0 & 0 & 0 & 0 & 0 & 0 & 0 & 0 \\
 \frac{1}{8} a_{12}^3 & 0 & 0 & \frac{1}{8} a_{12}^2 & 0 & 0 & -\frac{3}{64} a_{12}^4 & 0 \\
 0 & 0 & \frac{1}{8} a_{12}^2 & 0 & \frac{1}{8} a_{12}^3 & 0 & 0 & -\frac{3}{64} a_{12}^4 \\
\end{array}
\right)\,,\\
    A^{(2)}&=\left(
\begin{array}{cccccccc}
 \frac{1}{16} a_{1} a_{12}^3 & 0 & 0 & \frac{1}{8} a_{1} a_{12}^2 & 0 & 0 & -\frac{1}{64} a_{1} a_{12}^4 & 0 \\
 0 & 0 & 0 & 0 & 0 & 0 & 0 & 0 \\
 0 & 0 & -\frac{1}{4} a_{2}a_{12} & 0 & -\frac{1}{8} a_{2} a_{12}^2 & 0 & 0 & \frac{1}{32} a_{2} a_{12}^3 \\
 -\frac{1}{8} a_{1} a_{12}^2 & 0 & 0 & -\frac{1}{4} a_{1} a_{12} & 0 & 0 & \frac{1}{32} a_{1} a_{12}^3 & 0 \\
 0 & 0 & -\frac{1}{8} a_{2} a_{12}^2 & 0 & -\frac{1}{16} a_{2} a_{12}^3 & 0 & 0 & \frac{1}{64} a_{2} a_{12}^4 \\
 0 & 0 & 0 & 0 & 0 & 0 & 0 & 0 \\
 -\frac{1}{64} a_{1} a_{12}^4 & 0 & 0 & -\frac{1}{32} a_{1} a_{12}^3 & 0 & 0 & \frac{1}{256} a_{1} a_{12}^5 & 0 \\
 0 & 0 & -\frac{1}{32} a_{2} a_{12}^3 & 0 & -\frac{1}{64} a_{2} a_{12}^4 & 0 & 0 & \frac{1}{256} a_{2} a_{12}^5 \\
\end{array}
\right)\,.\label{mpex-res2}
\end{align}
The rank of these matrices are
\begin{align}
    \text{rank}\,A^{(1)}=4\,,\qquad  \text{rank}\,A^{(2)}=2\,.
\end{align}
Since the monodromy matrix~(\ref{exmp-mono}) contains a double pole in the spectral parameter $w$, the standard factorization procedure of Ref.~\cite{Katsimpouri:2013wka} cannot be applied directly.
Although a general method for treating such degenerate cases was discussed in Ref.~\cite{Camara:2017hez}, in the present situation the factorization can be carried out straightforwardly.
In what follows, we explicitly perform the factorization of the degenerate monodromy matrix~(\ref{exmp-mono}) and show that the resulting coset matrix correctly reproduces the metric of the five-dimensional extremal Myers-Perry black hole.

\medskip
To do this, we first define as in the charge matrix
\begin{align}
    Q^{(1)}=Y_{\rm flat}^{-1}A^{(1)}\,,\qquad 
     Q^{(2)}=Y_{\rm flat}^{-1}A^{(2)}\,.
\end{align}
We find that the charge matrices satisfy the relations
\begin{align}
    (Q^{(1)})^3=0\,,\qquad (Q^{(2)})^2=0\,,
\end{align}
and
\begin{align}
    [Q^{(1)},Q^{(2)}]=0\,,\qquad  [Q^{(1)},(Q^{(1)})^2]=0\,,\qquad  [Q^{(2)},(Q^{(1)})^2]=0\,.
\end{align}
Thanks to the algebraic properties, the monodromy matrix (\ref{exmp-mono}) can be rewritten in the exponential form
\begin{align}
    \cM_{\rm exMP}(w)=Y_{\rm flat}\exp\left[\frac{1}{w}Q^{(1)}+\frac{1}{w^2}\widetilde{Q}^{(2)}\right]\,.
\end{align}
where we defined
\begin{align}
    \widetilde{Q}^{(2)}=Q^{(2)}-\frac{1}{2}(Q^{(1)})^2\,.
\end{align}
The charge matrix $Q^{(1)}$ is obtained as the extremal limit of (\ref{mp-charge-expansion}) and belongs to $\mathfrak{so}(4,4)$. In contrast, $Q^{(2)}$ itself does not lie in $\mathfrak{so}(4,4)$, but the combination $\widetilde{Q}^{(2)}$ does.
These charge matrices are expanded as
\begin{align}
Q^{(1)}&=-\frac{M}{3 \pi}(H_1+H_2+H_3)-\frac{3}{64} a_{12}^4 E_0 +\frac{J_1-J_2}{2 \pi} (E_{p_0}+E_{q^0})+F_0\,,\\
\widetilde{Q}^{(2)}&=\frac{1}{8} \left(a_1^2-a_2^2\right)\Bigl(\frac{1}{8}  a_{12}^2(H_1+H_2+H_3)+\frac{1}{64} a_{12}^4 E_0-\frac{1}{16} a_{12}^3 (E_{p_0}+E_{q^0})+\frac{1}{2} a_{12} (F_{p_0}+F_{q^0})+F_0\Bigr)\,,
\end{align}
where $M$ and $J_{1,2}$ are the asymptotic quantities (\ref{mp-asy}) with $r_0^2=(a_1-a_2)^2$.
Next, we rewrite the spectral parameter $w$ in terms of another coordinate dependent spectral parameter $\la$ and the Weyl-Papapetrou coordinates $(\rho,z)$.
By denoting $\la_0$ by $\la(w=0;z, \rho)$ in $\la=\la(w;z,\rho)=\frac{1}{\rho}\left[(z-w)+ \sqrt{(z-w)^2+\rho^2}\right]$, we express the inverse of $w$ as 
\begin{align}\label{w-la-map}
\begin{split}
    \frac{1}{w}&=\nu_0\left( \frac{\la_0}{\la-\la_0}+\frac{1}{1+\la \la_0}\right)\,,\\
     \nu_0&=-\frac{2}{\rho\left(\la_0+\la_0^{-1}\right)}=-\frac{1}{\sqrt{\rho^2+z^2}}\,.
\end{split}
\end{align}
By using the commutativity of the charge matrices, we can factorize 
\begin{align}
    \cM_{\rm exMP}(w)=X_-(\la;z,\rho) M_{\rm exMP}(z,\rho)X_+(\la;z,\rho)\,.
\end{align}
Here, the coset matrix $M_{\rm exMP}(z,\rho)$ is
\begin{align}\label{exmp-coset}
   M_{\rm exMP}(z,\rho)= Y_{\rm flat}\exp\left[-\frac{1}{\sqrt{\rho^2+z^2}}Q^{(1)}-\frac{z}{(z^2+\rho^2)^{\frac{3}{2}}}\widetilde{Q}^{(2)}\right]\,,
\end{align}
and the matrix $X_{+}(\la;z,\rho)$ is given by
\begin{align}
\begin{split}
    X_+&=\exp\left(-\frac{\nu_0 \la \la_0}{1+\la \la_0}Q^{(1)}+\biggl[\frac{\nu_0 \la}{1+\la \la_0}\frac{\rho}{z^2+\rho^2}+\left(\frac{\nu_0\la \la_0}{1+\la \la_0}\right)^2\biggr]\widetilde{Q}^{(2)}\right)\,.
\end{split}
\end{align}

\medskip
By comparing the coset matrix with the parametrization (see Eq.\,(11) in \cite{Sakamoto:2025xbq}), we can read off the scalar fields 
\begin{align}\label{mrex-scalar2}
\begin{split}
 e^{2U}&=\frac{8(\rho^2+z^2)^{3/2}}{\sqrt{H_+H_-}}\,,\quad
    x^I=0\,,\quad y^I=\sqrt{\frac{H_-}{H_+}}\,,\\
   \zeta^0&=-\frac{a_{12}^2\left(a_1(z-\sqrt{z^2+\rho^2})+a_2(z+\sqrt{z^2+\rho^2})\right)}{2H_-}\,,\\
   \tilde{\zeta}_{0}&=\frac{a_{12}^2\left(a_1(z+\sqrt{z^2+\rho^2})+a_2(z-\sqrt{z^2+\rho^2})\right)}{2H_+}\,,\quad   \zeta^I=0\,,\quad \tilde{\zeta}_{I}=0\,,\\
\sigma&=\frac{4(\rho^2+z^2)^{3/2}\left(2(H_++H_-)-a_{12}^4\right)}{H_+H_-}\,.
\end{split}
\end{align}
where we defined the scalar functions
\begin{align}
    H_{\pm}=8 \left(\rho ^2+z^2\right)+z \left(a_1^2-a_2^2\right)\pm(a_1-a_2)^2 \sqrt{\rho^2+z^2}\,.
\end{align}
From the nilpotency of the charge matrices, it follows immediately that these matrices are orthogonal
\begin{align}
\Tr(Q^{(1)}Q^{(1)})=\Tr(Q^{(1)}\widetilde{Q}^{(2)})=\Tr(\widetilde{Q}^{(2)}\widetilde{Q}^{(2)})=0\,.
\end{align}
Since the equation of motion for the conformal factor is $\partial_{m}(\ln e^{2\nu})\propto \Tr(\partial_{m}MM^{-1}\partial_{m}MM^{-1})$, 
this means that the conformal factor $e^{2\nu}$ is trivial i.e.
\begin{align}
    e^{2\nu}=1\,,
\end{align}
where the normalization is fixed by the asymptotic flatness condition. 
The one-form fields $\hat{A}^0$ and $\omega_3$ in the corresponding 5D metric (\ref{new-angle}) can be constructed by solving the Hodge dual relations for the scalar fields $\tilde{\zeta}_{\Lambda}$ and $\sigma$ (For the details, see for example appendix A in \cite{Sakamoto:2025xbq}).
The resulting expressions of $\hat{A}^0$ and $\omega_3$ are
\begin{align}\label{exmp-one-form}
    \hat{A}^0=\frac{\rho ^2 (a_1-a_2)^2 (a_1+a_2)}{16 \left(\rho ^2+z^2\right)^{3/2}}d\phi\,,\qquad \omega_3=-\frac{8z(\rho^2+z^2)-\rho ^2 \left(a_1^2-a_2^2\right)}{8 \left(\rho ^2+z^2\right)^{3/2}}d\phi\,.
\end{align}
By constructing the corresponding five-dimensional metric from the scalar and one-form fields, and then performing the coordinate transformation\footnote{With the coordinate transformation (\ref{xytow-exmp}), we have $H_+=\frac{1}{2}(r^2+a_1a_2)\Delta$.}
\begin{align}\label{xytow-exmp}
    \rho=\frac{1}{4}(r^2+a_1a_2)\sin2\theta\,,\qquad z=\frac{1}{4}(r^2+a_1a_2)\cos2\theta\,,
\end{align}
we find that the constructed metric coincides with the one of the extremal Myers-Perry black hole.
Thus, the nilpotency condition of the charge matrix seems to be related to the extremal limit of the corresponding black hole solution. However, as we shall see in the black ring case, this relation turns out not to be so straightforward.

\medskip
Finally, let us comment on a possible extension an extremal Myers-Perry black hole to multi-center solutions. 
Taking the coset matrix~(\ref{exmp-coset}) as a starting point, we now consider the below more general coset matrix, 
\begin{align}\label{mexmp-coset}
   M_{\rm ex}(z,\rho)= Y_{\rm flat}\exp\left[f_1(z,\rho)\,Q^{(1)}+f_2(z,\rho)\,\widetilde{Q}^{(2)}\right]\,,
\end{align}
where the equation of motion for the conformal factor again holds by taking $e^{2\nu}=1$. 
It should be noted that the $ M_{\rm ex}(z,\rho)$ follows
\begin{align}
 \partial_{\rho}\left(\rho  (\partial_{\rho}M_{\rm ex}) M_{\rm ex}^{-1}\right)+ \partial_{z}\left(\rho  (\partial_{z}M_{\rm ex}) M_{\rm ex}^{-1}\right)=\rho \,Y_{\rm flat}\Bigl[\nabla^2f_1\, Q^{(1)}+\nabla^2f_2\, \widetilde{Q}^{(2)}\Bigr]Y_{\rm flat}^{-1}\,,
\end{align}
where $\nabla^2$ is the cylindrical Laplacian operator of a 3D Euclidean space 
\begin{align}
    \nabla^2=\frac{1}{\rho}\partial_{\rho}(\rho \partial_{\rho})+\partial_z^2+\frac{1}{\rho^2}\partial_{\varphi}^2\,.
\end{align}
Therefore, if the two functions $f_1$ and $f_2$ are chosen to be harmonic functions on 3D Euclidean space, the coset matrix~(\ref{mexmp-coset}) constitutes a solution of the theory~(\ref{2d-gravity-sigma}), and hence the 5D metric obtained through the dualization from the scalar fields to the metric is a solution to the 5D vacuum Einstein equations.
It then follows that a generalization to multi-center solutions can be obtained by extending these functions to include multiple sources
\begin{align}\label{eq:mutil-source}
    f_1= \sum_{i=1}^{n}\frac{1}{\sqrt{\rho^2+(z-z_i)^2}}\,,\qquad f_2= \sum_{i=1}^{n}\frac{z-z_i}{\sqrt{\rho^2+(z-z_i)^2}^3}\,.
\end{align}
This is similar to the earlier works~\cite{Clement:1986bt,Clement:1985gm} of Cl\'ement, who constructed a broad class of coset matrices on $SL(3,{\bm R})/SO(3)$ depending on two harmonic functions for 5D asymptotically Kaluza-Klein spacetime, in which the $Y_{\rm flat}$ takes a different form. 
In fact, the exact solution describing multi-centered rotating Kaluza-Klein black hole solutions were later constructed from the coset matrix for the extremal under-rotating Kaluza-Klein black hole solution~\cite{Teo:2023wfd,Tomizawa:2025tvb}. 
However, at this stage it remains unclear whether the solutions (\ref{mexmp-coset}) and (\ref{eq:mutil-source}) constructed in this way are indeed regular.
A detailed analysis of their regularity and related properties will be presented in our forthcoming paper.

\section{Unbalanced Pomeransky-Sen'kov black ring}\label{sec:ups}

Next, we present the explicit form of the monodromy matrix corresponding to the doubly rotating black ring solution~\cite{Pomeransky:2006bd,Tomizawa:2005wv,Chen:2011jb} and  show factorizing the monodromy matrix can reproduce the black ring solution.
The balanced doubly rotating black ring—free of conical singularities inside the ring—was first obtained by Pomeransky and Sen’kov~\cite{Pomeransky:2006bd}, and later generalized to unbalanced configurations~\cite{Morisawa:2007di,Chen:2011jb}.
The unbalanced doubly rotating black ring admits a number of physically significant limits, reducing to the Pomeransky-Sen'kov black ring (the balanced doubly rotating black ring), the Emparan-Reall black ring (the balanced $S^1$-rotating black ring), the Mishima-Iguchi-Figueras black ring (the unbalanced $S^2$-rotating black ring), and the five-dimensional Myers-Perry black hole (the doubly rotating black hole). 
After constructing the monodromy matrix corresponding to the most general case - the unbalanced doubly rotating black ring - we proceed, in the following section, to its various limiting cases of the monodromy matrix, which reproduce those of several physically important solutions: the Pomeransky-Sen'kov black ring, the extremal Pomeransky-Sen'kov black ring, the five-dimensional Myers-Perry black hole, the Emparan-Reall black ring, and the Mishima-Iguchi-Figueras black ring.

\subsection{Unbalanced Pomeransky-Sen'kov black ring solution}

\begin{figure}
\begin{center}
\begin{tikzpicture}[scale=0.65]
\node[font=\small ] at (-10,3) {$t$};
\node[font=\small ] at (-10,2) {$\tilde{\phi}$};
\node[font=\small ] at (-10,1) {$\tilde{\psi}$};
\node[font=\small ] at (-10,0) {$z$};
\node[font=\small ] at (-6.5,1.5) {$(0,0,1)$};
\node[font=\small ] at (2,1.5) {$(0,0,1)$};
\node[font=\small ] at (6.5,2.5) {$(0,1,0)$};
\node[font=\small ] at (-2,3.5) {$(1,\omega_{\tilde{\phi}},\omega_{\tilde{\psi}})$};
\node[font=\small ] at (-4,-0.5) {$w_1$};
\node[font=\small ] at (0,-0.5) {$w_2$};
\node[font=\small ] at (4,-0.5) {$w_3$};
\draw[gray,line width = 0.8] (-9,3) -- (9,3);
\draw[black,line width = 5] (-4,3) -- (0,3);
\draw[gray,line width = 0.8] (-9,3) -- (9,3);
\draw[gray,line width = 0.8] (-9,1) -- (9,1);
\draw[black,line width = 5] (4,2) -- (8.4,2);
\draw[black,line width = 5,dashed ] (8.5,2) -- (9,2);
\draw[gray,line width = 0.8] (-9,2) -- (9,2);
\draw[black,line width = 5,dashed ] (-9,1) -- (-8.5,1);
\draw[black,line width = 5] (-8.4,1) -- (-4,1);
\draw[black,line width = 5] (0,1) -- (4,1);
\draw[black,dashed ] (-4,0) -- (-4,3);
\draw[black,dashed ] (0,0) -- (0,3);
\draw[black,dashed ] (4,0) -- (4,3);
\draw[->,black,line width = 1] (-9,0) -- (9,0);
\end{tikzpicture}
\caption{Rod diagram for the 5D unbalanced Pomeransky-Sen’kov black ring. The intersection points $w_i$ of rods satisfy $w_1<w_2<w_3$. }
\label{fig:er-ring-rod}
\end{center}
\end{figure}
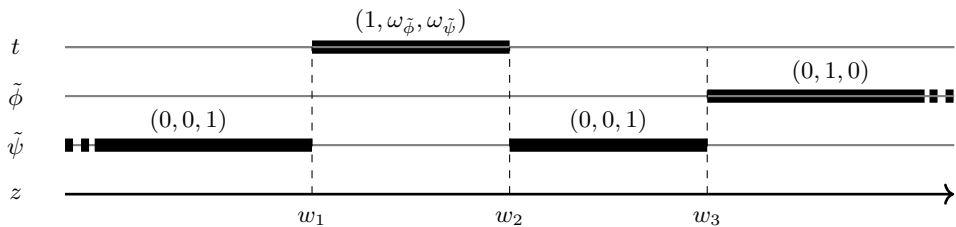

In the $C$-metric coordinates $u$ and $v$, the metric for the unbalanced Pomeransky-Sen'kov black ring~\cite{Chen:2011jb} is written as
\begin{align}\label{ups-ring}
    ds_5^2&=-\frac{H(v,u)}{H(u,v)}\left(dt-\Omega_{\tilde{\phi}}d\tilde{\phi}-\Omega_{\tilde{\psi}}d\tilde{\psi}\right)^2
    +\frac{2\tilde{\kappa}^2(1-\mu)^2(1-\nu)H(u,v)}{(1-\la)(1-\mu \nu)\Phi \Psi(u-v)^2}
    \left(\frac{du^2}{G(u)}-\frac{dv^2}{G(v)}\right)\no\\
   &\qquad +\frac{F(v,u)}{H(v,u)}d\tilde{\psi}^2
   -\frac{F(u,v)}{H(v,u)}d\tilde{\phi}^2+2\frac{J(u,v)}{H(v,u)}d\tilde{\phi}d\tilde{\psi}\,,
\end{align}
where $\Omega_{\tilde{\phi}}$ and $\Omega_{\tilde{\psi}}$ are given by
\begin{align}
    \Omega_{\tilde{\phi}}(u,v)&=\frac{\Omega_{0,\tilde{\phi}} (v+1) \tilde{\kappa} (\mu+\nu) \left(\Phi \left(\nu u^2 v+1\right)+ \nu(1-\mu) (-u v (\la+u)+\la u+1)\right)}{H(v,u)}\,,\\
    \Omega_{\tilde{\psi}}(u,v)&=\frac{\Omega_{\tilde{\psi}}(\mu+\nu)  v\left(1-u^2\right)  }{H(v,u)}\tilde{\kappa}\,,\\
  \Omega_{0,\tilde{\phi}}&= \frac{\sqrt{2 \Xi \Phi \Psi \la (1-\la) (\la+1) (\la-\mu) (1-\la \mu) (1-\mu \nu)}}{\Psi (1-\la) (1-\mu \nu)}\,,\\
   \Omega_{\tilde{\psi}}&=\frac{\sqrt{2 \Xi \Phi \Psi \la \left(1-\la^2\right) \nu (1-\mu \nu)}}{1-\mu \nu}\,,
\end{align}
and the functions $G\,,H\,, F\,,$ and $J$ are defined as
\begin{align}
    G(u)&=(1-u^2)(1+\mu u)(1+\nu u)\,,\\
    H(u,v)&=\Phi \Psi+ \Xi \Psi \nu u^2 v^2+\Phi (\la+1) \nu (\la-\mu)+\nu (\la-\mu) (1-\la \mu) (\mu+\nu) \left(1-\la \mu u^2 v^2\right)\no\\
    &\quad+\la (\mu+\nu) (\nu v (\la-\mu)+u (1-\la \mu)) (-\nu u v (\la-\mu)-\la \mu+1)\,,\\
    F(u,v)&=\frac{2 \tilde{\kappa}^2}{\Phi \mu \nu (1-\mu \nu) (u-v)^2}\biggl(\nu G(v) \Bigl(\Phi \mu \nu u^4 (\Xi \Psi-\la \mu (\la-\mu) (1-\la \mu) (\mu+\nu))\no\\
    &\quad+u^2 (u (\mu+\nu)+1) ((\Phi-1) \Phi \la \mu (\Xi+\Phi-\Psi)+\Xi \Psi)\no\\
    &\quad+(\la-\mu) (1-\la \mu) \left(u (-\mu \nu u+\mu+\nu) (\Psi+\nu (\nu+1) (\la-\mu))+\la (1-\la \mu) (\mu+\nu)^2\right)\Bigr)\no\\
    &\quad+\left(v^2-1\right) G(u) \Bigl(\left(1-\la^2\right) \mu (\Psi+\nu (\nu+1) (\la-\mu))^2\no\\
    &\quad-(1-\la \mu) (\mu+\nu) (\nu v+1) (\Xi \Psi-\la \mu (\la-\mu) (\Psi+\nu (\nu+1) (\la-\mu)))\Bigr)\biggr)\,,\\
    J(u,v)&=-\frac{2 \tilde{\kappa}^2 \sqrt{\nu(1-\la \mu)(\la-\mu)} \left(1-u^2\right) \left(1-v^2\right) (\mu+\nu)}{\Phi (1-\mu \nu) (u-v)}\no\\
    &\quad\times\biggl(\Phi \Psi -\Xi \Psi \nu u v+\Phi (\la+1) \nu (\la-\mu)+\nu (\la-\mu) (1-\la \mu) (\mu+\nu) (\la \mu u v+\la u+\la v+1)\biggr)\,,
\end{align}
where
\begin{align}
    \Phi &=-\la\mu-\la\nu+\mu \nu+1\,,\\
    \Psi &=-\la\mu^2-\la\nu+\mu \nu+\mu\,,\\
   \Xi &=-\la\mu^2+\la\nu-\mu \nu+\mu\,.
\end{align}
Here, $\tilde{\kappa}$ is a positive real parameter, and three real parameters $\mu\,,\nu\,,\la$ satisfy
\begin{align}\label{para-region}
    0\leq \nu\leq\mu\leq\la<1\,.
\end{align}
Within this parameter region, all three quantities $\Phi, \Psi, \Xi$ are positive.
The $C$-metric coordinates $u$ and $v$ run the ranges:
\begin{align}
    -1\leq u \leq 1\,,\quad -\infty < v \leq -1\,.
\end{align}
Since this is asymptotically flat, we can compute the ADM and two ADM angular momenta as
\begin{align}
\begin{split}\label{ps-con-charge}
    M&=\frac{3\pi\la(\mu+\nu)(1-\mu)\Phi\tilde{\kappa}^2}{2(1-\la)(1-\mu\nu)\Psi}\,,\\
    J_1&=\frac{\pi\tilde{\kappa}^3(\mu+\nu)(1-\mu)[2\nu(1-\la)(1-\mu)+(1-\nu)\Phi]  }{(1-\la)^{3/2}(1-\mu\nu)^{3/2}\Psi^{3/2}}\sqrt{\frac{2\la(\la-\mu)(1+\la)(1-\la\mu)\Xi}{\Phi}} \,,\\
    J_2&=\frac{2\pi \tilde{\kappa}^{3}(\mu+\nu)(1-\mu)}{(1-\mu\nu)^{3/2}}\sqrt{\frac{2\nu\la(1+\la)\Xi}{(1-\la)\Phi\Psi}}\,.
\end{split}
\end{align}
The figure \ref{fig:er-ring-rod} describes the rod structure corresponding to the solution (\ref{ups-ring}).
This structure~\cite{Harmark:2004rm} is characterized by the three intersection points on the $z$-axis
\begin{align}\label{pole-ups}
     w_1=-\frac{1}{2}\frac{\mu-\nu}{1-\mu\nu}\tilde{\kappa}^2\,,\qquad w_2=\frac{1}{2}\frac{\mu-\nu}{1-\mu\nu}\tilde{\kappa}^2\,,\qquad w_3=\frac{1}{2}\tilde{\kappa}^2\,.
\end{align}
which divide the axis into four rods: (i) the $\tilde{\psi}$-rotational axis: $I_1=\{(\rho,z) \lvert \rho=0,-\infty<z<w_1\}$, (ii) the horizon cross
section: $I_2=\{(\rho,z) \lvert \rho=0, w_1<z<w_2\}$, (iii) the inner rotational axis of the ring: $I_3=\{(\rho,z) \lvert \rho=0, w_2<z<w_3\}$, (iv) the $\tilde{\phi}$-rotational axis: $I_4=\{(\rho,z) \lvert \rho=0, w_2<z<\infty\}$. The rod
vector on the finite interval $I_2$ takes the form $v_2=(1,\omega_{\tilde{\phi}},\omega_{\tilde{\psi}})$ with the two angular velocities of the horizon given by
\begin{align}
    \omega_{\tilde{\phi}}&=\frac{1}{\tilde{\kappa}(1-\mu)}\sqrt{\frac{(\la-\mu)(1-\la)(1-\la\mu)(1-\mu\nu)\Psi}{2\la(1+\la)\Phi\Xi}}\,,\\
    \omega_{\tilde{\psi}}&=\frac{1+\mu}{\tilde{\kappa}(\mu+\nu)}\sqrt{\frac{\nu(1-\la)(1-\mu\nu)\Psi}{2\la(1+\la)\Phi\Xi}}\,.
\end{align}
The rod vectors on the semi-infinite intervals $I_1$ and $I_4$ are
$v_1=(0,0,1)$ and $v_4=(0,1,0)$, corresponding to the rotational axes along the $\tilde{\psi}$- and $\tilde{\phi}$-directions, respectively.
The finite rod $I_3$ with vector $v_3=(0,0,1)$ represents the $\tilde{\psi}$-rotational axis inside the black ring. 
There exist no conical singularities on $I_1$ and $I_4$, whereas in general one appears on $I_3$.

\subsection{Coset space description}

We then present the coset space description for the unbalanced Pomeransky-Sen’kov black ring solution (\ref{ups-ring}).
As in our previous work, we can obtain the corresponding coset matrix with only finite components by performing a dimensional reduction with the Euler angle (\ref{new-angle}).
To do this, we introduce the Weyl-Papapetrou coordinates $(\rho,z)$:
\begin{align}\label{weyl-er}
    \rho:=\sqrt{ -{\rm det}\left(g_{{\rm Killing}}\right)}=\frac{\tilde{\kappa}^2\sqrt{-G(u)G(v)}}{(u-v)^2(1-\mu\nu)}\,,\qquad 
    z=\frac{\tilde{\kappa}^2(1-uv)(2+(\mu+\nu)(u+v)+2\mu \nu u v)}{2(1-\mu\nu)(u-v)^2}\,.
\end{align}
with the Killing metric $g_{\rm Killing}=(g_{ij})\ (i,j=t,\phi,\psi)$, 
where for details on how $z$ is determined corresponding to this choice of $\rho$ (see the appendix H in \cite{Harmark:2004rm}). 
From this relation, the $C$-metric coordinate part in the metric is related by
\begin{align}
    d\rho^2+dz^2=K(u,v)\left(\frac{du^2}{G(u)}-\frac{dv^2}{G(v)}\right)\,,
\end{align}
where the scalar function $K(u,v)$ is
\begin{align}
    K(u,v)&=-\frac{\tilde{\kappa}^4}{4 (1-\mu \nu)^2 (u-v)^3}\Bigl[\mu^3 \Bigl(4 \nu^3 u^2 v^2 (u+v)+4 \nu^2 u v \left(u^2 \left(v^2+1\right)+3 u v+v^2\right)\no\\
    &\quad+\nu \left(u^3 \left(3 v^2+1\right)+3 u^2 v \left(v^2+3\right)+u \left(9 v^2-1\right)+v \left(v^2-1\right)\right)\no\\
    &\quad+u^3 \left(v-v^3\right)+u^2 \left(3 v^2+1\right)+u v \left(v^2+3\right)+v^2-1\Bigr)\no\\
    &\quad+\mu^2 \Bigl(4 \nu^3 u v \left(u^2 \left(v^2+1\right)+3 u v+v^2\right)+2 \nu^2 \left(u^3 \left(7 v^2+1\right)+u^2 v \left(7 v^2+9\right)+9 u v^2+u+v^3+v\right)\no\\
    &\quad+\nu \left(u^3 v \left(v^2+7\right)+7 u^2 \left(3 v^2+1\right)+7 u v \left(v^2+3\right)+7 v^2+1\right)\no\\
    &\quad-u^3 \left(v^2-1\right)-u^2 v \left(v^2-9\right)+u \left(9 v^2+3\right)+v \left(v^2+3\right)\Bigr)\no\\
    &\quad+\mu \Bigl(\nu^3 \left(u^3 \left(3 v^2+1\right)+3 u^2 v \left(v^2+3\right)+u \left(9 v^2-1\right)+v \left(v^2-1\right)\right)\no\\
    &\quad+\nu^2 \left(u^3 v \left(v^2+7\right)+7 u^2 \left(3 v^2+1\right)+7 u v \left(v^2+3\right)+7 v^2+1\right)\no\\
    &\quad+2 \nu \left(u^3 \left(v^2+1\right)+u^2 v \left(v^2+9\right)+u \left(9 v^2+7\right)+v \left(v^2+7\right)\right)+4 \left(u^2+3 u v+v^2+1\right)\Bigr)\no\\
    &\quad+\nu^3 \left(u^3 \left(v-v^3\right)+u^2 \left(3 v^2+1\right)+u v \left(v^2+3\right)+v^2-1\right)\no\\
    &\quad+\nu^2 \left(-\left(u^3 \left(v^2-1\right)\right)-u^2 v \left(v^2-9\right)+u \left(9 v^2+3\right)+v \left(v^2+3\right)\right)
    +4 \nu \left(u^2+3 u v+v^2+1\right)+4 (u+v)\Bigr]\,.
\end{align}

\medskip
The coset matrix $M_{\rm uPS}(z,\rho)$, as in the examples discussed above, can be obtained via dimensional reduction to three dimensions with the metric
\begin{align}
    ds_3^2=e^{2\nu}(d\rho^2+dz^2)+\rho^2d\phi^2\,.
\end{align}
The 16 scalar fields that parametrize the coset matrix are given by
\begin{align}
\begin{split}\label{ups-scalar}
 e^{2U}&=\frac{F(v,u)-F(u,v)-2J(u,v)}{4\sqrt{H(u,v)H(v,u)}}\,,\\
    x^I&=0\,,\quad y^I=\sqrt{\frac{H(v,u)}{H(u,v)}}\,,\\
   \zeta^{0}&=\frac{\Omega_{\tilde{\psi}}(u,v)-\Omega_{\tilde{\phi}}(u,v)}{2} \,,\qquad
   \tilde{\zeta}_{0}=-\frac{\Omega_{\tilde{\psi}}(v,u)-\Omega_{\tilde{\phi}}(v,u)}{2}\,,\\
   \zeta^{I}&=0\,,\qquad \tilde{\zeta}_{I}=0\,,\\
\sigma&=\frac{\tilde{\kappa}^2\sigma_0(u,v)}{2(1-\mu\nu)\Phi(u-v)H(u,v)H(v,u)}\,.
\end{split}
\end{align}
Here, $\sigma_0(u,v)$ is a symmetric polynomial of degree five in $u$ and $v$, and its explicit expression is presented in appendix~\ref{sec:twist-ups}.
We find that the coset matrix $M_{\rm uPS}(z,\rho)$ with the scalar field (\ref{ups-scalar}) approaches the constant matrix $Y_{\text{flat}}$ at the spacial infinity $r\to \infty$:
\begin{align}
    \lim_{r\to \infty}M_{\rm uPS}(z,\rho)= Y_{\text{flat}}\,.
\end{align}
Furthermore, the conformal factor $e^{2\nu}$ is given by
\begin{align}\label{ups-conf}
    e^{2\nu}
    &=-\frac{\tilde{\kappa}^2(1-\mu)^2(1-\nu)(F(v,u)-F(u,v)-2J(u,v))}{2(1-\la)(1-\mu\nu)\Phi\Psi(u-v)^2}\frac{1}{K(u,v)}\,.
\end{align}

\subsection{Monodromy matrix}

We now compute the monodromy matrix $\cM_{\rm uPS}(w)$ corresponding to the unbalanced Pomeransky-Sen'kov black ring solution, starting from the coset matrix $M_{\rm uPS}(z,\rho)$ parametrized by (\ref{ups-scalar}).
For this purpose, we express the $C$-metric coordinates $u$ and $v$ in terms of the Weyl-Papapetrou coordinates $(\rho,z)$ as
\begin{align}
    u&=-\frac{\mu+\nu}{2 \mu \nu+\mu-\nu}\frac{R_1-R_2+\frac{2  (\mu-\nu)}{\mu+\nu}R_3+\frac{\tilde{\kappa}^2 (\mu-\nu) (\mu \nu+1)}{(\mu+\nu) (\mu \nu-1)}}{R_1+\frac{(-2 \mu \nu+\mu-\nu)}{2 \mu \nu+\mu-\nu}R_2 +\frac{\tilde{\kappa}^2 \left(\mu^2-\nu^2\right)}{(\mu \nu-1) (2 \mu \nu+\mu-\nu)}}\,,\\
    v&=-\frac{\mu+\nu}{2 \mu \nu+\mu-\nu}\frac{R_1-R_2+\frac{2  (\mu-\nu)}{\mu+\nu}R_3-\frac{\tilde{\kappa}^2 (\mu-\nu) (\mu \nu+1)}{(\mu+\nu) (\mu \nu-1)}}{R_1+\frac{ (-2 \mu \nu+\mu-\nu)}{2 \mu \nu+\mu-\nu}R_2-\frac{\tilde{\kappa}^2 \left(\mu^2-\nu^2\right)}{(\mu \nu-1) (2 \mu \nu+\mu-\nu)}}\,,
\end{align}
where we introduced
\begin{align}
    R_i=\sqrt{\rho^2+(z-w_i)^2}\,.
\end{align}
By using the relation (\ref{sub-rule}), we can find that the monodromy matrix $\cM_{\rm uPS}(w)$ takes the form
\begin{align}\label{br-mono}
    \cM_{\rm uPS}(w)=Y_{\rm flat}+\sum_{i=1}^{3}\frac{A_i}{w-w_i}\,.
\end{align}
By left multiplying the residue matrices $A_j$ with $\eta'$, we cam write down $A_j$ in symmetric forms as
\begin{align}\label{sym-A-ups}
     \eta'A_i&=\left(
\begin{array}{cccccccc}
 F_{i,1}(F_{i,2})^2& 0 & 0 & F_{i,1} F_{i,2}& 0 & 0 & F_{i,1}F_{i,2}F_{i,3} & 0 \\
 0 & 0 & 0 & 0 & 0 & 0 & 0 & 0 \\
 0 & 0 & F_{i,4} & 0 & F_{i,4}F_{i,5}& 0 & 0 & F_{i,4}F_{i,6}  \\
 F_{i,1}F_{i,2} & 0 & 0 & F_{i,1} & 0 & 0 & F_{i,1}F_{i,3}& 0 \\
 0 & 0 & F_{i,4}F_{i,5} & 0 & F_{i,4}(F_{i,5})^2 & 0 & 0 & F_{i,4}F_{i,5} F_{i,6} \\
 0 & 0 & 0 & 0 & 0 & 0 & 0 & 0 \\
 F_{i,1}F_{i,2} F_{i,3} & 0 & 0 &F_{i,1}F_{i,3} & 0 & 0 &F_{i,1}(F_{i,3})^2 & 0 \\
 0 & 0 &F_{i,4}F_{i,6} & 0 & F_{i,4} F_{i,5} F_{i,6} & 0 & 0 &F_{i,4}(F_{i,6})^2 \\
\end{array}
\right)\,.
\end{align}
The building blocks $F_{i,j}$ of each residue matrix are given by
\begin{align}
\begin{split}
    F_{1,1}&=\frac{2 (1-\nu) (\la-\mu) (1-\la \mu) (\mu+\nu)}{\Phi (1-\la) (1+\mu) (\mu-\nu)}\,,\qquad
    F_{1,2}=\frac{ \Omega_{\tilde{\phi}} (1-\mu)\tilde{\kappa} }{2  (\la-\mu) (1-\la \mu)}\,,\\
    F_{1,3}&=-\frac{ (1-\mu)\tilde{\kappa}^2}{2\Phi \Psi(\la-\mu)(1-\la \mu)  (1-\la) (1-\mu \nu)}\left((\la-\mu)(1-\la \mu)\cF_1-\sqrt{\nu(\la-\mu)(1-\la\mu)} \,(1-\la)\Psi\cF_2\right)\,,\\
    F_{1,4}&=\frac{2 \Xi (\mu+\nu)}{\Phi (1+\mu) (\mu-\nu)}\,,\qquad
    F_{1,5}=-\frac{ \Omega_{\tilde{\psi}} \left(1-\mu^2\right)\tilde{\kappa}}{2 \Xi \Psi (1-\la)}\,,\\
    F_{1,6}&=\frac{ (1-\mu)\tilde{\kappa}^2}{2 \Phi \Psi (1-\la) (1-\mu \nu)}\left(\cF_2+\sqrt{\nu(\la-\mu)(1-\la\mu)}\,\cF_3 \right)\,,
\end{split}
\end{align}
\begin{align}
\begin{split}
    F_{2,1}&=-\frac{2\nu (1-\mu)  (\la-\mu) (1-\la \mu) (\mu+\nu)}{\Phi \Psi (\mu-\nu)}\,,\qquad
    F_{2,2}=\frac{ \Omega_{\tilde{\phi}} (1-\mu)\tilde{\kappa} }{2  (\la-\mu) (1-\la \mu)}\,,\\
    F_{2,3}&=-\frac{\tilde{\kappa}^2}{2 \Phi \Psi \nu(\la-\mu)(1-\la\mu)(1-\la) (1-\mu \nu)}\left(\nu (\la-\mu)(1-\la \mu)\cF_4-\sqrt{\nu(\la-\mu)(1-\la\mu)}(1-\la) \Psi \cF_5\right)\,,\\
    F_{2,4}&=-\frac{2\nu  (\la+1) (1-\mu) (\mu+\nu)}{\Phi (\nu+1) (\mu-\nu)}\,,\qquad
    F_{2,5}=-\frac{ \Omega_{\tilde{\psi}} (\nu+1)\tilde{\kappa}}{2 \Psi \left(1-\la^2\right) \nu}\,,\\
    F_{2,6}&=\frac{\tilde{\kappa}^2}{2 \Phi \Psi \nu(1-\la)  (1-\mu \nu)}\left(\nu\cF_5+ \sqrt{\nu(\la-\mu)(1-\la\mu)} \cF_6\right)\,,
\end{split}
\end{align}
\begin{align}
\begin{split}
    F_{3,1}&=-\frac{\Xi (1+\la) (1-\mu)}{\Psi (1-\la) (1+\mu)}\,,\qquad
    F_{3,2}=\frac{ \Omega_{\tilde{\phi}} (\mu+\nu)\tilde{\kappa}}{\Xi \left(1+\la\right) }\,,\\
    F_{3,3}&=-\frac{\tilde{\kappa}^2 (\mu+\nu)}{2 \Phi \Psi (1-\la) (1-\mu \nu)}\left(\cF_7-4\sqrt{\nu(\la-\mu)(1-\la\mu)}(1-\la)\Psi  \right)\,,\\
    F_{3,4}&=\frac{(1-\mu) (1-\nu)}{(1+\mu) (1+\nu)}\,,\qquad
    F_{3,5}=0\,,\\
    F_{3,6}&=\frac{\tilde{\kappa}^2 (\mu+\nu)}{2 \Phi \Psi (1-\la) (1-\mu \nu)}\left(\cF_7-4\sqrt{\nu(\la-\mu)(1-\la\mu)}(1-\la)\Psi \right)\,,
\end{split}
\end{align}
where we introduced
\begin{align}
    \cF_1&=\la \mu^3 \left(-\la^2\nu+\la \left(\nu^2-4 \nu+1\right)+3 \nu\right)+\la \nu \left(-\la^2 \nu+\la \left(\nu^2-4 \nu+1\right)+3 \nu\right)\no\\
    &\quad+\mu^2 \left(-\la^3 (\nu-2) \nu-2 \la^2 \nu^2+\la \left(\nu^2+6 \nu-2\right)+\nu \left(\nu^2-2 \nu-3\right)\right)\no\\
    &\quad+\mu \left(\left(2 \la^3+6 \la-3\right) \nu^2-\left(\la^3+2 \la^2-\la+2\right) \nu-2 \la \nu^3+1\right)\,,\\
    \cF_2&=3\Xi+(1-\la)(1-\nu \mu^2)+\la^2(1-\mu)(\mu+\nu)\,,\\
    \cF_3&=(1+\nu)\cF_2+2 \la (1-\mu) \Bigl(\Phi-\nu (\mu (\la-\nu+4)+\la \nu+3)\Bigr)\,,\\
    \cF_4&=\nu (\la \mu-1) \left(\la^2 (\mu ((\mu-3) \mu+5)-1)+\la \mu (\mu (2 \mu-5)-1)-(\mu-3) \mu^2\right)
    -\nu^3 (\la-\mu)^2 (2 \la+\mu-3)\no\\
    &\quad+\nu^2 (\la-\mu) (\la (\la (\mu ((\mu-5) \mu+3)-1)+\mu (\mu+5)-2)-3 \mu+1)+\mu (\la \mu-1)^2 ((2 \la-3) \mu+1)\,,\\
    \cF_5&=\mu^2 \left(-3 \left(\la^2+\la+1\right) \nu+(\la+1) (\la+2) \nu^2-2 \la+\nu^3\right)\no\\
    &\quad+\la \mu^3 \left((\la+3) \nu+\la-\nu^2\right)+\mu \left(\nu \left(\la \left(-3 (\la+1) \nu+\la-2 \nu^2+3\right)-3 \nu+2\right)+1\right)\no\\
    &\quad+\la \nu (\nu (\la \nu+\la+3)-1)\,,\\
    \cF_6&=-(\mu+1) \nu^2 \left(\la \left(\la \mu^2-(\la+2) \mu+\la\right)+\mu\right)+\nu^4 (\la-\mu)^2+(\mu-1) \nu^3 (\la (\mu-4)+3) (\mu-\la)\no\\
    &\quad-(\mu-1) \nu (\la \mu-1) (\la (4 \mu-1)-3 \mu)+\mu (\la \mu-1)^2\,,\\
    \cF_7&=-\nu^2 (\la-\mu)^2 (\la \mu+\la-2)+2 (1-2 \la) (1-\mu) \nu (\la-\mu) (1-\la \mu)+((\la-2) \mu+\la) (1-\la \mu)^2\,.
\end{align}
By definition, we find that the quantities $F_{i,j}$ satisfy the following algebraic relation:
\begin{align}\label{f-rel}
    F_{i,3}+F_{i,2}F_{i,5}+F_{i,6}=0\,.
\end{align}
Although the explicit expressions of the coset matrix are extremely complicated, all three residue matrices are of rank 2, as can be verified from Eq.~(\ref{sym-A-ups}).
Therefore, we can employ the Riemann-Hilbert approach developed in Refs.~\cite{Katsimpouri:2012ky,Katsimpouri:2013wka} to perform the factorization of the monodromy matrix.

\subsubsection*{Charge matrix}

As in the Myers-Perry black hole case, we introduce the charge matrix $Q$ to examine whether its nilpotent condition is related to the extremal limit.  The charge matrix $Q$ can be read off the large-$w$ behavior of the monodromy matrix defined as
\begin{align}\label{charge-ups}
     Q=Y_{\rm flat}^{-1}\left(\sum_{j=1}^{3}A_j\right)\,,
\end{align}
We can check that the charge matrix has the general expansion form (\ref{gQ-ex}) with the physical quantities $M,J_{1,2}$ given in (\ref{ps-con-charge}) and
\begin{align}
    Q_{E_0}=-\sum_{j=1}^{3}A_{j,77}=\sum_{j=1}^{3}A_{j,88}\,.
\end{align}
The charge matrix (\ref{charge-ups}) satisfies a modified cubic relation of the same form as that appearing in the Emparan-Reall black ring and the Chen-Teo black lens solution \cite{Sakamoto:2025xbq},
\begin{align}
    Q^3-\frac{1}{4}\Tr(Q^2)Q+q_{\rm PS}\left(H_1-\frac{1}{2}H_2-\frac{1}{2}H_3\right)=0\,,
\end{align}
where the constant $q_{\rm PS}$ is 
\begin{align}
    q_{\rm PS}&=\frac{\tilde{\kappa}^6 \la (1-\mu)^2 (1-\nu) (\mu+\nu)^2}{2 \Phi \Psi(1-\la) (1-\mu \nu)^3}
    \Bigl[4 \sqrt{\nu(\la-\mu)(1-\la\mu)} (\Phi+2 \nu (\la-\mu))\no\\
    &\quad+\la^2 \left(\mu^2-6 \mu \nu+\nu^2\right)+\la \left(6 \mu^2 \nu-2 \mu \left(\nu^2+1\right)+6 \nu\right)+\mu^2 \nu^2-6 \mu \nu+1\Bigr]\,,
\end{align}
and the trace of the square of the charge matrices $Q$ is
\begin{align}
    \Tr(Q^2)&=\frac{2\tilde{\kappa}^4 (1-\mu) (\mu+\nu)}{\Phi^2 \Psi (1-\la) (1-\mu \nu)^2}\Bigl[\nu (\la \mu-1)^2 \left(3 (\la-1) (\la+4) \mu^2+2 \la (1-5 \la) \mu+\la (7 \la+1)\right)\no\\
    &\quad-\left(\nu^4 (\la (\mu-1)-2) (\la-\mu)^3\right)-\nu^3 (\la-\mu)^2 (\la (\la (\mu-1) (7 \mu-3)+\mu (\mu+2)+9)-12)\no\\
    &\quad-(3 \la (\la+3)-2) (\mu-1) (\mu+1) \nu^2 (\la-\mu) (\la \mu-1)-(\la (\mu-1)+2 \mu) (\la \mu-1)^3\no\\
    &\quad+K_j \Bigl(-8 \bigl(\mu^2 \left(-2 \la \left(\la^2+\la+1\right) \nu+\la (\la+1)^2 \nu^2+2 \la-\nu^3+\nu\right)\no\\
    &\quad+\mu \left(\nu \left(\la \left(-2 \left(\la^2+\la+1\right) \nu+(\la+1)^2+2 \nu^2\right)+\nu\right)-1\right)\no\\
    &\quad+\la \mu^3 (\la (\nu (\la-\nu+2)-1)-\nu)+\la \nu (\la (\nu (\la-\nu+2)-1)-\nu)\bigr)\Bigr) \Bigr]\,.
\end{align}
The connection between the black ring’s extremal conditions and the nilpotent conditions of the charge matrix is discussed in a later section.

\subsection{Factorization of monodromy matrix}

Let us explicitly perform the factorization of the monodromy matrix (\ref{br-mono}).
To do this, we again express the residue matrices $A_j$ in terms of the eight-component vectors $a_j$ and $b_j$, which are taken as
\begin{align}
    a_i^T&=\left(\frac{F_{i,2}}{F_{i,3}},0,0,0,-\frac{1}{F_{i,3}},0,0,1\right)\,,\qquad
    b_i^T=\left(0,0,-\frac{1}{F_{i,6}},0,-\frac{F_{i,5}}{F_{i,6}},0,0,1\right)\eta\,,
\end{align}
and the constants $\alpha_i,\beta_i$ are given by
\begin{align}
    \alpha_i&=F_{i,1}F_{i,3}^2\,,\qquad
    \beta_i=-F_{i,4}F_{i,6}^2\,.
\end{align}
By using the relation~(\ref{f-rel}), we can check that this choice of the vectors satisfies the orthogonal relations~(\ref{ab-cond}).
In this choice of the vectors, the $3\times 3$ matrices $\Gamma^{(0)}$ and $\Gamma^{(a,b)}$ become
\begin{align}
\begin{split}\label{gamma-ups}
    \Gamma_{11}^{(0)}&=-\frac{1}{F_{1,3} F_{1,4}F_{1,6}}\biggl[1-\frac{F_{2,4}(F_{1,3}+F_{1,2} F_{2,5}+F_{2,6})}{w_1-w_2}-\frac{F_{3,4} (F_{1,3}+F_{1,2}F_{3,5}+F_{3,6})}{w_1-w_3}\biggr]\frac{1}{\la_1\nu_1}\,,\\
    \Gamma_{22}^{(0)}&=-\frac{1}{F_{2,3} F_{2,4}F_{2,6}}\biggl[1-\frac{F_{1,4} (F_{2,3}+F_{2,2}F_{1,5}+F_{1,6})}{w_2-w_1}-\frac{F_{3,4} (F_{2,3}+F_{2,2} F_{3,5}+F_{3,6})}{w_2-w_3}\biggr]\frac{1}{\la_2\nu_2}\,,\\
    \Gamma_{33}^{(0)}&=-\frac{1}{F_{3,3} F_{3,4}F_{3,6}}\biggl[1-\frac{F_{1,4} (F_{3,3}+F_{3,2}F_{1,5}+F_{1,6})}{w_3-w_1}-\frac{F_{2,4} (F_{3,3}+F_{3,2}F_{2,5}+F_{2,6})}{w_3-w_2}\biggr]\frac{1}{\la_3\nu_3}\,,\\
    \Gamma_{12}^{(0)}&=-\frac{F_{1,3}+F_{1,2} F_{2,5}+F_{2,6}}{F_{1,3} F_{2,6}}\frac{1}{\la_1-\la_2}\,,\qquad 
    \Gamma_{21}^{(0)}=\frac{F_{1,6}+F_{1,5} F_{2,2}+F_{2,3}}{F_{1,6}F_{2,3}}\frac{1}{\la_1-\la_2}\,,\\
    \Gamma_{13}^{(0)}&=-\frac{F_{1,3}+F_{1,2} F_{3,5}+F_{3,6}}{F_{1,3} F_{3,6}}\frac{1}{\la_1-\la_3}\,,\qquad
    \Gamma_{31}^{(0)}=\frac{F_{1,6}+F_{1,5} F_{3,2}+F_{3,3}}{F_{1,6}F_{3,3}}\frac{1}{\la_1-\la_3}\,,\\
    \Gamma_{23}^{(0)}&=-\frac{F_{2,3}+F_{2,2} F_{3,5}+F_{3,6}}{F_{2,3} F_{3,6}}\frac{1}{\la_2-\la_3}\,,\qquad
    \Gamma_{32}^{(0)}=\frac{F_{2,6}+F_{2,5} F_{3,2}+F_{3,3}}{F_{2,6}F_{3,3}}\frac{1}{\la_2-\la_3}\,,
\end{split}
\end{align}
and
\begin{align}
    \Gamma^{(a)}=\Gamma^{(a)}=0_{3\times 3}\,.
\end{align}
Then, by computing the matrix $X_+$ from the $\Gamma$ matrix (\ref{gamma-ups}), we can show that the monodromy matrix $\cM_{\rm uPS}(w)$ are factorized as
\begin{align}\label{mon-fac-er}
    \cM_{\rm uPS}(w)=X_-(\la,z,\rho)M_{\rm uPS}(z,\rho)X_+(\la,z,\rho)\,.
\end{align}
Extracting the 16 scalar fields from the coset matrix obtained in the factorized form~(\ref{mon-fac-er}), we can confirm that they all coincide with the scalar fields (\ref{ups-scalar}).

\medskip
Finally, we compute the conformal factor $e^{2\nu}$.
Since $\Gamma^{(a,b)}=0_{3\times 3}$, we can again use the formula~(\ref{conf}).
From the expression (\ref{gamma-ups}) of $\Gamma^{(0)}$, the right-hand side of (\ref{conf}) can be computed.
By fixing the overall normalization constant $k_{\rm BM}$ as
\begin{align}
    &k_{\rm BM}^{-1}=-\frac{(1+\nu) (\la-\mu) (1-\la \mu)}{ (1-\la)\Psi}\prod_{i=1}^{3}(F_{i,3}F_{i,4}F_{i,6})\,,
\end{align}
the resulting expression precisely reproduces the conformal factor (\ref{ups-conf}).
This confirms that the monodromy matrix (\ref{br-mono}) correctly describes the unbalanced Pomeransky-Sen'kov black ring.

\section{Various limits of unbalanced Pomeransky-Sen'kov black ring}\label{ps-lim}

\begin{figure}
\begin{center}

\begin{tikzpicture}[x=0.75pt,y=0.75pt,yscale=-1,xscale=1]
\draw  [dash pattern={on 0.84pt off 2.51pt}] (110.33,72.46) .. controls (110.33,59.96) and (120.46,49.83) .. (132.96,49.83) -- (571.71,49.83) .. controls (604.2,49.83) and (614.33,59.96) .. (614.33,72.46) -- (614.33,265.21) .. controls (614.33,277.7) and (604.2,287.83) .. (571.71,287.83) -- (132.96,287.83) .. controls (120.46,287.83) and (110.33,277.7) .. (110.33,265.21) -- cycle ;
\draw  [dash pattern={on 0.84pt off 2.51pt}] (110.33,341.17) .. controls (110.33,333.99) and (116.15,328.17) .. (123.33,328.17) -- (360.33,328.17) .. controls (367.51,328.17) and (373.33,333.99) .. (373.33,341.17) -- (373.33,417.17) .. controls (373.33,424.35) and (367.51,430.17) .. (360.33,430.17) -- (123.33,430.17) .. controls (116.15,430.17) and (110.33,424.35) .. (110.33,417.17) -- cycle ;
\draw [line width=1.5]    (233.33,114) -- (233.33,144) ;
\draw [shift={(233.33,148.17)}, rotate = 270] [fill={rgb, 255:red, 0; green, 0; blue, 0 }  ][line width=0.08]  [draw opacity=0] (11.61,-5.58) -- (0,0) -- (11.61,5.58) -- cycle    ;
\draw [line width=1.5]    (233.33,190) -- (233.33,220) ;
\draw [shift={(233.33,225.17)}, rotate = 270] [fill={rgb, 255:red, 0; green, 0; blue, 0 }  ][line width=0.08]  [draw opacity=0] (11.61,-5.58) -- (0,0) -- (11.61,5.58) -- cycle    ;
\draw [line width=1.5]    (480,190) -- (480,220) ;
\draw [shift={(480,230.17)}, rotate = 270] [fill={rgb, 255:red, 0; green, 0; blue, 0 }  ][line width=0.08]  [draw opacity=0] (11.61,-5.58) -- (0,0) -- (11.61,5.58) -- cycle    ;
\draw [line width=1.5]    (72.33,90) -- (72.33,400) ;
\draw [line width=1.5]    (72.33,364) -- (152.33,364) ;
\draw [shift={(156.33,364.83)}, rotate = 180.56] [fill={rgb, 255:red, 0; green, 0; blue, 0 }  ][line width=0.08]  [draw opacity=0] (11.61,-5.58) -- (0,0) -- (11.61,5.58) -- cycle    ;
\draw [line width=1.5]    (72.33,400) -- (152.33,400) ;
\draw [shift={(156.33,400)}, rotate = 180.37] [fill={rgb, 255:red, 0; green, 0; blue, 0 }  ][line width=0.08]  [draw opacity=0] (11.61,-5.58) -- (0,0) -- (11.61,5.58) -- cycle    ;
\draw [line width=1.5]    (72.33,91.83) -- (140.33,91.83) ;
\draw [line width=1.5]    (330,93) .. controls (358.88,91.86) and (429.67,114.65) .. (462.86,154.72) ;
\draw [shift={(465.33,157.83)}, rotate = 232.7] [fill={rgb, 255:red, 0; green, 0; blue, 0 }  ][line width=0.08]  [draw opacity=0] (11.61,-5.58) -- (0,0) -- (11.61,5.58) -- cycle    ;
\draw [line width=1.5]    (330,170) .. controls (334.88,168.86) and (385.67,189.65) .. (418.86,229.72) ;
\draw [shift={(421.33,232.83)}, rotate = 232.7] [fill={rgb, 255:red, 0; green, 0; blue, 0 }  ][line width=0.08]  [draw opacity=0] (11.61,-5.58) -- (0,0) -- (11.61,5.58) -- cycle    ;

\draw (160,84) node [anchor=north west][inner sep=0.75pt]   [align=left] {\fontsize{10pt}{12pt}\selectfont Unbalanced PS black ring};
\draw (168,163) node [anchor=north west][inner sep=0.75pt]   [align=left] {\fontsize{10pt}{12pt}\selectfont PS black ring (Sec.\ref{sec:PS})};
\draw (141,240) node [anchor=north west][inner sep=0.75pt]   [align=left] {\fontsize{10pt}{12pt}\selectfont Extremal PS black ring (Sec.\ref{sec:ePS})};
\draw (400,240) node [anchor=north west][inner sep=0.75pt]   [align=left] {\fontsize{10pt}{12pt}\selectfont Extremal MP black hole (Sec.\ref{sec:MP})};
\draw (422,163) node [anchor=north west][inner sep=0.75pt]   [align=left] {\fontsize{10pt}{12pt}\selectfont MP black hole (Sec.\ref{sec:MP})};
\draw (164,393) node [anchor=north west][inner sep=0.75pt]   [align=left] {\fontsize{10pt}{12pt}\selectfont MIF black ring (Sec.\ref{sec:MIF})};
\draw (164,358) node [anchor=north west][inner sep=0.75pt]   [align=left] {\fontsize{10pt}{12pt}\selectfont ER black ring (Sec.\ref{sec:ER})};
\draw (160,120) node [anchor=north west][inner sep=0.75pt]   [align=left] {\fontsize{10pt}{12pt}\selectfont $\la\to \frac{2\mu}{1+\mu^2}$};
\draw (170,200) node [anchor=north west][inner sep=0.75pt]   [align=left] {\fontsize{10pt}{12pt}\selectfont $\mu\to\nu$};
\draw (390,90) node [anchor=north west][inner sep=0.75pt]   [align=left] {\fontsize{10pt}{12pt}\selectfont (\ref{mp-lim})};
\draw (360,160) node [anchor=north west][inner sep=0.75pt]   [align=left] {\fontsize{10pt}{12pt}\selectfont (\ref{exmp-lim})};
\draw (430,200) node [anchor=north west][inner sep=0.75pt]   [align=left] {\fontsize{10pt}{12pt}\selectfont $\alpha\to0$};
\draw (93,345) node [anchor=north west][inner sep=0.75pt]   [align=left] {\fontsize{10pt}{12pt}\selectfont $\nu\to 0$};
\draw (93,380) node [anchor=north west][inner sep=0.75pt]   [align=left] {\fontsize{10pt}{12pt}\selectfont $\mu\to 0$};
\draw  [color={rgb, 255:red, 255; green, 255; blue, 255 }  ,draw opacity=1 ][fill={rgb, 255:red, 255; green, 255; blue, 255 }  ,fill opacity=1 ]  (180,316) -- (282,316) -- (282,340) -- (180,340) -- cycle  ;
\draw (183,320) node [anchor=north west][inner sep=0.75pt]  [color={rgb, 255:red, 0; green, 0; blue, 0 }  ,opacity=1 ] [align=left] {\fontsize{10pt}{12pt}\selectfont Singly rotating};
\draw  [color={rgb, 255:red, 255; green, 255; blue, 255 }  ,draw opacity=1 ][fill={rgb, 255:red, 255; green, 255; blue, 255 }  ,fill opacity=1 ]  (176,38) -- (284,38) -- (284,62) -- (176,62) -- cycle  ;
\draw (179,42) node [anchor=north west][inner sep=0.75pt]   [align=left] {\fontsize{10pt}{12pt}\selectfont Doubly rotating};
\end{tikzpicture}
\caption{Degenerate limits of the unbalanced Pomeransky-Sen’kov black ring. Through various limiting procedures, the unbalanced Pomeransky-Sen’kov (PS) black ring reduces to the balanced PS black ring, extremal PS black ring, non-extremal and extremal Myers-Perry (MP) black holes, Emperan-Reall (ER) black ring, and Mishima-Iguchi-Figueras (MIF) black ring.}
\label{fig:summary}
\end{center}
\end{figure}
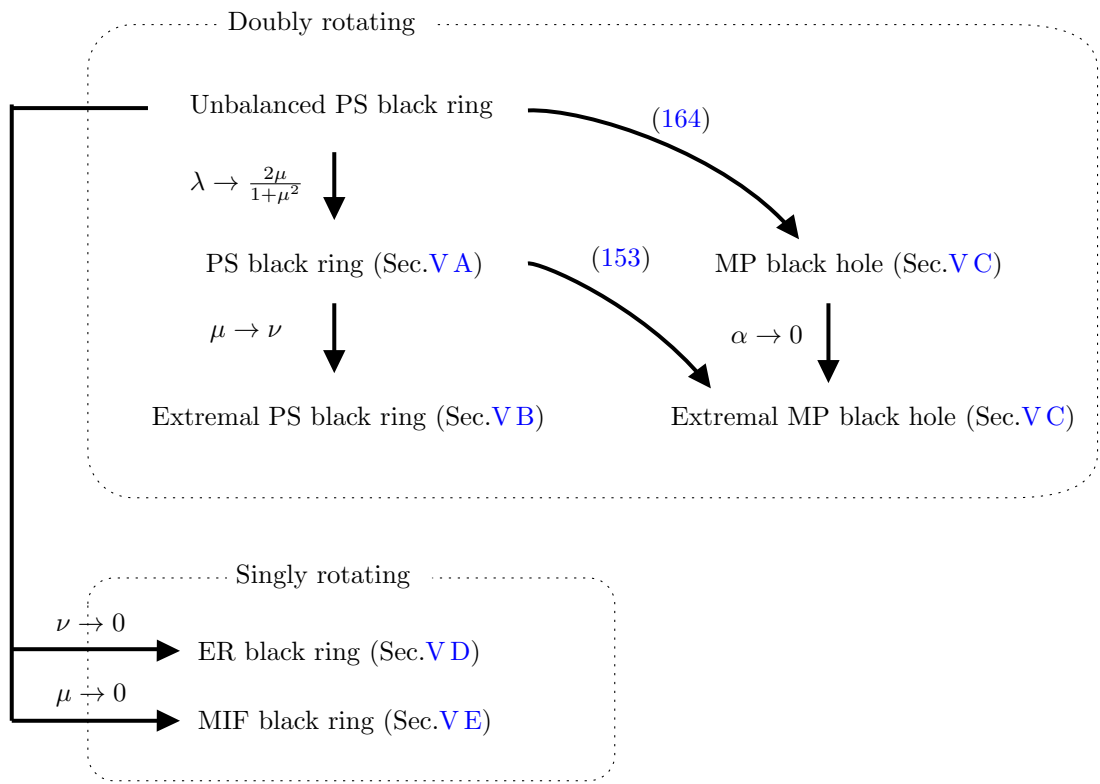

In this section, we investigate several limiting cases of the monodromy matrix (\ref{br-mono}) associated with the unbalanced Pomeransky-Sen'kov black ring:
\begin{itemize}
\item[(A)] the (balanced) Pomeransky-Sen'kov black ring,
\item[(B)] the extremal Pomeransky-Sen'kov black ring,
\item[(C)] the Myers-Perry black hole,
\item[(D)]  the Emparan-Reall black ring, and
\item[(E)] the Mishima-Iguchi-Figueras black ring.
\end{itemize}
The interrelations among the corresponding gravitational solutions obtained in these limits are summarized in Fig.,\ref{fig:summary}.

\subsection{(Balanced) Pomeransky-Sen’kov black ring}\label{sec:PS}

We begin by examining the (balanced) Pomeransky-Sen'kov black ring solution \cite{Pomeransky:2006bd} which is obtained by imposing the balance condition 
\begin{align}\label{bc-ups}
    \la=\frac{2\mu}{1+\mu^2}\,.
\end{align}
This condition is obtained by requiring the absence of conical singularities at $u=1$ \cite{Chen:2011jb}. The positions of the poles for the monodromy matrix remain invariant.
The algebraic structure (\ref{sym-A-ups}) of the residue matrices $A_i$ is preserved under the constraint (\ref{bc-ups}), and all residue matrices are still rank 2.
The building blocks $F_{i,j}$ of $A_i$ are largely simplified, and denoting them as $F_{i,j}^{\text{b}}$, they take the following form:
\begin{align}
\begin{split}
    F_{1,1}^{\text{b}}&=\frac{2 \mu (1-\nu) (\mu+\nu)}{(1-\mu) (\mu-\nu) (1-\mu \nu)}\,,\qquad
    F_{1,2}^{\text{b}}=\frac{\tilde{\kappa} \sqrt{\left(1-\mu^2\right) \left(1-\nu^2\right)}}{(1-\mu) (1-\nu)}\,,\\
    F_{1,3}^{\text{b}}&=-\frac{\tilde{\kappa}^2}{2 \sqrt{\mu} (1-\mu) (1-\nu) \left(1-\sqrt{\mu \nu}\right)^2}\biggl[4 \mu^{3/2} \nu+\left(1-\mu \left(\mu^2+\mu+3\right)\right) \nu^{3/2}+\left(\mu^2+1\right) \sqrt{\mu} \nu^2\\
    &\qquad\qquad\qquad\qquad\qquad\qquad\qquad\qquad\qquad+\left(\mu^2-3 \mu-1\right) \mu \sqrt{\nu}+\left(\mu^2+1\right) \sqrt{\mu}-\sqrt{\nu}\biggr]\,,\\
    F_{1,4}^{\text{b}}&=\frac{2 \mu (1+\nu) (\mu+\nu)}{(1+\mu) (\mu-\nu) (1-\mu \nu)}\,,\qquad
    F_{1,5}^{\text{b}}=-\frac{\tilde{\kappa} (\mu+1)^2 \nu}{\sqrt{\mu\nu \left(1-\mu^2\right)\left(1-\nu^2\right)}}\,,\\
    F_{1,6}^{\text{b}}&=\frac{\tilde{\kappa}^2}{2 \sqrt{\mu} (1-\mu) (1-\nu) \left(1-\sqrt{\mu \nu}\right)^2}\biggl[\mu^{5/2}+\left(\mu^2+1\right) \sqrt{\mu} \nu^2-4 \sqrt{\mu} \left(\mu^2+\mu+1\right) \nu\\
    &\qquad\qquad\qquad\qquad\qquad\qquad\qquad+(\mu (\mu (\mu+3)-1)+1) \nu^{3/2}+(\mu ((\mu-1) \mu+3)+1) \sqrt{\nu}+\sqrt{\mu}\biggr]\,,
\end{split}
\end{align}
\begin{align}
\begin{split}
    F_{2,1}^{\text{b}}&=-\frac{2 \nu(1-\mu)  (\mu+\nu)}{(1-\nu) (1-\mu \nu) (\mu-\nu)}\,,\qquad
    F_{2,2}^{\text{b}}=\frac{\tilde{\kappa} \sqrt{\left(1-\mu^2\right) \left(1-\nu^2\right)}}{(1-\mu) (1-\nu)}\,,\\
    F_{2,3}^{\text{b}}&=-\frac{\tilde{\kappa}^2}{2 \sqrt{\nu} (1-\mu) (1-\nu) \left(1-\sqrt{\mu \nu}\right)^2}\biggl[\mu^{3/2} \left(1-\nu \left(\nu^2+\nu+3\right)\right)+\mu^2 \left(\nu^2+1\right) \sqrt{\nu}+4 \mu \nu^{3/2}\\
    &\qquad\qquad\qquad\qquad\qquad\qquad+\sqrt{\mu} (\nu ((\nu-3) \nu-1)-1)+\nu^{5/2}+\sqrt{\nu}\biggr]\,,\\
    F_{2,4}^{\text{b}}&=-\frac{2\nu (1+\mu)  (\mu+\nu)}{(1+\nu) (1-\mu \nu) (\mu-\nu)}\,,\qquad
    F_{2,5}^{\text{b}}=-\frac{\tilde{\kappa} \mu (1+\nu)^2}{\sqrt{\mu \nu\left(1-\mu^2\right)  \left(1-\nu^2\right)}}\,,\\
   F_{2,6}^{\text{b}}&=\frac{\tilde{\kappa}^2}{2\sqrt{\nu} (1-\mu)  (1-\nu)\left(1-\sqrt{\mu \nu}\right)^2} \biggl[\mu^{3/2} (\nu (\nu (\nu+3)-1)+1)+\mu^2 \left(\nu^2+1\right) \sqrt{\nu}\\
   &\qquad\qquad\qquad\qquad\qquad\qquad-4 \mu \sqrt{\nu} \left(\nu^2+\nu+1\right)+\sqrt{\mu} (\nu ((\nu-1) \nu+3)+1)+\nu^{5/2}+\sqrt{\nu}\biggr]\,,
\end{split}
\end{align}
\begin{align}
\begin{split}
    F_{3,1}^{\text{b}}&=-\frac{(1+\mu) (1+\nu)}{(1-\mu) (1-\nu)}\,,\qquad
    F_{3,2}^{\text{b}}=\frac{2 \tilde{\kappa} (\mu+\nu)}{\sqrt{\left(1-\mu^2\right) \left(1-\nu^2\right)}}\,,\qquad
    F_{3,3}^{\text{b}}=-\frac{\tilde{\kappa}^2 (\mu+\nu) \left(\mu+\nu-2 \sqrt{\mu \nu}\right)}{(1-\mu) (1-\nu) \left(1-\sqrt{\mu \nu}\right)^2}\,,\\
    F_{3,4}^{\text{b}}&=\frac{(1-\mu) (1-\nu)}{(1+\mu) (1+\nu)}\,,\qquad
    F_{3,5}^{\text{b}}=0\,,\qquad
    F_{3,6}^{\text{b}}=\frac{\tilde{\kappa}^2 (\mu+\nu) \left(\mu+\nu-2 \sqrt{\mu \nu}\right)}{(1-\mu) (1-\nu) \left(1-\sqrt{\mu \nu}\right)^2}\,.
\end{split}
\end{align}
The corresponding metric takes the same expression with the one in \cite{Pomeransky:2006bd} by defining the new parameters $\tilde{\la}$ and $\tilde{\nu}$ by
\begin{align}
    \tilde{\la}=\mu+\nu\,,\qquad \tilde{\nu}=\mu\nu\,.
\end{align}

\subsection{Extremal Pomeransky-Sen'kov black ring}\label{sec:ePS}

Next, we consider the standard extremal limit of the (balanced) Pomeransky-Sen’kov black ring \cite{Elvang:2007hs}. This limit is realized as $\mu\to\nu$, corresponding to the case where the timelike rod degenerates to a single point. In this limit, the two simple poles $w_1$ and $w_2$ collide at a point $w=0$, and this degeneration gives rise to a simple pole and a double pole at $w=0$.
Relabeling these poles as
\begin{align}
    w_1^{\text{ex}}=\lim_{\mu \to \nu} w_1=\lim_{\mu \to \nu}w_2=0\,,\qquad w_2^{\text{ex}}=w_3=\frac{1}{2}\tilde{\kappa}^2\,,
\end{align}
the reduced monodromy matrix takes the form
\begin{align}\label{rps-mono}
    \cM^{\text{ex}}(w)=\lim_{\mu\to\nu}\cM(w)\lvert_{(\ref{bc-ups})}=Y_{\rm flat}+\frac{A_1^{(1)\text{ex}}}{w}+\frac{A_1^{(2)\text{ex}}}{w^2}+\frac{A_2^{(1)\text{ex}}}{w-w_2^{\text{ex}}}\,.
\end{align}
where the residue matrices for the simple poles are obtained as
\begin{align}
    A_1^{(1)\text{ex}}=\lim_{\mu\to\nu}(A_1+A_2)\lvert_{(\ref{bc-ups})}\,,\qquad A_2^{(1)\text{ex}}=\lim_{\mu\to\nu}A_{3}\lvert_{(\ref{bc-ups})}\,.
\end{align}
The explicit expressions of the residue matrices are 
\begin{align}
     \eta'A_1^{(1)\text{ex}}&=\left(
\begin{array}{cccccccc}
 \frac{4 \nu (\nu+1)^2\tilde{\kappa}^2}{(1-\nu)^4} & 0 & 0 & \frac{4 \nu (\nu+1) \tilde{\kappa}}{(1-\nu)^3} & 0 & 0 & -\frac{2 \nu (\nu+1) }{(1-\nu)^3}\tilde{\kappa}^3 & 0 \\
 0 & 0 & 0 & 0 & 0 & 0 & 0 & 0 \\
 0 & 0 & \frac{4 \nu}{(\nu+1)^2} & 0 & 0 & 0 & 0 & \frac{(2 \nu) \tilde{\kappa}^2}{(1-\nu)^2} \\
 \frac{4 \nu (\nu+1) \tilde{\kappa}}{(1-\nu)^3} & 0 & 0 & \frac{4 \nu}{(1-\nu)^2} & 0 & 0 & -\frac{2 \nu \tilde{\kappa}^2}{(1-\nu)^2} & 0 \\
 0 & 0 & 0 & 0 & -\frac{4 \nu \tilde{\kappa}^2}{(1-\nu)^2} & 0 & 0 & \frac{2 \nu (\nu+1) \tilde{\kappa}^3}{(1-\nu)^3} \\
 0 & 0 & 0 & 0 & 0 & 0 & 0 & 0 \\
 -\frac{2 \nu (\nu+1) \tilde{\kappa}^3}{(1-\nu)^3} & 0 & 0 & -\frac{2 \tilde{\kappa}^2 \nu}{(1-\nu)^2} & 0 & 0 & 0 & 0 \\
 0 & 0 & \frac{2 \nu \tilde{\kappa}^2}{(1-\nu)^2} & 0 & \frac{2 \nu (\nu+1) \tilde{\kappa}^3}{(1-\nu)^3} & 0 & 0 & 0 \\
\end{array}
\right)\,,\\
  \eta'A_1^{(2)\text{ex}}&=\frac{4 \nu^2}{(1-\nu)^2}\left(
\begin{array}{cccccccc}
 -\frac{\tilde{\kappa}^4}{(1-\nu)^2} & 0 & 0 & -\frac{\tilde{\kappa}^3}{1-\nu^2} & 0 & 0 & 0 & 0 \\
 0 & 0 & 0 & 0 & 0 & 0 & 0 & 0 \\
 0 & 0 & -\frac{\tilde{\kappa}^2}{(\nu+1)^2} & 0 & \frac{\tilde{\kappa}^3}{1-\nu^2} & 0 & 0 & -\frac{\tilde{\kappa}^4}{(1-\nu)^2} \\
 -\frac{\tilde{\kappa}^3}{1-\nu^2} & 0 & 0 & -\frac{\tilde{\kappa}^2}{(\nu+1)^2} & 0 & 0 & 0 & 0 \\
 0 & 0 & \frac{\tilde{\kappa}^3}{1-\nu^2} & 0 & -\frac{\tilde{\kappa}^4}{(1-\nu)^2} & 0 & 0 & \frac{\tilde{\kappa}^5 (\nu+1)}{(1-\nu)^3} \\
 0 & 0 & 0 & 0 & 0 & 0 & 0 & 0 \\
 0 & 0 & 0 & 0 & 0 & 0 & 0 & 0 \\
 0 & 0 & -\frac{\tilde{\kappa}^4}{(1-\nu)^2} & 0 & \frac{\tilde{\kappa}^5 (\nu+1)}{(1-\nu)^3} & 0 & 0 & -\frac{\tilde{\kappa}^6 (\nu+1)^2}{(1-\nu)^4} \\
\end{array}
\right)\,,\\
     \eta'A_2^{(1)\text{ex}}&=\left(
\begin{array}{cccccccc}
 -\frac{16 \tilde{\kappa}^2 \nu^2}{(1-\nu)^4}& 0 & 0 & -\frac{4  \nu (\nu+1)\tilde{\kappa}}{(1-\nu)^3}& 0 & 0 &0 & 0 \\
 0 & 0 & 0 & 0 & 0 & 0 & 0 & 0 \\
 0 & 0 & \frac{(1-\nu)^2}{(\nu+1)^2} & 0 & 0& 0 & 0 & 0 \\
-\frac{ 4 \nu (\nu+1)\tilde{\kappa}}{(1-\nu)^3} & 0 & 0 &-\frac{(\nu+1)^2}{(1-\nu)^2} & 0 & 0 &0& 0 \\
 0 & 0 & 0& 0 & 0 & 0 & 0 & 0 \\
 0 & 0 & 0 & 0 & 0 & 0 & 0 & 0 \\
0 & 0 & 0 &0 & 0 & 0 &0 & 0 \\
 0 & 0 &0 & 0 & 0 & 0 & 0 &0 \\
\end{array}
\right)\,,
\end{align}
and the rank of these matrices are
\begin{align}
    \text{Rank}\,A_1^{(1)\text{ex}}=4\,,\qquad 
    \text{Rank}\,A_1^{(2)\text{ex}}=2\,,\qquad 
    \text{Rank}\,A_2^{(1)\text{ex}}=2\,.
\end{align}
As in the case of the Myers-Perry black hole, one might suspect that the nilpotency of the charge matrix is related to the extremality of black ring solutions.
However, it is easy to verify that the charge matrices defined as
\begin{align}
    Q^{(1)\text{ex}}=Y_{\rm flat}^{-1}(A_1^{(1)\text{ex}}+A_2^{(1)\text{ex}})\,,\qquad 
    Q^{(2)\text{ex}}=Y_{\rm flat}^{-1}A_1^{(2)\text{ex}}
\end{align}
are not nilpotent. 
Therefore, the nilpotency of the charge matrix is not necessarily equivalent to the extremality of black ring solutions.
Moreover, since the reduced monodromy matrix (\ref{rps-mono}) does not exhibit any particular simplification, its factorization is considerably more involved than in the case of the extremal Myers-Perry black hole.
A systematic procedure for factorizing monodromy matrices containing double poles has been discussed, for example, in Ref.~\cite{Camara:2017hez}. Their method may prove useful for handling the factorization of the monodromy matrix (\ref{rps-mono}); however, we leave this issue for future investigation.

\subsection{ Myers-Perry black hole}\label{sec:MP}

In the analysis of the phase diagram of the balanced Pomeransky-Sen'kov black ring by Elvang and Rodriguez~\cite{Elvang:2007hs}, it was shown that another extremal limit exists in addition to the one discussed in the previous section.
In this limit, the black ring solution continuously degenerates into the extremal Myers-Perry black hole.
It is therefore expected that the charge matrix becomes nilpotent in this extremal limit.
In what follows, we examine whether this expectation is indeed realized.

\medskip
This limit is performed by considering the parameter region
\begin{align}\label{exmp-lim}
    \tilde{\la}\to 2\,,\qquad \tilde{\nu}\to 1\,,
\end{align}
while keeping the following quantities fixed:
\begin{align}
    \tilde{\sigma}=\frac{1+\tilde{\nu}-\tilde{\la}}{(1-\tilde{\nu})^2}\,,\qquad r_0^2=\frac{\tilde{\kappa}^2}{(1-\tilde{\nu})^2\tilde{\sigma}}\,.
\end{align}
This limit is realized by taking $\epsilon\to 0$ together with the redefinition
\begin{align}
    \tilde{\la}=2-\epsilon-\tilde{\sigma}\epsilon^2\,,\qquad \tilde{\nu}=1-\epsilon\,,\qquad \tilde{\kappa}=\epsilon\,\tilde{k}\,.
\end{align}
By combining the coordinate transformations of the $C$-metric coordinates $u$ and $v$ (see Eq.~(4.16) in \cite{Elvang:2007hs}),
\begin{align}\label{bps-coord}
    u=-1+\frac{16\sqrt{a_2}\tilde{\kappa}^3\cos^2\theta}{(a_1+a_2)^{3/2}(r^2-a_1a_2)}\,,\qquad v=-1-\frac{16\sqrt{a_2}\tilde{\kappa}^3\sin^2\theta}{(a_1+a_2)^{3/2}(r^2-a_1a_2)}\,,
\end{align}
the metric of the Pomeransky-Sen’kov black ring becomes the extremal Myers-Perry black hole solution with the extremal condition
\begin{align}
    r_0=\frac{4\tilde{k}}{\sqrt{\tilde{\sigma}}}=a_1+a_2\,,\qquad a_1=r_0(1-\tilde{\sigma})\,,\qquad a_2=r_0\tilde{\sigma}\,.
\end{align}
In this case, the monodromy matrix for the balanced Pomeransky-Sen'kov black ring reduces to one with a double pole at $w=0$ :
\begin{align}\label{exmpp-mono}
    \cM_{\text{exMP}}(w)=Y_{\rm flat}+\frac{A_1^{(1)\text{ex}}}{w}+\frac{A_1^{(2)\text{ex}}}{w^2}\,.
\end{align}
The residue matrices are
\footnotesize
\begin{align}
    A_1^{(1)\text{ex}}&=
 \left(
\begin{array}{cccccccc}
 -\frac{(a_{12}^+)^2}{4} & 0 & 0 & 0 & 0 & 0 & \frac{1}{8} (a_{12}^+)^2 a_{12} & 0 \\
 0 & 0 & 0 & 0 & 0 & 0 & 0 & 0 \\
 0 & 0 & -1 & 0 & 0 & 0 & 0 & -\frac{(a_{12}^+)^2}{8} \\
 0 & 0 & 0 & -1 & 0 & 0 & -\frac{(a_{12}^+)^2}{8} & 0 \\
 0 & 0 & 0 & 0 & \frac{(a_{12}^+)^2}{4} & 0 & 0 & -\frac{1}{8} (a_{12}^+)^2 a_{12} \\
 0 & 0 & 0 & 0 & 0 & 0 & 0 & 0 \\
 \frac{1}{8} (a_{12}^+)^2 a_{12} & 0 & 0 & \frac{(a_{12}^+)^2}{8} & 0 & 0 & -(a_{12}^+)^2A_{1+}A_{1-} & 0 \\
 0 & 0 & \frac{(a_{12}^+)^2}{8} & 0 & \frac{1}{8} (a_{12}^+)^2 a_{12} & 0 & 0 & -(a_{12}^+)^2A_{1+} A_{1-} \\
\end{array}
\right)\,,\\
    A_1^{(2)\text{ex}}&=
  \left(
\begin{array}{cccccccc}
 \frac{a_{1} (a_{12}^+)^3}{16} & 0 & 0 & \frac{a_{1} (a_{12}^+)^2}{8} & 0 & 0 & -\frac{1}{8} a_{1} (a_{12}^+)^3 A_{1-} & 0 \\
 0 & 0 & 0 & 0 & 0 & 0 & 0 & 0 \\
 0 & 0 & \frac{a_{2} (a_{12}^+)}{4} & 0 & -\frac{a_{2} (a_{12}^+)^2}{8} & 0 & 0 & \frac{1}{4} a_{2} (a_{12}^+)^2 A_{1+} \\
 -\frac{a_{1} (a_{12}^+)^2}{8} & 0 & 0 & -\frac{a_{1} (a_{12}^+)}{4} & 0 & 0 & \frac{1}{4} a_{1} (a_{12}^+)^2 A_{1-} & 0 \\
 0 & 0 & -\frac{a_{2} (a_{12}^+)^2}{8} & 0 & \frac{a_{2} (a_{12}^+)^3}{16} & 0 & 0 & \frac{1}{8} (-a_{2}) (a_{12}^+)^3 A_{1+} \\
 0 & 0 & 0 & 0 & 0 & 0 & 0 & 0 \\
 -\frac{1}{8} a_{1} (a_{12}^+)^3 A_{1-} & 0 & 0 & -\frac{1}{4} a_{1} (a_{12}^+)^2 A_{1-} & 0 & 0 & \frac{1}{4} a_{1} (a_{12}^+)^3 A_{1-}^2 & 0 \\
 0 & 0 & -\frac{1}{4} a_{2}(a_{12}^+)^2 A_{1+} & 0 & \frac{1}{8} a_{2} (a_{12}^+)^3 A_{1+} & 0 & 0 & -\frac{1}{4}a_{2} (a_{12}^+)^3 A_{1+}^2 \\
\end{array}
\right)\,,
\end{align}
\normalsize
where we set $a_{12}^+=a_1+a_2$, $A_{1+}=\frac{1}{8}(3a_1-a_2), A_{1-}=\frac{1}{8}(a_1-3a_2)$. 
It should be noted that, since the monodromy matrix is independent of the coordinates, 
the coordinate transformation (\ref{bps-coord}) is not required before taking the limit (\ref{exmp-lim}) when one is concerned only with the relation between the monodromy matrices of the two black hole solutions.

\medskip
The charge matrices $Q^{(1)}=Y_{\rm flat}^{-1}A_1^{(1)\text{ex}}$ and $\widetilde{Q}^{(2)}=Q^{(2)}-\frac{1}{2} (Q^{(1)})^2\,(Q^{(2)}=Y_{\rm flat}^{-1}A_1^{(2)\text{ex}})$ are expanded as
\begin{align}
Q^{(1)}&=-\frac{M}{3 \pi}(H_1+H_2+H_3)-\frac{1}{64} (a_1+a_2)^2 \left(3a_1^2-10 a_1a_2+3 a_2^2\right) E_0 +\frac{J_1-J_2}{2 \pi} (E_{p_0}+E_{q^0})+F_0\,,\\
\widetilde{Q}^{(2)}&=\frac{1}{8} \left(a_1^2-a_2^2\right)\Bigl(\frac{1}{8}(a_{12}^+)^2(H_1+H_2+H_3)+\frac{1}{64} (a_{12}^+)^2 \left(a_1^2-14 a_1a_2+a_2^2\right) E_0\no\\
&\qquad-\frac{(a_1+a_2)^2 \left(a_1^2-6 a_1 a_2+a_2^2\right)}{16 (a_1-a_2)}(E_{p_0}+E_{q^0})+\frac{(a_1+a_2)^2}{2 (a_1-a_2)} (F_{p_0}+F_{q^0})+F_0\Bigr)\,,
\end{align}
where $M$ and $J_{1,2}$ are the asymptotic quantities (\ref{mp-asy}) with $r_0^2=(a_1+a_2)^2$.
These charge matrices are not nilpotent, but they satisfy the cubic relation (\ref{mp-q-rel}) and the quadratic relation\footnote{We can check the trace of the square of the charge matrix satisfies $\frac{1}{4}\Tr(Q^2)=\left(\frac{1}{2} \Tr(\widetilde{Q}^{(2)}\widetilde{Q}^{(2)})\right)^2$.}
\begin{align}
    \left(\widetilde{Q}^{(2)}\right)^2-\frac{1}{4}\Tr(Q^2)\widetilde{Q}^{(2)}=0\,,
\end{align}
respectively.
This follows from choosing the branch $r_0=a_1+a_2$ rather than $r_0=a_1-a_2$.
Hence, the reduced monodromy matrix (\ref{exmpp-mono}) clearly exhibits a different algebraic structure from (\ref{exmp-mono}) which corresponds to another branch $r_0=a_1-a_2$. 
At a conceptual level, factorizing this reduced monodromy matrix should yield the same extremal doubly rotating Myers-Perry black hole black hole solution as obtained from the factorization of (\ref{exmp-mono}).
A more detailed analysis of this issue will be presented elsewhere.

\medskip
For completeness, we also comment on the limit from the unbalanced Pomeransky-Sen'kov black ring to the five-dimensional doubly rotating non-extremal Myers-Perry black hole.
The monodromy matrix (\ref{mr-mono}) for the Myers-Perry black hole can be obtained by taking a limit $\mu\to 1$ after imposing the condition \cite{Chen:2011jb}
\begin{align}\label{mp-lim}
    \la=1-c(1-\mu)\,,\qquad 0<c\leq 1\,.
\end{align}
The physical parameters $r_0, a_1, a_2$ in the Myers-Perry black hole are expressed in terms of $c, \tilde{\kappa}, \nu$ as
\begin{align}\label{mp-ring-para}
    r_0^2&=\frac{4\tilde{\kappa}^2(1+\nu)}{1-\nu}\,,\quad
    a_1=\frac{2\tilde{\kappa}\sqrt{(1-c^2)(1-\nu)(1+\nu+c-c\nu)}}{\sqrt{c}(1-\nu+c+c \nu)}\,,\quad
    a_2=\frac{4\tilde{\kappa}\sqrt{c\nu(1+\nu+c-c\nu)}}{\sqrt{1-\nu}(1-\nu+c+c \nu)}\,.
\end{align}
Since from these identifications we have
\begin{align}
    \alpha=\tilde{\kappa}^2\,,
\end{align}
the simple poles (\ref{pole-ups}) for the unbalanced Pomeransky-Sen’kov black ring precisely reduces to those of the non-extremal Myers-Perry black hole
\begin{align}
   \lim_{\mu\to1} w_1=-\frac{\tilde{\kappa}^2}{2}=-\frac{\alpha}{2}\,,\qquad 
   \lim_{\mu\to1} w_2=\lim_{\mu\to1} w_3=\frac{\tilde{\kappa}^2}{2}=\frac{\alpha}{2}\,.
\end{align}
To rigorously realize the above limit for the metric, it is necessary to perform the coordinate transformation \cite{Chen:2011jb}
\begin{align}\label{coord-tr}
    u=-1+\frac{8\alpha \cos^2\theta(1-\mu)}{2r^2+a_1^2+a_2^2-r_0^2-4\alpha\cos 2\theta}\,,\qquad 
    v=-1-\frac{8\alpha\sin^2\theta(1-\mu)}{2r^2+a_1^2+a_2^2-r_0^2-4\alpha\cos 2\theta}
\end{align}
before taking the limit $\mu\to1$. 
Similarly, the monodromy matrix (\ref{br-mono}) for the unbalanced Pomeransky-Sen'kov black ring reduces to that (\ref{mr-mono}) of the non-extremal Myers-Perry black hole in the limit $\mu\to1$ without the need to perform the coordinate transformation (\ref{coord-tr}).

\subsection{Emperan-Reall black ring}\label{sec:ER}

The (unbalanced) Emparan-Reall black ring~\cite{Emparan:2001wn}, which rotates only along the $\tilde{\phi}$ direction, can be obtained by taking the limit:
\begin{align}\label{er-limit}
    \nu\to 0
\end{align}
of the unbalanced Pomeransky-Sen'kov black ring.
While the building blocks $F_{2,3},F_{2,5},F_{2,6}$ of the residue matrix $A_2$ diverges in this limit, the residue matrix $A_2$ itself has only finite entries and reduces to a simple expression
\begin{align}
   \eta'A_2^{\rm ER}= \left(
\begin{array}{cccccccc}
 0 & 0 & 0 & 0 & 0 & 0 & 0 & 0 \\
 0 & 0 & 0 & 0 & 0 & 0 & 0 & 0 \\
 0 & 0 & 0 & 0 & 0 & 0 & 0 & 0 \\
 0 & 0 & 0 & 0 & 0 & 0 & 0 & 0 \\
 0 & 0 & 0 & 0 & -\frac{\tilde{\kappa}^2 \la (1-\mu)}{1-\la} & 0 & 0 & \sqrt{\frac{\la(\la-\mu)}{2(1-\la^2)}}\frac{\tilde{\kappa}^3 (1+\la) (1-\mu) }{1-\la} \\
 0 & 0 & 0 & 0 & 0 & 0 & 0 & 0 \\
 0 & 0 & 0 & 0 & 0 & 0 & -\frac{1}{2} \tilde{\kappa}^4 \mu(1-\mu)  & 0 \\
 0 & 0 & 0 & 0 &\sqrt{\frac{\la(\la-\mu)}{2(1-\la^2)}}\frac{\tilde{\kappa}^3 (1+\la) (1-\mu) }{1-\la} & 0 & 0 & -\frac{\tilde{\kappa}^4 (1+\la) (1-\mu) (\la-\mu)}{2 (1-\la)^2} \\
\end{array}
\right)\,.
\end{align}
On the other hand, all $F_{1,j}$ and $F_{3,j}$ are finite under the limit, and hence the reduced matrix $A_1^{\rm ER}$ and $A_3^{\rm ER}$ are expressed in terms of $F_{i,j}^{\text{ER}}$ which are given by
\begin{align}
    F_{1,1}^{\text{ER}}&=\frac{2 (\la-\mu)}{(1-\la) (\mu+1)}\,,\quad
    F_{1,2}^{\text{ER}}=\frac{\tilde{\kappa}(1-\mu)}{(1-\la) }\sqrt{\frac{\la(1-\la^2)}{2(\la-\mu)}}\,,\quad
    F_{1,3}^{\text{ER}}=-F_{1,6}^{\text{ER}}=-\frac{\tilde{\kappa}^2 (1-\mu)}{2 (1-\la)}\,,\quad F_{1,4}^{\text{ER}}=\frac{2 \mu}{1+\mu}\,,\quad
    F_{1,5}^{\text{ER}}=0\,,\\
    F_{3,1}^{\text{ER}}&=-\frac{(1+\la) (1-\mu)}{(1-\la) (1+\mu)}\,,\quad
    F_{3,2}^{\text{ER}}=\tilde{\kappa}\sqrt{\frac{  2\la(\la-\mu)}{1-\la^2}}\,,\quad
    F_{3,3}^{\text{ER}}=-F_{3,6}^{\text{ER}}=-\frac{\tilde{\kappa}^2 (\la \mu+\la-2 \mu)}{2 (1-\la)}\,,\quad
    F_{3,4}^{\text{ER}}=\frac{1-\mu}{1+\mu}\,,\quad
    F_{3,5}^{\text{ER}}=0\,.
\end{align}
By replacing the parameters $(\la,\mu)$ with $(b,c)$, the resulting monodromy matrix coincides with the one obtained in our previous work~\cite{Sakamoto:2025xbq}.

\subsection{Mishima-Iguchi-Figueras black ring}\label{sec:MIF}

Finally, we consider the limit of the unbalanced Pomeransky-Sen'kov black ring to the $S^2$-rotating black ring that carries a single angular momentum along the $\tilde{\psi}$ direction~ \cite{Mishima:2005id,Figueras:2005zp}. 
This occurs when, instead of imposing Eq.~(\ref{er-limit}), we take the limit:
\begin{align}
    \mu\to0\,.
\end{align}
As in the previous example, some $F_{1,j}$ and $F_{2,j}$ diverge in this limit, but the resulting residue matrices $A_{1}^{\text{MIF}}$ and $A_{2}^{\text{MIF}}$ remain finite and take the expressions
\begin{align}
   \eta'A_1^{\text{MIF}}&=\left(
\begin{array}{cccccccc}
 F_{1,1}^{\text{MIF}} & 0 & 0 & 0 & 0 & 0 & F_{1,1}^{\text{MIF}} F_{1,2}^{\text{MIF}} & 0 \\
 0 & 0 & 0 & 0 & 0 & 0 & 0 & 0 \\
 0 & 0 & F_{1,4}^{\text{MIF}} & 0 & F_{1,4}^{\text{MIF}}F_{1,5}^{\text{MIF}} & 0 & 0 & F_{1,4}^{\text{MIF}}F_{1,6}^{\text{MIF}} \\
 0 & 0 & 0 & 0 & 0 & 0 & 0 & 0 \\
 0 & 0 &  F_{1,4}^{\text{MIF}} F_{1,5}^{\text{MIF}} & 0 & F_{1,4}^{\text{MIF}}(F_{1,5}^{\text{MIF}})^2 & 0 & 0 &F_{1,4}^{\text{MIF}} F_{1,5}^{\text{MIF}}F_{1,6}^{\text{MIF}} \\
 0 & 0 & 0 & 0 & 0 & 0 & 0 & 0 \\
 F_{1,1}^{\text{MIF}} F_{1,2}^{\text{MIF}} & 0 & 0 & 0 & 0 & 0 & F_{1,1}^{\text{MIF}} (F_{1,2}^{\text{MIF}})^2 & 0 \\
 0 & 0 & F_{1,4}^{\text{MIF}}F_{1,6}^{\text{MIF}}& 0 & F_{1,4}^{\text{MIF}} F_{1,5}^{\text{MIF}}F_{1,6}^{\text{MIF}} & 0 & 0 &  F_{1,4}^{\text{MIF}}( F_{1,6}^{\text{MIF}})^2 \\
\end{array}
\right)\,,\\
   \eta'A_2^{\text{MIF}}&= \eta'A_1^{\text{MIF}}\biggl\lvert_{\mu\leftrightarrow\nu}\,,
\end{align}
where
\begin{align}
\begin{split}
    F_{1,1}^{\text{MIF}}&=\frac{\tilde{\kappa}^2 \mu (1-\nu) (\mu+\nu)}{(\mu-\nu) (1-\mu \nu)}\,,\quad
    F_{1,2}^{\text{MIF}}=-F_{1,5}^{\text{MIF}}=\frac{\tilde{\kappa} (\mu+1) \sqrt{\mu \nu}}{\mu \sqrt{2 (1-\mu \nu)}}\,,\\
    F_{1,4}^{\text{MIF}}&=\frac{2 \mu (\mu+\nu)}{(\mu+1) (\mu-\nu)}\,,\quad
    F_{1,6}^{\text{MIF}}=\frac{\tilde{\kappa}^2 (1-(\mu+2) \nu)}{2 (1-\mu \nu)}\,.
\end{split}
\end{align}
The third residue matrix $A_3^{\text{MIF}}$ is expressed as (\ref{sym-A-ups}) with 
\begin{align}
    F_{3,1}^{\text{MIF}}&=-1\,,\quad
    F_{3,2}^{\text{MIF}}=0\,,\quad
    F_{3,3}^{\text{MIF}}=-F_{3,6}^{\text{MIF}}=\frac{\tilde{\kappa}^2 (\mu+\nu)}{2 (1-\mu \nu)}\,,\quad
    F_{3,4}^{\text{MIF}}=\frac{(1-\mu) (1-\nu)}{(1+\mu) (1+\nu)}\,,\quad
    F_{3,5}^{\text{MIF}}=0\,.
\end{align}

\section{Conclusion and discussion}\label{sec:dis}

We have presented an extension of the Breitenlohner-Maison (BM) linear system approach to encompass doubly rotating configurations in five-dimensional vacuum Einstein gravity, thereby generalizing previous our work that focused on single-angular-momentum black holes~\cite{Sakamoto:2025xbq}. 
The analysis is based on the integrability of a two-dimensional non-linear sigma model whose target space is the symmetric coset space $SO(4,4)/(SO(2,2)\times SO(2,2))$, allowing the reduction of the Einstein equations for asymptotically flat, stationary, and bi-axisymmetric spacetimes into a linear system amenable to algebraic methods. 
Within this unified framework, we have constructed the explicit monodromy matrices corresponding to two fundamental examples of five-dimensional black objects: 
the doubly rotating Myers-Perry black hole with the horizon-cross section of the topology $S^3$, and 
the Pomeransky-Sen'kov black ring with the horizon-cross section of the topology $S^1\times S^2$. By analyzing these monodromy matrices, we have showed how the algebraic data encoded in the poles and residues of the monodromy matrix faithfully reproduce the geometric characteristics of the underlying spacetime, including the rod structure encoding the horizon topologies, the topologies of the domain of communication, and the angular velocities of the horizon. 
This might help us establish a clear correspondence between the algebraic structures in the BM framework and the physical parameters characterizing the gravitational solutions.
In addition, our results reveal that the extremal limits of these black holes correspond to special degenerations in the pole structure of the monodromy matrix. 
For the Myers-Perry black hole, the charge matrix $Q$ satisfies the nilpotent condition, leading to compact exponential representations consistent with known extremal solutions, whereas no such nilpotent behavior arises for the Pomeransky-Sen'kov black ring.

\medskip
We further have explored the inverse procedure, showing that the complete spacetime geometries can be recovered from the factorization of the monodromy matrix through the Riemann-Hilbert problem, following the formalism developed by Katsimpouri and collaborators. 
By explicitly solving this problem for both the Myers-Perry and Pomeransky-Sen'kov cases, we have confirmed that the monodromy matrix reproduces the exact metrics of these solutions.
These results not only validate the BM method as a robust, unifying algebraic approach for constructing higher-dimensional rotating black holes but also highlight its potential for systematic generalization. 
In particular, the work opens a pathway toward generating new classes of solutions---such as multi-black-hole systems, black lenses, and other non-spherical topologies---by engineering appropriate monodromy data subject to regularity and asymptotic constraints, thereby providing a powerful algebraic foundation for future explorations in higher-dimensional gravitational physics.

\section*{Acknowledgements}

S.T.\ was supported by JSPS KAKENHI Grant Number 21K03560.

\newpage

\appendix

\section{Explicit expression of the twist potential}\label{sec:twist-ups}

In this appendix, we present the explicit expression of the numerator $\sigma_0(u,v)$ of the twist potential
\begin{align}
    \sigma=\frac{\tilde{\kappa}^2\sigma_0(u,v)}{2(1-\mu\nu)\Phi(u-v)H(u,v)H(v,u)}\,.
\end{align}
The numerator $\sigma_0(u,v)$ is a symmetric polynomial with degree 5 of $u$ and $v$, and takes the form
\begin{align}
    \sigma_0(u,v)=f(u,v)+\sqrt{\nu(\la-\mu)(1-\la\mu)}(\mu+\nu)h(u,v)\,,
\end{align}
where
\begin{align}
    f(u,v)=\sum_{i,j=1}^{5}f^{i,j}u^iv^j\,,\qquad  h(u,v)=\sum_{i,j=1}^{5}h^{i,j}u^iv^j\,.
\end{align}

\subsubsection*{$f$ part}

\begin{align}
   f^{5,5}&=2 \nu^2 (\la \mu-1) (\mu+\nu) \Bigl(\la^4 \nu^4 (\la \nu+1)+\la^3 \mu \nu^3 \left(4 \la^2 \nu-5 \la \nu^2+\la-4 \nu\right)\no\\
   &\quad+\la \mu^6 \left(\la^5-2 \la^3 \left(2 \nu^2+1\right)+\la^2 \nu \left(4 \nu^2-1\right)+\la \left(-\nu^4+4 \nu^2+1\right)-\nu^3+\nu\right)\no\\
   &\quad+\la^2 \mu^2 \nu^2 \left(\la^4 \left(-\nu^2\right)+4 \la^3 \nu+\la^2 \left(2-16 \nu^2\right)+5 \la \nu \left(2 \nu^2-1\right)+6 \nu^2-2\right)\no\\
   &\quad+\mu^5 \Bigl(\la^5 \left(8 \nu^2-2\right)+\la^4 \left(2 \nu-12 \nu^3\right)+\la^3 \left(4 \nu^4-6 \nu^2+4\right)-\la^2 \left(\nu^3+\nu\right)+\la \left(4 \nu^4-2 \nu^2-2\right)-\nu \left(\nu^2-1\right)^2\Bigr)\no\\
   &\quad+\la \mu^3 \nu \Bigl(-4 \la^5 \nu^2+2 \la^4 \left(2 \nu^3+\nu\right)+\la^3 \left(1-11 \nu^2\right)+6 \la^2 \nu \left(4 \nu^2-1\right)
   +\la \left(-10 \nu^4+9 \nu^2-1\right)-4 \nu \left(\nu^2-1\right)\Bigr)\no\\
   &\quad+\mu^4 \Bigl(-4 \la^6 \nu^2+\la^5 \nu \left(12 \nu^2-1\right)+\la^4 \left(1-6 \nu^4\right)+\la^3 \nu \left(9 \nu^2-1\right)\no\\
   &\quad-2 \la^2 \left(8 \nu^4-3 \nu^2+1\right)+\la \nu \left(5 \nu^4-7 \nu^2+2\right)+\left(\nu^2-1\right)^2\Bigr)\Bigr)\,,\\ 
   f^{4,5}&=-\nu^2\Bigl(\la^2 \left(-2 \nu \la^5+\left(6 \nu^2+1\right) \la^4+\left(10 \nu-6 \nu^3\right) \la^3+\left(2 \nu^4-14 \nu^2-1\right) \la^2+2 \nu \left(\nu^2-4\right) \la+\nu^4+\nu^2\right) \mu^8\no\\
   &\quad+\la \Bigl(-3 \left(2 \nu^2+1\right) \la^6+2 \left(6 \nu^3+\nu\right) \la^5+\left(-6 \nu^4+16 \nu^2+2\right) \la^4+\left(8 \nu^3-22 \nu\right) \la^3+\left(-21 \nu^4+25 \nu^2+1\right) \la^2\no\\
   &\quad+2 \nu \left(3 \nu^4-9 \nu^2+10\right) \la+3 \nu^2 \left(\nu^2-1\right)\Bigr) \mu^7\no\\
   &\quad+\Bigl(-2 \left(3 \nu^3+\nu\right) \la^7+\left(6 \nu^4-13 \nu^2+9\right) \la^6-8 \nu^3 \la^5+4 \left(14 \nu^4-9 \nu^2-3\right) \la^4\no\\
   &\quad+\left(-28 \nu^5+34 \nu^3+18 \nu\right) \la^3+\left(\nu^6+2 \nu^4-8 \nu^2+3\right) \la^2-8 \nu \left(\nu^4-3 \nu^2+2\right) \la+2 \nu^2 \left(\nu^2-1\right)^2\Bigr) \mu^6\no\\
   &\quad+\Bigl(\left(11 \nu^2-2 \nu^4\right) \la^7-2 \nu \left(8 \nu^2-5\right) \la^6+\left(-52 \nu^4+25 \nu^2-9\right) \la^5+\left(52 \nu^5-42 \nu^3-4 \nu\right) \la^4\no\\
   &\quad+\left(-5 \nu^6-9 \nu^4+3 \nu^2+14\right) \la^3+2 \nu \left(15 \nu^4-22 \nu^2-5\right) \la^2+\left(-11 \nu^6+15 \nu^4+\nu^2-5\right) \la+4 \nu \left(\nu^2-1\right)^2\Bigr) \mu^5\no\\
   &\quad+\Bigl(14 \nu^3 \la^7+15 \nu^2 \left(\nu^2-1\right) \la^6-2 \nu \left(24 \nu^4-27 \nu^2+7\right) \la^5+\left(10 \nu^6-15 \nu^4+12 \nu^2+3\right) \la^4\no\\
   &\quad+\left(-40 \nu^5+24 \nu^3+10 \nu\right) \la^3+\left(25 \nu^6-32 \nu^4+2 \nu^2-5\right) \la^2+\left(-12 \nu^5+8 \nu^3+4 \nu\right) \la+2 \left(\nu^2-1\right)^2\Bigr) \mu^4\no\\
   &\quad+\la \nu \Bigl(\nu^3 \la^6+\left(22 \nu^4-28 \nu^2\right) \la^5-2 \nu \left(5 \nu^4-17 \nu^2+2\right) \la^4+2 \left(10 \nu^4-6 \nu^2+3\right) \la^3\no\\
   &\quad+\left(-30 \nu^5+46 \nu^3-13 \nu\right) \la^2+2 \left(4 \nu^4+7 \nu^2-3\right) \la-9 \nu \left(\nu^2-1\right)\Bigr) \mu^3\no\\
   &\quad+\la^2 \nu^2 \left(-4 \nu^3 \la^5+5 \nu^2 \left(\nu^2-3\right) \la^4+8 \nu \la^3+4 \left(5 \nu^4-9 \nu^2+2\right) \la^2+4 \nu \left(2 \nu^2-5\right) \la+15 \nu^2-5\right) \mu^2\no\\
   &\quad-\la^3 \nu^3 \left(\nu^3 \la^4+2 \nu^2 \la^3+\nu \left(7 \nu^2-11\right) \la^2+6 \left(2 \nu^2-1\right) \la+11 \nu\right) \mu+\la^4 \nu^4 \left(\la^2 \nu^2+4 \la \nu+3\right)\Bigr)\,,\\ 
   f^{3,5}&=\nu^2 \Bigl(\mu (\mu+\nu)^3 \left(2 \mu^4+2 \mu^3+\left(-2 \nu^2+5 \nu+1\right) \mu^2+2 \nu (2 \nu-3) \mu+\nu^2 (\nu+1)\right) \la^7\no\\
   &\quad-(\mu+\nu)^2 \Bigl((4 \nu+2) \mu^6+\left(-4 \nu^2+14 \nu+4\right) \mu^5+\left(-8 \nu^3+27 \nu^2-6 \nu+7\right) \mu^4\no\\
   &\quad+\left(20 \nu^3-14 \nu^2+19 \nu+3\right) \mu^3+\nu \left(5 \nu^3+2 \nu^2+18 \nu-11\right) \mu^2+\nu^2 \left(-2 \nu^2+7 \nu-7\right) \mu+(\nu-1) \nu^3\Bigr) \la^6\no\\
   &\quad+\Bigl(7 (\nu-1) \mu^8+\left(-12 \nu^3+41 \nu^2-15 \nu+6\right) \mu^7+\left(-24 \nu^4+91 \nu^3-15 \nu^2+48 \nu-2\right) \mu^6\no\\
   &\quad+\left(-12 \nu^5+97 \nu^4-17 \nu^3+131 \nu^2-20 \nu+9\right) \mu^5+\left(50 \nu^5-20 \nu^4+170 \nu^3-52 \nu^2+35 \nu+3\right) \mu^4\no\\
   &\quad+\nu \left(10 \nu^5-20 \nu^4+114 \nu^3-48 \nu^2+53 \nu-1\right) \mu^3+\nu^2 \left(-10 \nu^4+38 \nu^3-18 \nu^2+39 \nu-11\right) \mu^2\no\\
   &\quad+\nu^3 \left(5 \nu^3-4 \nu^2+14 \nu-11\right) \mu+2 (\nu-2) \nu^4\Bigr) \la^5\no\\
   &\quad+\Bigl(\left(4 \nu^3-9 \nu^2+10 \nu+1\right) \mu^8+\left(12 \nu^4-38 \nu^3+4 \nu^2-13 \nu+21\right) \mu^7\no\\
   &\quad+\left(8 \nu^5-59 \nu^4-8 \nu^3-80 \nu^2+57 \nu-6\right) \mu^6+\left(-40 \nu^5+18 \nu^4-159 \nu^3+51 \nu^2-42 \nu+10\right) \mu^5\no\\
   &\quad+\left(-10 \nu^6+40 \nu^5-145 \nu^4+7 \nu^3-95 \nu^2+24 \nu-5\right) \mu^4+\left(20 \nu^6-62 \nu^5-2 \nu^4-96 \nu^3+26 \nu^2-15 \nu-1\right) \mu^3\no\\
   &\quad-\nu \left(10 \nu^5-6 \nu^4+45 \nu^3-18 \nu^2+16 \nu+1\right) \mu^2-\nu^2 \left(8 \nu^3-2 \nu^2+7 \nu+1\right) \mu-\nu^3 \left(4 \nu^2+\nu+1\right)\Bigr) \la^4\no\\
   &\quad+\mu \Bigl(\left(-2 \nu^4+5 \nu^3+3 \nu^2-2 \nu+4\right) \mu^7+\left(-2 \nu^5+15 \nu^4+9 \nu^3+11 \nu^2-22 \nu-3\right) \mu^6\no\\
   &\quad+\left(15 \nu^5-9 \nu^4+56 \nu^3-26 \nu^2+3 \nu-21\right) \mu^5+\left(5 \nu^6-35 \nu^5+81 \nu^4+28 \nu^3+42 \nu^2-53 \nu+2\right) \mu^4\no\\
   &\quad+\left(-20 \nu^6+48 \nu^5+24 \nu^4+75 \nu^3-47 \nu^2+12 \nu-8\right) \mu^3+\left(10 \nu^6-4 \nu^5+51 \nu^4-9 \nu^3+21 \nu^2-10 \nu+1\right) \mu^2\no\\
   &\quad+2 \left(6 \nu^5+11 \nu^4+7 \nu^3+\nu\right) \mu+\nu^2 \left(16 \nu^3+3 \nu^2+2 \nu+1\right)\Bigr) \la^3\no\\
   &\quad+\mu^2 \Bigl(-\nu^2 \left(\nu^2+2 \nu-1\right) \mu^6+\left(-2 \nu^5+2 \nu^4-5 \nu^3+\nu^2+4 \nu-12\right) \mu^5\no\\
   &\quad+\left(-\nu^6+14 \nu^5-18 \nu^4-15 \nu^3+\nu^2+10 \nu+3\right) \mu^4\no\\
   &\quad+\left(10 \nu^6-17 \nu^5-15 \nu^4-18 \nu^3+44 \nu^2+5 \nu+7\right) \mu^3+\nu \left(-5 \nu^5+\nu^4-23 \nu^3+8 \nu^2-2 \nu+15\right) \mu^2\no\\
   &\quad+\left(-8 \nu^5-38 \nu^4-7 \nu^3+15 \nu^2+2\right) \mu-24 \nu^5-3 \nu^4+7 \nu^3+2 \nu\Bigr) \la^2\no\\
   &\quad-\mu^3 \left(\nu^2-1\right) \Bigl(\nu^2 (2 \nu-1) \mu^4+2 \left(\nu^4-\nu^3-\nu+6\right) \mu^3\no\\
   &\quad-\left(\nu^4+4 \nu^2-6 \nu+1\right) \mu^2-2 \nu \left(\nu^2+11 \nu+1\right) \mu-\nu^2 (16 \nu+1)\Bigr) \la-4 \mu^4 (\mu+\nu) \left(\nu^2-1\right)^2\Bigr)\,,\\ 
   f^{2,5}&=\la \nu^2 (\mu+\nu)^2\Bigl(\la^2 \mu-\la \left(\mu^2+1\right)+\mu\Bigr) \no\\
   &\quad\times \Bigl(\la^4 \left(3 \mu^4-\mu^3 (\nu-1)+\mu^2 \nu (3-5 \nu)-\mu \nu \left(\nu^2-3 \nu+2\right)+(\nu-2) \nu^2\right)\no\\
   &\quad+\la^3 \Bigl(\mu^4 (\nu-1)+5 \mu^3 \left(2 \nu^2-\nu-1\right)+\mu^2 \left(3 \nu^3-7 \nu^2+2 \nu-2\right)+\mu \nu \left(-3 \nu^2+7 \nu-4\right)+2 \nu \left(2 \nu^2-\nu+1\right)\Bigr)\no\\
   &\quad+\la^2 \Bigl(\mu^4 \left(-5 \nu^2+2 \nu-4\right)+\mu^3 \left(-3 \nu^3+5 \nu^2+4 \nu+2\right)\no\\
   &\quad+\mu^2 \left(3 \nu^3-8 \nu^2+6 \nu+1\right)-\mu \left(12 \nu^3-4 \nu^2+\nu-1\right)+\nu (4 \nu+1)\Bigr)\no\\
   &\quad+\la \mu \left(\mu^3 \nu \left(\nu^2-\nu-4\right)-\mu^2 \left(\nu^3-3 \nu^2+2 \nu-8\right)+\mu \left(12 \nu^3-2 \nu^2-5 \nu-1\right)-8 \nu^2-\nu+1\right)\no\\
   &\quad-4 \mu^2 \left(\nu^2-1\right) (\mu \nu-1)\Bigr)\,,\\ 
   f^{1,5}&=-4 \la^2 \nu^3 (\la-\mu)^3 (\la \mu-1)^2 (\mu+\nu)^3\,,
\end{align}

\begin{align}
    f^{4,4}&=2 (\la-\mu) \nu^2 (\mu+\nu) \Bigl(\mu (\mu+\nu)^2 \left((7 \nu-2) \mu^3-\left(\nu^2+5 \nu-2\right) \mu^2-\nu (4 \nu+1) \mu+(3-\nu) \nu^2\right) \la^6\no\\
   &\quad+\Bigl(\left(-13 \nu^2+7 \nu-4\right) \mu^6+\left(-10 \nu^3+25 \nu^2-24 \nu+7\right) \mu^5+\left(3 \nu^4+33 \nu^3-35 \nu^2+26 \nu-6\right) \mu^4\no\\
   &\quad+\nu \left(19 \nu^3-33 \nu^2+37 \nu-9\right) \mu^3+\nu^2 \left(4 \nu^3-17 \nu^2+25 \nu-6\right) \mu^2+\nu^3 \left(\nu^2+8 \nu-5\right) \mu+(\nu-2) \nu^4\Bigr) \la^5\no\\
   &\quad+\Bigl(\left(5 \nu^3-9 \nu^2-7 \nu+1\right) \mu^6+\left(-3 \nu^4-24 \nu^3+26 \nu^2-16 \nu+11\right) \mu^5+\left(-21 \nu^4+52 \nu^3-52 \nu^2+33 \nu-9\right) \mu^4\no\\
   &\quad+\left(-6 \nu^5+33 \nu^4-56 \nu^3+31 \nu^2-26 \nu+6\right) \mu^3+\nu \left(-4 \nu^4-25 \nu^3+29 \nu^2-27 \nu+9\right) \mu^2\no\\
   &\quad+\nu^2 \left(-4 \nu^3+11 \nu^2-12 \nu+5\right) \mu+\nu^4 (3 \nu-2)\Bigr) \la^4\no\\
   &\quad+\Bigl(\left(\nu^4+5 \nu^3+6 \nu^2-2 \nu+3\right) \mu^6+\left(9 \nu^4-25 \nu^3+14 \nu^2+15 \nu-3\right) \mu^5\no\\
   &\quad+\left(4 \nu^5-27 \nu^4+37 \nu^3-17 \nu^2+10 \nu-9\right) \mu^4+\left(6 \nu^5+27 \nu^4-38 \nu^3+32 \nu^2-22 \nu+5\right) \mu^3\no\\
   &\quad+\left(6 \nu^5-21 \nu^4+25 \nu^3-15 \nu^2+10 \nu-2\right) \mu^2-\nu \left(12 \nu^4-6 \nu^3+\nu^2-6 \nu+3\right) \mu+\nu^2 \left(3 \nu^2+\nu-2\right)\Bigr) \la^3\no\\
   &\quad-\mu \Bigl(\nu \left(\nu^3-\nu^2-\nu-3\right) \mu^5+\left(\nu^5-8 \nu^4+6 \nu^3+8 \nu^2-4 \nu+9\right) \mu^4+\left(4 \nu^5+11 \nu^4-9 \nu^3+4 \nu^2+9 \nu-3\right) \mu^3\no\\
   &\quad+\left(4 \nu^5-17 \nu^4+14 \nu^3-14 \nu^2-1\right) \mu^2+\left(-18 \nu^5+6 \nu^4+4 \nu^3+5 \nu^2-6 \nu+1\right) \mu+\nu \left(9 \nu^3+2 \nu^2-6 \nu+1\right)\Bigr) \la^2\no\\
   &\quad+\mu^2 \left(\nu^2-1\right) \left(\nu \left(\nu^2+\nu+6\right) \mu^3+\left(\nu^3-5 \nu^2+2 \nu-9\right) \mu^2-\left(12 \nu^3-2 \nu^2+\nu-1\right) \mu+9 \nu^2+\nu-1\right) \la\no\\
   &\quad+3 \mu^3 (\mu \nu-1) \left(\nu^2-1\right)^2\Bigr)\,,\\
    f^{3,4}&=\la \nu (\mu+\nu)^2 \Bigl(\la \left(3 \la^5+3 \nu \la^4+\left(4 \nu^2-\nu-3\right) \la^3+\nu \left(-4 \nu^2+2 \nu-7\right) \la^2-\nu^2 \left(5 \nu^2+\nu+9\right) \la-\nu^3 \left(\nu^2+5\right)\right) \mu^6\no\\
   &\quad+\Bigl(-3 \nu \la^6-2 \left(4 \nu^2-\nu+6\right) \la^5+\nu \left(12 \nu^2-7 \nu+3\right) \la^4+\left(20 \nu^4+6 \nu^3+11 \nu^2+3 \nu+12\right) \la^3\no\\
   &\quad+\nu \left(5 \nu^4-\nu^3+22 \nu^2-4 \nu+21\right) \la^2+\nu^2 \left(-5 \nu^2+\nu+18\right) \la-5 \nu^3 \left(\nu^2-1\right)\Bigr) \mu^5\no\\
   &\quad+\Bigl(\nu (4 \nu-1) \la^6+\nu \left(-12 \nu^2+8 \nu+3\right) \la^5-6 \left(5 \nu^4+2 \nu^3-\nu^2+\nu-3\right) \la^4\no\\
   &\quad+\nu \left(-10 \nu^4+4 \nu^3-37 \nu^2+15 \nu-27\right) \la^3+\left(21 \nu^4-8 \nu^3-31 \nu^2-3 \nu-18\right) \la^2\no\\
   &\quad+\nu \left(25 \nu^4+\nu^3-11 \nu^2+2 \nu-21\right) \la+9 \nu^2 \left(\nu^2-1\right)\Bigr) \mu^4\no\\
   &\quad+\Bigl(\nu \left(4 \nu^2-3 \nu+1\right) \la^6+\nu \left(20 \nu^3+10 \nu^2-9 \nu+3\right) \la^5+\nu \left(10 \nu^4-6 \nu^3+29 \nu^2-18 \nu+9\right) \la^4\no\\
   &\quad+\left(-34 \nu^4+18 \nu^3+6 \nu^2+6 \nu-12\right) \la^3+\nu \left(-50 \nu^4-4 \nu^3+4 \nu^2-9 \nu+33\right) \la^2\no\\
   &\quad+\left(-36 \nu^4+2 \nu^3+13 \nu^2+\nu+12\right) \la-7 \nu \left(\nu^2-1\right)\Bigr) \mu^3\no\\
   &\quad+\Bigl(\left(-5 \nu^4-3 \nu^3+\nu^2\right) \la^6+\nu \left(-5 \nu^4+4 \nu^3-10 \nu^2+7 \nu-3\right) \la^5+\nu \left(26 \nu^3-16 \nu^2+9 \nu-3\right) \la^4\no\\
   &\quad+\nu \left(50 \nu^4+6 \nu^3+4 \nu^2+12 \nu-15\right) \la^3+\left(54 \nu^4-6 \nu^3+2 \nu^2-2 \nu+3\right) \la^2+\nu \left(21 \nu^2+\nu-12\right) \la+3 \left(\nu^2-1\right)\Bigr) \mu^2\no\\
   &\quad+\la \nu \Bigl(\nu^2 \left(\nu^2-\nu+1\right) \la^5+\nu \left(-9 \nu^2+5 \nu-2\right) \la^4-\left(25 \nu^4+4 \nu^3+\nu^2+5 \nu-3\right) \la^3\no\\
   &\quad+\left(-36 \nu^3+6 \nu^2-7 \nu+1\right) \la^2+\left(-21 \nu^2-2 \nu+6\right) \la-6 \nu\Bigr) \mu\no\\
   &\quad+\la^2 \nu \left(\nu^3 \la^4+\nu^2 \left(5 \nu^2+\nu-1\right) \la^3+\nu \left(9 \nu^2-2 \nu+1\right) \la^2+\left(7 \nu^2+\nu-1\right) \la+3 \nu\right)\Bigr)\,,\\
    f^{2,4}&=\nu \Bigl(\mu^2 (\mu+\nu)^3 \left(2 \nu \mu^3+\left(-2 \nu^2+\nu+1\right) \mu^2+2 (\nu-2) \nu \mu+\nu^2 (\nu+7)\right) \la^7\no\\
   &\quad-(\mu+\nu)^2 \Bigl(\left(2 \nu^2+\nu-1\right) \mu^6-\left(6 \nu^3-6 \nu^2+\nu-1\right) \mu^5+\left(9 \nu^3+15 \nu^2+6 \nu+4\right) \mu^4\no\\
   &\quad+\nu \left(4 \nu^3+21 \nu^2+9 \nu-10\right) \mu^3-2 \nu^2 \left(\nu^2-2 \nu-2\right) \mu^2+16 \nu^3 \mu-2 \nu^4\Bigr) \la^6\no\\
   &\quad+\Bigl(\left(-2 \nu^3+3 \nu^2-10 \nu+1\right) \mu^8+\left(-8 \nu^4+15 \nu^3+9 \nu^2+10 \nu-6\right) \mu^7\no\\
   &\quad+\left(-6 \nu^5+27 \nu^4+63 \nu^3+36 \nu^2-26 \nu+4\right) \mu^6+\left(21 \nu^5+59 \nu^4+58 \nu^3-8 \nu^2+20 \nu+6\right) \mu^5\no\\
   &\quad+\nu^2 \left(6 \nu^4+7 \nu^3+43 \nu^2+94 \nu+36\right) \mu^4+\nu^2 \left(-8 \nu^4+12 \nu^3+126 \nu^2+27 \nu-17\right) \mu^3\no\\
   &\quad+\nu^3 \left(36 \nu^2+5 \nu-3\right) \mu^2+\nu^4 \left(-8 \nu^2-3 \nu+15\right) \mu+(7-\nu) \nu^5\Bigr) \la^5\no\\
   &\quad+\Bigl(\nu \left(2 \nu^3-3 \nu^2-5 \nu-2\right) \mu^8+\left(2 \nu^5-10 \nu^4-25 \nu^3-15 \nu^2+30 \nu-4\right) \mu^7\no\\
   &\quad-\left(11 \nu^5+21 \nu^4+36 \nu^3-4 \nu^2+26 \nu-14\right) \mu^6-\left(4 \nu^6-11 \nu^5+35 \nu^4+128 \nu^3+58 \nu^2-42 \nu+6\right) \mu^5\no\\
   &\quad+\left(12 \nu^6-12 \nu^5-150 \nu^4-51 \nu^3+35 \nu^2-22 \nu-4\right) \mu^4-\nu \left(36 \nu^4+8 \nu^3+33 \nu^2+27 \nu+2\right) \mu^3\no\\
   &\quad+\nu^2 \left(12 \nu^4+11 \nu^3-63 \nu^2-10 \nu+6\right) \mu^2+\nu^3 \left(4 \nu^3-27 \nu^2+3 \nu+6\right) \mu+2 \nu^4 \left(-2 \nu^2+\nu+1\right)\Bigr) \la^4\no\\
   &\quad+\Bigl(\nu \left(\nu^3+2 \nu^2+\nu+4\right) \mu^8+2 \nu \left(\nu^4+3 \nu^2+6 \nu+3\right) \mu^7+\left(\nu^6-10 \nu^5+9 \nu^4+46 \nu^3+24 \nu^2-30 \nu+6\right) \mu^6\no\\
   &\quad+\left(-8 \nu^6+4 \nu^5+58 \nu^4+27 \nu^3-25 \nu^2+28 \nu-16\right) \mu^5+\left(12 \nu^5+49 \nu^3+41 \nu^2-30 \nu+4\right) \mu^4\no\\
   &\quad+\left(-8 \nu^6-15 \nu^5+71 \nu^4+16 \nu^3-18 \nu^2+11 \nu+1\right) \mu^3+\left(-6 \nu^6+43 \nu^5-9 \nu^4-8 \nu^3+9 \nu^2+\nu\right) \mu^2\no\\
   &\quad+\nu^2 \left(16 \nu^4-6 \nu^3-4 \nu^2+\nu+1\right) \mu+(1-\nu) \nu^3\Bigr) \la^3\no\\
   &\quad+\mu \Bigl(2 \nu \left(\nu^4+\nu^2-\nu-6\right) \mu^6+\nu \left(2 \nu^5-3 \nu^2-13 \nu-6\right) \mu^5+\left(4 \nu^4-\nu^3-15 \nu^2+10 \nu-4\right) \mu^4\no\\
   &\quad+\left(2 \nu^6+9 \nu^5-13 \nu^4-6 \nu^3+12 \nu^2-13 \nu+9\right) \mu^3+\left(4 \nu^6-37 \nu^5+9 \nu^4+12 \nu^3-12 \nu^2+11 \nu-1\right) \mu^2\no\\
   &\quad-\nu \left(24 \nu^5-6 \nu^4-10 \nu^3+\nu^2+\nu+2\right) \mu+\nu^2 \left(2 \nu^2-3 \nu-1\right)\Bigr) \la^2\no\\
   &\quad-\mu^2 \left(\nu^2-1\right) \Bigl(\nu (\nu+12) \mu^4+2 \nu \left(\nu^2+5 \nu+1\right) \mu^3+\left(\nu^4-18 \nu^3+4 \nu^2+1\right) \mu^2\no\\
   &\quad+2 \left(-8 \nu^4+\nu^3+\nu-1\right) \mu+(\nu-2) \nu\Bigr) \la-4 \mu^4 \nu (\mu+\nu) \left(\nu^2-1\right)^2\Bigr)\,,\\
    f^{1,4}&=\la \nu (\mu+\nu)^2 \Bigl(\la^6 (\mu+\nu)^2 \left(\mu^4+\mu^3 (1-3 \nu)+2 \mu^2 \nu+\mu (\nu-3) \nu+\nu^2\right)\no\\
   &\quad+\la^5 \Bigl(\mu^6 (\nu-1)+\mu^5 \left(10 \nu^2-7 \nu-3\right)+\mu^4 \left(9 \nu^3-15 \nu^2+10 \nu-4\right)-13 \mu^3 (\nu-1)^2 \nu\no\\
   &\quad+\mu^2 \nu \left(-4 \nu^3+10 \nu^2-15 \nu+9\right)+\mu \nu^2 \left(-3 \nu^2-7 \nu+10\right)-(\nu-1) \nu^3\Bigr)\no\\
   &\quad+\la^4 \Bigl(\mu^6 \left(-5 \nu^2+3 \nu-2\right)+\mu^5 \left(-9 \nu^3+12 \nu^2-7 \nu+4\right)+\mu^4 \left(15 \nu^3-35 \nu^2+22 \nu+2\right)\no\\
   &\quad+6 \mu^3 (\nu-1)^4+\mu^2 \nu \left(2 \nu^3+22 \nu^2-35 \nu+15\right)+\mu \nu \left(4 \nu^3-7 \nu^2+12 \nu-9\right)+\nu^2 \left(-2 \nu^2+3 \nu-5\right)\Bigr)\no\\
   &\quad+\la^3 (\mu-1) \Bigl(3 \mu^5 (\nu-1) \nu^2+\mu^4 \left(-4 \nu^3+9 \nu^2-9 \nu+8\right)+\mu^3 \left(-4 \nu^4+18 \nu^3-18 \nu^2+6 \nu+2\right)\no\\
   &\quad-2 \mu^2 \left(\nu^4+3 \nu^3-9 \nu^2+9 \nu-2\right)+\mu \nu \left(-8 \nu^3+9 \nu^2-9 \nu+4\right)+3 (\nu-1) \nu\Bigr)\no\\
   &\quad+\la^2 \Bigl(\mu^6 \nu^2 (\nu+2)+\mu^5 \nu^2 \left(\nu^2-7 \nu+6\right)+\mu^4 \left(-3 \nu^4+10 \nu^3-7 \nu^2+9 \nu-12\right)\no\\
   &\quad+\mu^3 (\nu-1)^2 \left(4 \nu^2-5 \nu+4\right)+\mu^2 \left(-12 \nu^4+9 \nu^3-7 \nu^2+10 \nu-3\right)+\mu \left(6 \nu^2-7 \nu+1\right)+\nu (2 \nu+1)\Bigr)\no\\
   &\quad+\la \mu \left(\mu^4 \nu^2 \left(\nu^2-\nu-4\right)-\mu^3 \nu^2 \left(\nu^2-4 \nu+3\right)+\mu^2 \left(8 \nu^4-3 \nu^3-2 \nu^2-3 \nu+8\right)+\mu \left(-3 \nu^2+4 \nu-1\right)-4 \nu^2-\nu+1\right)\no\\
   &\quad-2 \mu^2 \left(\nu^2-1\right) \left(\mu^2 \nu^2-1\right)\Bigr)\,,\\
    f^{0,4}&=-4 \la^2 \nu^2 (\la-\mu)^2 (\la \mu-1)^3 (\mu+\nu)^3\,,
\end{align}
\begin{align}
    f^{3,3}&=2 \nu (\mu+\nu) \Bigl((\mu+\nu)^2 \left(\nu \mu^5-\left(\nu^2+\nu-4\right) \mu^4-2 \nu (\nu+6) \mu^3-(\nu-5) \nu^2 \mu^2-3 \nu^3 \mu+2 \nu^4\right) \la^7\no\\
   &\quad+\Bigl(\left(-\nu^2+\nu+1\right) \mu^7+\left(2 \nu^3+7 \nu^2+3 \nu+1\right) \mu^6+\left(3 \nu^4+15 \nu^3+16 \nu^2+7 \nu-16\right) \mu^5\no\\
   &\quad+\nu \left(13 \nu^3+16 \nu^2+15 \nu+5\right) \mu^4+\nu^2 \left(4 \nu^3+5 \nu^2+13 \nu+37\right) \mu^3\no\\
   &\quad+\nu^3 \left(-7 \nu^2+4 \nu+18\right) \mu^2+\left(5 \nu^4-10 \nu^6\right) \mu+3 \nu^5\Bigr) \la^6\no\\
   &\quad-\Bigl(\left(\nu^3+3 \nu^2+4 \nu+1\right) \mu^7+\left(3 \nu^4+12 \nu^3-8 \nu^2+9 \nu+5\right) \mu^6+\left(15 \nu^4+19 \nu^3+27 \nu^2+4 \nu+4\right) \mu^5\no\\
   &\quad+\left(6 \nu^5+19 \nu^4+31 \nu^3+41 \nu^2+16 \nu-24\right) \mu^4+\nu \left(-18 \nu^4+12 \nu^3+38 \nu^2+20 \nu-9\right) \mu^3\no\\
   &\quad+\nu^2 \left(-20 \nu^4+35 \nu^2+7 \nu+25\right) \mu^2+\nu^3 \left(17 \nu^2-2 \nu+24\right) \mu+(4-\nu) \nu^4\Bigr) \la^5\no\\
   &\quad+\Bigl(\left(\nu^4+3 \nu^3-9 \nu^2+2 \nu-4\right) \mu^7+\left(7 \nu^4+10 \nu^3+13 \nu^2+13 \nu+4\right) \mu^6+\left(4 \nu^5+17 \nu^4+23 \nu^3-19 \nu^2+22 \nu+10\right) \mu^5\no\\
   &\quad+\left(-22 \nu^5+12 \nu^4+15 \nu^3+36 \nu^2-16 \nu+6\right) \mu^4+\left(-20 \nu^6+75 \nu^4+15 \nu^3+29 \nu^2+16 \nu-16\right) \mu^3\no\\
   &\quad+\nu \left(38 \nu^4-7 \nu^3+44 \nu^2+11 \nu-17\right) \mu^2-\nu^2 \left(4 \nu^3-17 \nu^2+\nu+1\right) \mu+\nu^3 \left(4 \nu^2-2 \nu+11\right)\Bigr) \la^4\no\\
   &\quad-\Bigl(\nu \left(\nu^3+2 \nu^2+\nu+4\right) \mu^7+\left(\nu^5+2 \nu^4+5 \nu^3-23 \nu^2+6 \nu-16\right) \mu^6\no\\
   &\quad+\left(-13 \nu^5+4 \nu^4-12 \nu^3+18 \nu^2+15 \nu+6\right) \mu^5+\left(-10 \nu^6+65 \nu^4+9 \nu^3-7 \nu^2+22 \nu+10\right) \mu^4\no\\
   &\quad+\left(42 \nu^5-9 \nu^4-\nu^3+19 \nu^2-27 \nu+4\right) \mu^3+\left(-6 \nu^5+27 \nu^4-3 \nu^3+4 \nu^2+7 \nu-4\right) \mu^2\no\\
   &\quad+\nu \left(16 \nu^4-6 \nu^3+21 \nu^2+2 \nu-7\right) \mu-\nu^2 \left(4 \nu^2+\nu+4\right)\Bigr) \la^3\no\\
   &\quad+\mu \Bigl(-2 \nu^4 \mu^6+\nu \left(-3 \nu^4-7 \nu^2+2 \nu+12\right) \mu^5+\left(-2 \nu^6+20 \nu^4+\nu^3-11 \nu^2+6 \nu-24\right) \mu^4\no\\
   &\quad+\left(23 \nu^5-5 \nu^4-38 \nu^3+9 \nu^2+7 \nu+4\right) \mu^3+\left(-4 \nu^5+19 \nu^4-3 \nu^3+16 \nu^2+9 \nu+5\right) \mu^2\no\\
   &\quad+\left(24 \nu^5-6 \nu^4+\nu^3+3 \nu^2-11 \nu+1\right) \mu-\nu \left(12 \nu^3+2 \nu^2+\nu-1\right)\Bigr) \la^2\no\\
   &\quad-\mu^2 \left(\nu^2-1\right) \left(\nu \left(5 \nu^2-\nu-12\right) \mu^3-\left(\nu^3-5 \nu^2+2 \nu-16\right) \mu^2+\left(16 \nu^3-2 \nu^2-\nu-1\right) \mu-12 \nu^2-\nu-1\right) \la\no\\
   &\quad+4 \mu^3 (\mu \nu-1) \left(\nu^2-1\right)^2\Bigr)\,,\\
    f^{2,3}&=\nu \Bigl((\mu+\nu)^3 \left(\mu^5+(10 \nu-1) \mu^4+\left(-9 \nu^2-3 \nu+2\right) \mu^3+\nu \left(2 \nu^2-3 \nu-9\right) \mu^2-(\nu-10) \nu^2 \mu+\nu^3\right) \la^7\no\\
   &\quad-(\mu+\nu)^2 \Bigl((7 \nu-1) \mu^6+\left(6 \nu^2-7 \nu+1\right) \mu^5-\left(21 \nu^3+15 \nu^2-34 \nu+4\right) \mu^4+(\nu-1)^2 \left(8 \nu^2+3 \nu+8\right) \mu^3\no\\
   &\quad-\nu \left(4 \nu^3-34 \nu^2+15 \nu+21\right) \mu^2+\nu^2 \left(\nu^2-7 \nu+6\right) \mu-(\nu-7) \nu^3\Bigr) \la^6\no\\
   &\quad-\Bigl(\left(3 \nu^2+3 \nu+2\right) \mu^8+\left(15 \nu^3+18 \nu^2-13 \nu+4\right) \mu^7+\left(21 \nu^4+42 \nu^3-83 \nu^2+30 \nu+6\right) \mu^6\no\\
   &\quad+\left(-3 \nu^5+48 \nu^4-127 \nu^3+84 \nu^2-32 \nu+6\right) \mu^5+\left(-12 \nu^6+27 \nu^5-97 \nu^4+116 \nu^3-97 \nu^2+27 \nu-12\right) \mu^4\no\\
   &\quad+\nu \left(6 \nu^5-32 \nu^4+84 \nu^3-127 \nu^2+48 \nu-3\right) \mu^3+\nu^2 \left(6 \nu^4+30 \nu^3-83 \nu^2+42 \nu+21\right) \mu^2\no\\
   &\quad+\nu^3 \left(4 \nu^3-13 \nu^2+18 \nu+15\right) \mu+\nu^4 \left(2 \nu^2+3 \nu+3\right)\Bigr) \la^5\no\\
   &\quad+\Bigl(\nu \left(7 \nu^2+3 \nu-4\right) \mu^8+\left(2 \nu^4+13 \nu^3-4 \nu^2+9 \nu+4\right) \mu^7+3 \left(-3 \nu^5+7 \nu^4-14 \nu^3+15 \nu^2+\nu+2\right) \mu^6\no\\
   &\quad+\left(-8 \nu^6+15 \nu^5-50 \nu^4+87 \nu^3-118 \nu^2+36 \nu+14\right) \mu^5+\left(4 \nu^6-10 \nu^5+81 \nu^4-210 \nu^3+81 \nu^2-10 \nu+4\right) \mu^4\no\\
   &\quad+\left(14 \nu^6+36 \nu^5-118 \nu^4+87 \nu^3-50 \nu^2+15 \nu-8\right) \mu^3+3 \nu \left(2 \nu^5+\nu^4+15 \nu^3-14 \nu^2+7 \nu-3\right) \mu^2\no\\
   &\quad+\nu^2 \left(4 \nu^4+9 \nu^3-4 \nu^2+13 \nu+2\right) \mu+\nu^3 \left(-4 \nu^2+3 \nu+7\right)\Bigr) \la^4\no\\
   &\quad+\Bigl(\nu^2 \left(2 \nu^2-\nu+6\right) \mu^8+\nu \left(4 \nu^4-3 \nu^3-3 \nu^2-6 \nu+20\right) \mu^7\no\\
   &\quad+\left(2 \nu^6-3 \nu^5+5 \nu^4-22 \nu^3+19 \nu^2-9 \nu+4\right) \mu^6-\left(\nu^6-3 \nu^5+30 \nu^4-73 \nu^3+36 \nu^2+17 \nu+4\right) \mu^5\no\\
   &\quad-\left(11 \nu^6+18 \nu^5-53 \nu^4+54 \nu^3-53 \nu^2+18 \nu+11\right) \mu^4-\left(4 \nu^6+17 \nu^5+36 \nu^4-73 \nu^3+30 \nu^2-3 \nu+1\right) \mu^3\no\\
   &\quad+\left(4 \nu^6-9 \nu^5+19 \nu^4-22 \nu^3+5 \nu^2-3 \nu+2\right) \mu^2+\nu \left(20 \nu^4-6 \nu^3-3 \nu^2-3 \nu+4\right) \mu+\nu^2 \left(6 \nu^2-\nu+2\right)\Bigr) \la^3\no\\
   &\quad+\mu \Bigl(\nu^2 \left(3 \nu^2+\nu-16\right) \mu^6+\nu \left(-2 \nu^4+3 \nu^3+8 \nu^2+3 \nu-36\right) \mu^5\no\\
   &\quad+\left(3 \nu^6+3 \nu^5+16 \nu^4+9 \nu^3-6 \nu^2+3 \nu-16\right) \mu^4+\left(\nu^6+8 \nu^5+9 \nu^4+12 \nu^3+9 \nu^2+8 \nu+1\right) \mu^3\no\\
   &\quad+\left(-16 \nu^6+3 \nu^5-6 \nu^4+9 \nu^3+16 \nu^2+3 \nu+3\right) \mu^2+\nu \left(-36 \nu^4+3 \nu^3+8 \nu^2+3 \nu-2\right) \mu+\nu^2 \left(-16 \nu^2+\nu+3\right)\Bigr) \la^2\no\\
   &\quad-14 \mu^2 (\mu \nu+1)^2 \left(\mu^2-\nu^2\right) \left(\nu^2-1\right) \la-4 \mu^3 (\mu \nu+1)^2 \left(\nu^2-1\right)^2\Bigr)\,,\\
    f^{1,3}&=\nu \Bigl((\mu+\nu)^3 \left((7 \nu+1) \mu^3+2 \nu (1-2 \nu) \mu^2+\nu \left(\nu^2+\nu-2\right) \mu+2 \nu^2\right) \la^7\no\\
   &\quad+(\mu+\nu)^2 \Bigl(2 \mu^6-16 \nu \mu^5+\left(-4 \nu^2-4 \nu+2\right) \mu^4+\left(10 \nu^3-9 \nu^2-21 \nu-4\right) \mu^3\no\\
   &\quad-\nu \left(4 \nu^3+6 \nu^2+15 \nu+9\right) \mu^2+\nu \left(-\nu^3+\nu^2-6 \nu+6\right) \mu+\nu^2 \left(\nu^2-\nu-2\right)\Bigr) \la^6\no\\
   &\quad+\Bigl((7 \nu-1) \mu^8+\left(15 \nu^2-3 \nu-8\right) \mu^7+\nu \left(-3 \nu^2+5 \nu+36\right) \mu^6+\left(-17 \nu^4+27 \nu^3+126 \nu^2+12 \nu-8\right) \mu^5\no\\
   &\quad+\left(36 \nu^4+94 \nu^3+43 \nu^2+7 \nu+6\right) \mu^4+\nu \left(6 \nu^5+20 \nu^4-8 \nu^3+58 \nu^2+59 \nu+21\right) \mu^3\no\\
   &\quad+\nu \left(4 \nu^5-26 \nu^4+36 \nu^3+63 \nu^2+27 \nu-6\right) \mu^2\no\\
   &\quad+\nu^2 \left(-6 \nu^4+10 \nu^3+9 \nu^2+15 \nu-8\right) \mu+\nu^3 \left(\nu^3-10 \nu^2+3 \nu-2\right)\Bigr) \la^5\no\\
   &\quad+\Bigl(2 \left(\nu^2+\nu-2\right) \mu^8+\left(6 \nu^3+3 \nu^2-27 \nu+4\right) \mu^7+\left(6 \nu^4-10 \nu^3-63 \nu^2+11 \nu+12\right) \mu^6\no\\
   &\quad-\nu \left(2 \nu^4+27 \nu^3+33 \nu^2+8 \nu+36\right) \mu^5-\left(4 \nu^6+22 \nu^5-35 \nu^4+51 \nu^3+150 \nu^2+12 \nu-12\right) \mu^4\no\\
   &\quad-\left(6 \nu^6-42 \nu^5+58 \nu^4+128 \nu^3+35 \nu^2-11 \nu+4\right) \mu^3+\nu \left(14 \nu^5-26 \nu^4+4 \nu^3-36 \nu^2-21 \nu-11\right) \mu^2\no\\
   &\quad-\nu \left(4 \nu^5-30 \nu^4+15 \nu^3+25 \nu^2+10 \nu-2\right) \mu-\nu^2 \left(2 \nu^3+5 \nu^2+3 \nu-2\right)\Bigr) \la^4\no\\
   &\quad+\Bigl((\nu-1) \nu^2 \mu^8+\left(\nu^4+\nu^3-4 \nu^2-6 \nu+16\right) \mu^7+\left(\nu^5+9 \nu^4-8 \nu^3-9 \nu^2+43 \nu-6\right) \mu^6\no\\
   &\quad+\left(\nu^6+11 \nu^5-18 \nu^4+16 \nu^3+71 \nu^2-15 \nu-8\right) \mu^5+\nu \left(4 \nu^5-30 \nu^4+41 \nu^3+49 \nu^2+12\right) \mu^4\no\\
   &\quad+\left(-16 \nu^6+28 \nu^5-25 \nu^4+27 \nu^3+58 \nu^2+4 \nu-8\right) \mu^3+\left(6 \nu^6-30 \nu^5+24 \nu^4+46 \nu^3+9 \nu^2-10 \nu+1\right) \mu^2\no\\
   &\quad+2 \left(3 \nu^5+6 \nu^4+3 \nu^3+\nu\right) \mu+\nu^2 \left(4 \nu^3+\nu^2+2 \nu+1\right)\Bigr) \la^3\no\\
   &\quad-\mu \Bigl(\nu^2 \left(\nu^2+3 \nu-2\right) \mu^6+\left(2 \nu^5+\nu^4+\nu^3-10 \nu^2-6 \nu+24\right) \mu^5\no\\
   &\quad+\left(\nu^6-11 \nu^5+12 \nu^4-12 \nu^3-9 \nu^2+37 \nu-4\right) \mu^4+\left(-9 \nu^6+13 \nu^5-12 \nu^4+6 \nu^3+13 \nu^2-9 \nu-2\right) \mu^3\no\\
   &\quad+\nu^2 \left(4 \nu^4-10 \nu^3+15 \nu^2+\nu-4\right) \mu^2+\left(6 \nu^5+13 \nu^4+3 \nu^3-2\right) \mu+2 \nu \left(6 \nu^4+\nu^3-\nu^2-1\right)\Bigr) \la^2\no\\
   &\quad-\mu^2 \left(\nu^2-1\right) \Bigl(\nu^2 (2 \nu-1) \mu^4+2 \left(\nu^4-\nu^3-\nu+8\right) \mu^3-\left(\nu^4+4 \nu^2-18 \nu+1\right) \mu^2\no\\
   &\quad-2 \nu \left(\nu^2+5 \nu+1\right) \mu-\nu^2 (12 \nu+1)\Bigr) \la-4 \mu^3 (\mu+\nu) \left(\nu^2-1\right)^2\Bigr)\,,\\
    f^{0,3}&=-\la \nu \left(\la^2 \mu-\la \left(\mu^2+1\right)+\mu\right) (\mu+\nu)^2 \Bigl(\la^4 \left(\mu^4 (2 \nu-1)+\mu^3 \left(2 \nu^2-3 \nu+1\right)+\mu^2 \nu (5-3 \nu)-\mu (\nu-1) \nu^2-3 \nu^3\right)\no\\
   &\quad+\la^3 \left(-2 \mu^4 \left(\nu^2-\nu+2\right)+\mu^3 \left(4 \nu^2-7 \nu+3\right)+\mu^2 \left(2 \nu^3-2 \nu^2+7 \nu-3\right)+5 \mu \nu \left(\nu^2+\nu-2\right)+(\nu-1) \nu^2\right)\no\\
   &\quad-\la^2 \Bigl(\mu^4 \nu (\nu+4)+\mu^3 \left(\nu^3-\nu^2+4 \nu-12\right)+\mu^2 \left(\nu^3+6 \nu^2-8 \nu+3\right)\no\\
   &\quad+\mu \left(2 \nu^3+4 \nu^2+5 \nu-3\right)+\nu \left(-4 \nu^2+2 \nu-5\right)\Bigr)\no\\
   &\quad+\la \Bigl(\mu^3 \nu \left(-\nu^2+\nu+8\right)+\mu^2 \left(\nu^3+5 \nu^2+2 \nu-12\right)\no\\
   &\quad+\mu \left(-8 \nu^3+2 \nu^2-3 \nu+1\right)+4 \nu^2+\nu-1\Bigr)+4 \mu \left(\nu^2-1\right) (\mu \nu-1)\Bigr)\,,
\end{align}

\begin{align}
    f^{2,2}&=2 (\mu+\nu) \Bigl((\mu+\nu)^2 \left(2 \mu^5-3 \nu \mu^4+\nu (5 \nu-1) \mu^3-2 \nu^2 (6 \nu+1) \mu^2+\nu^2 \left(4 \nu^2-\nu-1\right) \mu+\nu^3\right) \la^7\no\\
   &\quad+\Bigl(3 \nu \mu^7+5 \left(\nu^2-2\right) \mu^6+\nu \left(18 \nu^2+4 \nu-7\right) \mu^5+\nu \left(37 \nu^3+13 \nu^2+5 \nu+4\right) \mu^4\no\\
   &\quad+\nu^2 \left(5 \nu^3+15 \nu^2+16 \nu+13\right) \mu^3+\nu^2 \left(-16 \nu^4+7 \nu^3+16 \nu^2+15 \nu+3\right) \mu^2\no\\
   &\quad+\nu^3 \left(\nu^3+3 \nu^2+7 \nu+2\right) \mu+\nu^4 \left(\nu^2+\nu-1\right)\Bigr) \la^6\no\\
   &\quad-\Bigl(\nu (4 \nu-1) \mu^7+\nu \left(24 \nu^2-2 \nu+17\right) \mu^6+\left(25 \nu^4+7 \nu^3+35 \nu^2-20\right) \mu^5\no\\
   &\quad+\nu \left(-9 \nu^4+20 \nu^3+38 \nu^2+12 \nu-18\right) \mu^4+\nu \left(-24 \nu^5+16 \nu^4+41 \nu^3+31 \nu^2+19 \nu+6\right) \mu^3\no\\
   &\quad+\nu^2 \left(4 \nu^4+4 \nu^3+27 \nu^2+19 \nu+15\right) \mu^2+\nu^2 \left(5 \nu^4+9 \nu^3-8 \nu^2+12 \nu+3\right) \mu+\nu^3 \left(\nu^3+4 \nu^2+3 \nu+1\right)\Bigr) \la^5\no\\
   &\quad+\Bigl(\nu \left(11 \nu^2-2 \nu+4\right) \mu^7-\nu \left(\nu^3+\nu^2-17 \nu+4\right) \mu^6+\nu \left(-17 \nu^4+11 \nu^3+44 \nu^2-7 \nu+38\right) \mu^5\no\\
   &\quad+\left(-16 \nu^6+16 \nu^5+29 \nu^4+15 \nu^3+75 \nu^2-20\right) \mu^4+\nu \left(6 \nu^5-16 \nu^4+36 \nu^3+15 \nu^2+12 \nu-22\right) \mu^3\no\\
   &\quad+\nu \left(10 \nu^5+22 \nu^4-19 \nu^3+23 \nu^2+17 \nu+4\right) \mu^2+\nu^2 \left(4 \nu^4+13 \nu^3+13 \nu^2+10 \nu+7\right) \mu\no\\
   &\quad-4 \nu^6+2 \nu^5-9 \nu^4+3 \nu^3+\nu^2\Bigr) \la^4\no\\
   &\quad+\Bigl(\nu^2 \left(4 \nu^2+\nu+4\right) \mu^7+\nu \left(7 \nu^4-2 \nu^3-21 \nu^2+6 \nu-16\right) \mu^6+\nu \left(4 \nu^5-7 \nu^4-4 \nu^3+3 \nu^2-27 \nu+6\right) \mu^5\no\\
   &\quad+\nu \left(-4 \nu^5+27 \nu^4-19 \nu^3+\nu^2+9 \nu-42\right) \mu^4-\left(10 \nu^6+22 \nu^5-7 \nu^4+9 \nu^3+65 \nu^2-10\right) \mu^3\no\\
   &\quad-\nu \left(6 \nu^5+15 \nu^4+18 \nu^3-12 \nu^2+4 \nu-13\right) \mu^2+\nu \left(16 \nu^5-6 \nu^4+23 \nu^3-5 \nu^2-2 \nu-1\right) \mu\no\\
   &\quad-\nu^2 \left(4 \nu^3+\nu^2+2 \nu+1\right)\Bigr) \la^3\no\\
   &\quad+\Bigl(\nu^2 \left(\nu^3-\nu^2-2 \nu-12\right) \mu^6+\nu \left(\nu^5-11 \nu^4+3 \nu^3+\nu^2-6 \nu+24\right) \mu^5\no\\
   &\quad+\nu \left(5 \nu^5+9 \nu^4+16 \nu^3-3 \nu^2+19 \nu-4\right) \mu^4+\nu \left(4 \nu^5+7 \nu^4+9 \nu^3-38 \nu^2-5 \nu+23\right) \mu^3\no\\
   &\quad+\left(-24 \nu^6+6 \nu^5-11 \nu^4+\nu^3+20 \nu^2-2\right) \mu^2+\nu \left(12 \nu^4+2 \nu^3-7 \nu^2-3\right) \mu-2 \nu^2\Bigr) \la^2\no\\
   &\quad-\mu^2 \nu \left(\nu^2-1\right) \left(\nu \left(\nu^2+\nu+12\right) \mu^3+\left(\nu^3+\nu^2+2 \nu-16\right) \mu^2+\left(-16 \nu^3+2 \nu^2-5 \nu+1\right) \mu+12 \nu^2+\nu-5\right) \la\no\\
   &\quad-4 \mu^3 \nu (\mu \nu-1) \left(\nu^2-1\right)^2\Bigr)\,,\\
    f^{1,2}&=\la (\mu+\nu)^2 \Bigl(\left(\nu \mu^6+\left(\nu^2-\nu+1\right) \mu^5+\nu \left(\nu^2-3 \nu-5\right) \mu^4+\nu^2 \left(\nu^2-3 \nu+4\right) \mu^3-(\nu-4) \nu^3 \mu^2-3 \nu^4 \mu+3 \nu^5\right) \la^6\no\\
   &\quad+\Bigl(\left(-\nu^2+\nu+5\right) \mu^6+\nu \left(-2 \nu^2+5 \nu-9\right) \mu^5+\left(-3 \nu^4+7 \nu^3-10 \nu^2+4 \nu-5\right) \mu^4\no\\
   &\quad+\nu \left(3 \nu^3-9 \nu^2+10 \nu+20\right) \mu^3+\nu^2 \left(3 \nu^2+8 \nu-12\right) \mu^2-2 \nu^3 \left(6 \nu^2-\nu+4\right) \mu+3 \nu^4\Bigr) \la^5\no\\
   &\quad+\Bigl(\nu \left(\nu^2-2 \nu+9\right) \mu^6+\left(3 \nu^4-5 \nu^3-\nu^2-4 \nu-25\right) \mu^5+\nu \left(-3 \nu^3+9 \nu^2-16 \nu+26\right) \mu^4\no\\
   &\quad+\left(9 \nu^4-18 \nu^3+29 \nu^2-6 \nu+10\right) \mu^3+6 \nu \left(3 \nu^4-\nu^3+\nu^2-2 \nu-5\right) \mu^2\no\\
   &\quad+\nu^2 \left(3 \nu^2-7 \nu+12\right) \mu-\nu^3 \left(3 \nu^2+\nu-4\right)\Bigr) \la^4\no\\
   &\quad+\Bigl(\nu^2 \left(-\nu^2+\nu+7\right) \mu^6+\nu \left(\nu^3-7 \nu^2+6 \nu-36\right) \mu^5+\left(-15 \nu^4+12 \nu^3+4 \nu^2+6 \nu+50\right) \mu^4\no\\
   &\quad+2 \nu \left(-6 \nu^4+3 \nu^3+3 \nu^2+9 \nu-17\right) \mu^3+\left(-27 \nu^4+15 \nu^3-37 \nu^2+4 \nu-10\right) \mu^2\no\\
   &\quad+\nu \left(12 \nu^4+3 \nu^3+11 \nu^2+6 \nu+20\right) \mu+\nu^2 \left(-7 \nu^2+2 \nu-4\right)\Bigr) \la^3\no\\
   &\quad+\Bigl(3 \nu^3 \mu^6+\nu^2 \left(6 \nu^2-2 \nu-21\right) \mu^5+\nu \left(3 \nu^4-2 \nu^3+2 \nu^2-6 \nu+54\right) \mu^4\no\\
   &\quad+\left(33 \nu^4-9 \nu^3+4 \nu^2-4 \nu-50\right) \mu^3-\nu \left(18 \nu^4+3 \nu^3+31 \nu^2+8 \nu-21\right) \mu^2\no\\
   &\quad+\left(21 \nu^4-4 \nu^3+22 \nu^2-\nu+5\right) \mu-\nu \left(9 \nu^2+\nu+5\right)\Bigr) \la^2\no\\
   &\quad+\Bigl(-6 \nu^3 \mu^5+\nu^2 \left(-12 \nu^2+\nu+21\right) \mu^4+\nu \left(12 \nu^4+\nu^3+13 \nu^2+2 \nu-36\right) \mu^3\no\\
   &\quad+\left(-21 \nu^4+2 \nu^3-11 \nu^2+\nu+25\right) \mu^2+\nu \left(18 \nu^2+\nu-5\right) \mu-5 \nu^2-1\Bigr) \la\no\\
   &\quad-\mu \left(\nu^2-1\right) \left(3 \mu^3 \nu^3-7 \mu^2 \nu^2+9 \mu \nu-5\right)\Bigr)\,,\\
    f^{0,2}&=(\mu+\nu)^3 \Bigl((\nu+1) \mu^4+2 \nu (2-3 \nu) \mu^3+\nu \left(\nu^2+5 \nu-2\right) \mu^2+2 \nu^3 \mu+2 \nu^3\Bigr) \la^7\no\\
   &\quad+(\mu+\nu)^2 \Bigl((\nu-1) \mu^6+\left(7 \nu^2-7 \nu+2\right) \mu^5+\left(11 \nu^3-18 \nu^2-2 \nu-5\right) \mu^4-\nu \left(3 \nu^3+19 \nu^2-14 \nu+20\right) \mu^3\no\\
   &\quad+\nu \left(-7 \nu^3+6 \nu^2-27 \nu+8\right) \mu^2-2 \nu^2 \left(2 \nu^2+7 \nu-2\right) \mu-2 \nu^3 (\nu+2)\Bigr) \la^6\no\\
   &\quad+\Bigl(2 \nu (1-2 \nu) \mu^8+\left(-11 \nu^3+14 \nu^2-4 \nu+5\right) \mu^7+\left(-11 \nu^4+39 \nu^3-18 \nu^2+38 \nu-10\right) \mu^6\no\\
   &\quad-\left(\nu^5-53 \nu^4+48 \nu^3-114 \nu^2+20 \nu-10\right) \mu^5+\nu \left(3 \nu^5+35 \nu^4-52 \nu^3+170 \nu^2-20 \nu+50\right) \mu^4\no\\
   &\quad+\nu \left(9 \nu^5-20 \nu^4+131 \nu^3-17 \nu^2+97 \nu-12\right) \mu^3+\nu^2 \left(-2 \nu^4+48 \nu^3-15 \nu^2+91 \nu-24\right) \mu^2\no\\
   &\quad+\nu^3 \left(6 \nu^3-15 \nu^2+41 \nu-12\right) \mu-7 (\nu-1) \nu^5\Bigr) \la^5\no\\
   &\quad-\Bigl(\nu \left(\nu^2+\nu+4\right) \mu^8+\nu \left(\nu^3+7 \nu^2-2 \nu+8\right) \mu^7+\left(\nu^5+16 \nu^4-18 \nu^3+45 \nu^2-6 \nu+10\right) \mu^6\no\\
   &\quad+\left(\nu^6+15 \nu^5-26 \nu^4+96 \nu^3+2 \nu^2+62 \nu-20\right) \mu^5+\left(5 \nu^6-24 \nu^5+95 \nu^4-7 \nu^3+145 \nu^2-40 \nu+10\right) \mu^4\no\\
   &\quad+\nu \left(-10 \nu^5+42 \nu^4-51 \nu^3+159 \nu^2-18 \nu+40\right) \mu^3+\nu \left(6 \nu^5-57 \nu^4+80 \nu^3+8 \nu^2+59 \nu-8\right) \mu^2\no\\
   &\quad+\nu^2 \left(-21 \nu^4+13 \nu^3-4 \nu^2+38 \nu-12\right) \mu-\nu^3 \left(\nu^3+10 \nu^2-9 \nu+4\right)\Bigr) \la^4\no\\
   &\quad+\Bigl(\nu \left(\nu^3+2 \nu^2+3 \nu+16\right) \mu^7+2 \nu \left(\nu^4+7 \nu^2+11 \nu+6\right) \mu^6+\left(\nu^6-10 \nu^5+21 \nu^4-9 \nu^3+51 \nu^2-4 \nu+10\right) \mu^5\no\\
   &\quad+\left(-8 \nu^6+12 \nu^5-47 \nu^4+75 \nu^3+24 \nu^2+48 \nu-20\right) \mu^4+\left(2 \nu^6-53 \nu^5+42 \nu^4+28 \nu^3+81 \nu^2-35 \nu+5\right) \mu^3\no\\
   &\quad+\nu \left(-21 \nu^5+3 \nu^4-26 \nu^3+56 \nu^2-9 \nu+15\right) \mu^2+\nu \left(-3 \nu^5-22 \nu^4+11 \nu^3+9 \nu^2+15 \nu-2\right) \mu\no\\
   &\quad+\nu^2 \left(4 \nu^4-2 \nu^3+3 \nu^2+5 \nu-2\right)\Bigr) \la^3\no\\
   &\quad+\Bigl(\nu \left(2 \nu^4+7 \nu^2-3 \nu-24\right) \mu^6+\nu \left(2 \nu^5+15 \nu^3-7 \nu^2-38 \nu-8\right) \mu^5+\left(15 \nu^5-2 \nu^4+8 \nu^3-23 \nu^2+\nu-5\right) \mu^4\no\\
   &\quad+\left(7 \nu^6+5 \nu^5+44 \nu^4-18 \nu^3-15 \nu^2-17 \nu+10\right) \mu^3+\left(3 \nu^6+10 \nu^5+\nu^4-15 \nu^3-18 \nu^2+14 \nu-1\right) \mu^2\no\\
   &\quad+\nu \left(-12 \nu^5+4 \nu^4+\nu^3-5 \nu^2+2 \nu-2\right) \mu+\nu^2 \left(\nu^2-2 \nu-1\right)\Bigr) \la^2\no\\
   &\quad-\mu \left(\nu^2-1\right) \Bigl(\nu (\nu+16) \mu^4+2 \nu \left(\nu^2+11 \nu+1\right) \mu^3+\left(\nu^4-6 \nu^3+4 \nu^2+1\right) \mu^2\no\\
   &\quad+2 \left(-6 \nu^4+\nu^3+\nu-1\right) \mu+(\nu-2) \nu\Bigr) \la-4 \mu^3 \nu (\mu+\nu) \left(\nu^2-1\right)^2\,,
\end{align}

\begin{align}
    f^{1,1}&=2 (\la \mu-1) (\mu+\nu) \Bigl((\mu+\nu)^2 \left((3 \nu-1) \mu^3-\nu (\nu+4) \mu^2+\nu \left(2 \nu^2-5 \nu-1\right) \mu+\nu^2 (7-2 \nu)\right) \la^6\no\\
   &\quad+\Bigl((1-2 \nu) \mu^6+\left(-5 \nu^2+8 \nu+1\right) \mu^5+\left(-6 \nu^3+25 \nu^2-17 \nu+4\right) \mu^4\no\\
   &\quad+\nu \left(-9 \nu^3+37 \nu^2-33 \nu+19\right) \mu^3+\nu \left(-6 \nu^4+26 \nu^3-35 \nu^2+33 \nu+3\right) \mu^2\no\\
   &\quad+\nu^2 \left(7 \nu^3-24 \nu^2+25 \nu-10\right) \mu+\nu^3 \left(-4 \nu^2+7 \nu-13\right)\Bigr) \la^5\no\\
   &\quad+\Bigl((3-2 \nu) \mu^6+\left(5 \nu^3-12 \nu^2+11 \nu-4\right) \mu^5+\left(9 \nu^4-27 \nu^3+29 \nu^2-25 \nu-4\right) \mu^4\no\\
   &\quad+\left(6 \nu^5-26 \nu^4+31 \nu^3-56 \nu^2+33 \nu-6\right) \mu^3+\nu \left(-9 \nu^4+33 \nu^3-52 \nu^2+52 \nu-21\right) \mu^2\no\\
   &\quad+\nu \left(11 \nu^4-16 \nu^3+26 \nu^2-24 \nu-3\right) \mu+\nu^2 \left(\nu^3-7 \nu^2-9 \nu+5\right)\Bigr) \la^4\no\\
   &\quad+\Bigl(\nu \left(-2 \nu^2+\nu+3\right) \mu^6-\left(3 \nu^4-6 \nu^3+\nu^2-6 \nu+12\right) \mu^5+\left(-2 \nu^5+10 \nu^4-15 \nu^3+25 \nu^2-21 \nu+6\right) \mu^4\no\\
   &\quad+\left(5 \nu^5-22 \nu^4+32 \nu^3-38 \nu^2+27 \nu+6\right) \mu^3+\left(-9 \nu^5+10 \nu^4-17 \nu^3+37 \nu^2-27 \nu+4\right) \mu^2\no\\
   &\quad+\nu \left(-3 \nu^4+15 \nu^3+14 \nu^2-25 \nu+9\right) \mu+\nu \left(3 \nu^4-2 \nu^3+6 \nu^2+5 \nu+1\right)\Bigr) \la^3\no\\
   &\quad-\Bigl(\nu \left(\nu^3-6 \nu^2+2 \nu+9\right) \mu^5+\left(\nu^5-6 \nu^4+5 \nu^3+4 \nu^2+6 \nu-18\right) \mu^4-\left(\nu^5+14 \nu^3-14 \nu^2+17 \nu-4\right) \mu^3\no\\
   &\quad+\left(-3 \nu^5+9 \nu^4+4 \nu^3-9 \nu^2+11 \nu+4\right) \mu^2+\left(9 \nu^5-4 \nu^4+8 \nu^3+6 \nu^2-8 \nu+1\right) \mu-\nu \left(3 \nu^3+\nu^2+\nu-1\right)\Bigr) \la^2\no\\
   &\quad+\mu \left(\nu^2-1\right) \left(\nu \left(\nu^2-\nu-9\right) \mu^3+\left(-\nu^3+\nu^2-2 \nu+12\right) \mu^2+\left(9 \nu^3-2 \nu^2+5 \nu-1\right) \mu-6 \nu^2-\nu-1\right) \la\no\\
   &\quad-3 \mu^2 (\mu \nu-1) \left(\nu^2-1\right)^2\Bigr)\,,\\
    f^{0,1}&=-\la^4 \left(\la^2+4 \nu \la+3 \nu^2\right) \mu^8+\la^3 \left(\la^4+2 \nu \la^3+\left(7-11 \nu^2\right) \la^2-6 \nu \left(\nu^2-2\right) \la+11 \nu^2\right) \mu^7\no\\
   &\quad+\la^2 \left(4 \nu \la^5+5 \left(3 \nu^2-1\right) \la^4-8 \nu^3 \la^3-4 \left(2 \nu^4-9 \nu^2+5\right) \la^2+4 \nu \left(5 \nu^2-2\right) \la+5 \nu^2 \left(\nu^2-3\right)\right) \mu^6\no\\
   &\quad-\la \Bigl(\nu^2 \la^6+\left(22 \nu-28 \nu^3\right) \la^5-2 \left(2 \nu^4-17 \nu^2+5\right) \la^4+2 \nu \left(3 \nu^4-6 \nu^2+10\right) \la^3\no\\
   &\quad+\left(-13 \nu^4+46 \nu^2-30\right) \la^2+\left(-6 \nu^5+14 \nu^3+8 \nu\right) \la+9 \nu^2 \left(\nu^2-1\right)\Bigr) \mu^5\no\\
   &\quad-\Bigl(14 \nu^3 \la^7-15 \nu^2 \left(\nu^2-1\right) \la^6-2 \nu \left(7 \nu^4-27 \nu^2+24\right) \la^5+\left(3 \nu^6+12 \nu^4-15 \nu^2+10\right) \la^4\no\\
   &\quad+2 \nu \left(5 \nu^4+12 \nu^2-20\right) \la^3+\left(-5 \nu^6+2 \nu^4-32 \nu^2+25\right) \la^2+4 \nu \left(\nu^4+2 \nu^2-3\right) \la+2 \nu^2 \left(\nu^2-1\right)^2\Bigr) \mu^4\no\\
   &\quad+\Bigl(\left(2 \nu^2-11 \nu^4\right) \la^7-2 \nu^3 \left(5 \nu^2-8\right) \la^6+\nu^2 \left(9 \nu^4-25 \nu^2+52\right) \la^5+\left(4 \nu^5+42 \nu^3-52 \nu\right) \la^4\no\\
   &\quad+\left(-14 \nu^6-3 \nu^4+9 \nu^2+5\right) \la^3+2 \nu \left(5 \nu^4+22 \nu^2-15\right) \la^2+\left(5 \nu^6-\nu^4-15 \nu^2+11\right) \la-4 \nu \left(\nu^2-1\right)^2\Bigr) \mu^3\no\\
   &\quad+\Bigl(2 \nu^3 \left(\nu^2+3\right) \la^7+\left(-9 \nu^6+13 \nu^4-6 \nu^2\right) \la^6+8 \nu^3 \la^5+4 \nu^2 \left(3 \nu^4+9 \nu^2-14\right) \la^4\no\\
   &\quad-2 \nu \left(9 \nu^4+17 \nu^2-14\right) \la^3-\left(3 \nu^6-8 \nu^4+2 \nu^2+1\right) \la^2+8 \left(2 \nu^5-3 \nu^3+\nu\right) \la-2 \left(\nu^2-1\right)^2\Bigr) \mu^2\no\\
   &\quad+\la \nu \Bigl(3 \nu^3 \left(\nu^2+2\right) \la^6-2 \nu^2 \left(\nu^2+6\right) \la^5-2 \nu \left(\nu^4+8 \nu^2-3\right) \la^4+\left(22 \nu^4-8 \nu^2\right) \la^3\no\\
   &\quad-\nu \left(\nu^4+25 \nu^2-21\right) \la^2-2 \left(10 \nu^4-9 \nu^2+3\right) \la+3 \nu \left(\nu^2-1\right)\Bigr) \mu\no\\
   &\quad+\la^2 \nu^2 \left(2 \nu^3 \la^5-\nu^2 \left(\nu^2+6\right) \la^4+\left(6 \nu-10 \nu^3\right) \la^3+\left(\nu^4+14 \nu^2-2\right) \la^2+\left(8 \nu^3-2 \nu\right) \la-\nu^2-1\right)\,,\\
    f^{0,0}&=2 (\la-\mu) (\mu+\nu) \Bigl(\la^4 \mu^6 (\la+\nu)+\la^3 \mu^5 \left(4 \la^2 \nu+\la \left(\nu^2-5\right)-4 \nu\right)\no\\
   &\quad+\la \nu \left(\la^5 \nu^4-2 \la^3 \nu^2 \left(\nu^2+2\right)-\la^2 \nu \left(\nu^2-4\right)+\la \left(\nu^4+4 \nu^2-1\right)+\nu \left(\nu^2-1\right)\right)\no\\
   &\quad+\la^2 \mu^4 \left(-\la^4 \nu+4 \la^3 \nu^2+2 \la^2 \nu \left(\nu^2-8\right)-5 \la \left(\nu^2-2\right)-2 \nu \left(\nu^2-3\right)\right)\no\\
   &\quad+\la \mu^3 \left(-4 \la^5 \nu^2+2 \la^4 \nu \left(\nu^2+2\right)+\la^3 \nu^2 \left(\nu^2-11\right)-6 \la^2 \nu \left(\nu^2-4\right)-\la \left(\nu^4-9 \nu^2+10\right)+4 \nu \left(\nu^2-1\right)\right)\no\\
   &\quad-\mu \left(2 \la^5 \nu^3 \left(\nu^2-4\right)-2 \la^4 \nu^2 \left(\nu^2-6\right)+\la^3 \left(-4 \nu^5+6 \nu^3-4 \nu\right)+\la^2 \left(\nu^4+\nu^2\right)+2 \la \nu \left(\nu^4+\nu^2-2\right)+\left(\nu^2-1\right)^2\right)\no\\
   &\quad+\mu^2 \Bigl(-4 \la^6 \nu^3-\la^5 \nu^2 \left(\nu^2-12\right)+\la^4 \nu \left(\nu^4-6\right)-\la^3 \nu^2 \left(\nu^2-9\right)-2 \la^2 \nu \left(\nu^4-3 \nu^2+8\right)\no\\
   &\quad+\la \left(2 \nu^4-7 \nu^2+5\right)+\nu \left(\nu^2-1\right)^2\Bigr)\Bigr)\,,
\end{align}

\subsubsection*{$h$ part}

\begin{align}
    h^{5,5}&=-4 \nu^2 \left(\la^2 \nu^2+\mu^3 \left(-\la^3+\la^2 \nu+\la\right)+\mu^2 \left(\la^3 (-\nu)+\la^2-\la \nu+\nu^2-1\right)+\la \mu \nu (\la-2 \nu)\right)^2\,,\\
    h^{4,5}&=2 \la \nu^2 (\la-\mu) (\mu+\nu) \Bigl(\la \mu \nu \left(2 \la^2 \nu+3 \la \left(\nu^2-1\right)+6 \nu\right)\no\\
   &\quad-\la^2 \nu^2 (\la \nu+3)+\la \mu^4 \left(-3 \la^3+2 \la^2 \nu+\la \left(\nu^2+3\right)+\nu\right)\no\\
   &\quad-\mu^3 \left(2 \la^4 \nu+2 \la^3 \left(\nu^2-3\right)+6 \la^2 \nu-2 \la \left(\nu^2-3\right)-\nu^3+\nu\right)\no\\
   &\quad+\mu^2 \left(\la^4 \nu^2+5 \la^3 \nu-\la^2 \left(4 \nu^2+3\right)+\la \left(4 \nu-3 \nu^3\right)-3 \nu^2+3\right)\Bigr)\,,\\
    h^{3,5}&=\nu \Bigl(\la^3 \nu^3 \left(\la^3 \nu^2-\la^2 \nu (\nu+2)+\la \left(-2 \nu^2+2 \nu-3\right)-2 \nu-1\right)\no\\
   &\quad-\la^2 \mu^6 \left(\la^4 (2 \nu-1)+\la^3 \left(-4 \nu^2+3 \nu-2\right)+\la^2 \left(2 \nu^3-3 \nu^2-\nu+1\right)+\la \left(\nu^3+8 \nu^2-2 \nu+2\right)+\nu \left(-\nu^2+\nu+2\right)\right)\no\\
   &\quad+2 \la^2 \mu \nu^2 \left(3 \la^4 \nu^2-\la^3 \nu \left(2 \nu^2+2 \nu+1\right)+\la^2 \left(2 \nu^3+\nu^2+3 \nu-2\right)+\la \left(4 \nu^3-3 \nu^2+3 \nu-1\right)+3 \nu^2+\nu-1\right)\no\\
   &\quad+\la \mu^5 \Bigl(-4 \la^5 (\nu-1) \nu+\la^4 \left(4 \nu^3-9 \nu^2+10 \nu-3\right)+\la^3 \left(6 \nu^3+4 \nu^2+4 \nu-6\right)\no\\
   &\quad-\la^2 \left(\nu^4+4 \nu^3+2 \nu^2+2 \nu-3\right)-2 \la \left(\nu^4-\nu^3-5 \nu^2+2 \nu-3\right)+\nu \left(-\nu^3-4 \nu^2+\nu+4\right)\Bigr)\no\\
   &\quad+\la \mu^2 \nu \Bigl(\la^5 \nu^2 (\nu+6)+\la^4 \nu \left(-16 \nu^2-9 \nu+6\right)+\la^3 \left(6 \nu^4+13 \nu^3+6 \nu^2+9 \nu-3\right)\no\\
   &\quad+\la^2 \left(-6 \nu^4+6 \nu^3-13 \nu^2+8 \nu-2\right)+\la \left(-12 \nu^4+6 \nu^3+\nu^2+\nu\right)-6 \nu^3-\nu^2+6 \nu+1\Bigr)\no\\
   &\quad+\mu^4 \Bigl(\la^6 \nu \left(-2 \nu^2+6 \nu-3\right)+\la^5 \nu \left(-9 \nu^2+6 \nu-10\right)+\la^4 \left(3 \nu^4+11 \nu^3+16 \nu^2-14 \nu+3\right)\no\\
   &\quad+\la^3 \left(-13 \nu^3-8 \nu^2+2 \nu+6\right)+\la^2 \left(\nu^5+7 \nu^4+10 \nu^3-2 \nu^2+\nu-3\right)\no\\
   &\quad-\la \left(\nu^5-4 \nu^4+\nu^3-2 \nu^2-2 \nu+6\right)-2 \nu \left(\nu^2-1\right)^2\Bigr)\no\\
   &\quad+\mu^3 \Bigl(2 \la^6 \nu^2 (2 \nu-1)-\la^5 \nu \left(3 \nu^3+14 \nu^2+13 \nu-6\right)+4 \la^4 \nu \left(3 \nu^3+5 \nu^2-2 \nu+2\right)\no\\
   &\quad-\la^3 \left(4 \nu^5+15 \nu^4+10 \nu^3+8 \nu^2-6 \nu+1\right)+2 \la^2 \left(2 \nu^5-5 \nu^4+4 \nu^3-3 \nu^2-2 \nu-1\right)\no\\
   &\quad+\la \left(8 \nu^5-2 \nu^4-8 \nu^3+\nu^2+1\right)+2 \left(\nu^2-1\right)^2\Bigr)\Bigr)\,,\\
    h^{2,5}&=-\la \nu (\mu+\nu) \Bigl(-\left(\la^5 \left(\mu^5+\mu^4 (1-2 \nu)-\mu^3 (\nu-3) \nu+\mu^2 \nu \left(2 \nu^2+3 \nu-1\right)+\mu \nu^3+\nu^3\right)\right)\no\\
   &\quad+\la^4 \Bigl(\mu^5 (1-2 \nu)+\mu^4 \left(-2 \nu^2+5 \nu+2\right)+\mu^3 \left(6 \nu^3+7 \nu^2-4 \nu+3\right)+\mu^2 \nu \left(3 \nu^2-2 \nu+7\right)\no\\
   &\quad+\mu \nu \left(2 \nu^2+5 \nu-2\right)+\nu^3 (2 \nu+1)\Bigr)\no\\
   &\quad+\la^3 \Bigl(\mu^5 \left(\nu^2-2 \nu+2\right)-\mu^4 \left(6 \nu^3+5 \nu^2+\nu+3\right)+\mu^3 \nu \left(-3 \nu^2+6 \nu-11\right)\no\\
   &\quad+\mu^2 \left(-11 \nu^2+2 \nu-3\right)-\mu \nu \left(8 \nu^3+3 \nu^2+\nu+5\right)+\nu \left(4 \nu^2-2 \nu+1\right)\Bigr)\no\\
   &\quad+\la^2 \Bigl(\mu^5 \nu \left(2 \nu^2+\nu+4\right)+\mu^4 \left(\nu^3-6 \nu^2+4 \nu-6\right)+\mu^3 \left(-2 \nu^3+7 \nu^2+8 \nu+3\right)\no\\
   &\quad+\mu^2 \left(12 \nu^4+3 \nu^3+7 \nu-2\right)+\mu \left(-12 \nu^3+4 \nu^2+1\right)+\nu (2 \nu+1)\Bigr)\no\\
   &\quad+\la \mu \Bigl(2 \mu^4 \nu^2+\mu^3 \nu \left(\nu^2-\nu-8\right)-\mu^2 \left(8 \nu^4+\nu^3-3 \nu^2+2 \nu-6\right)+\mu \left(12 \nu^3-2 \nu^2-5 \nu-1\right)\no\\
   &\quad-4 \nu^2-\nu+1\Bigr)+2 \mu^2 \left(\nu^2-1\right) (\mu \nu-1)^2\Bigr)\,,\\
    h^{1,5}&=-2 \la^2 \nu^2 (\la-\mu)^2 (\la \mu-1) (\mu+\nu)^2 (\la (\mu-\nu)+\mu \nu-1)\,,
\end{align}

\begin{align}
    h^{4,4}&=-2 \nu \Bigl(\la^6 \left(-(\mu+\nu)^2\right) \left(\mu^4 (6 \nu-1)-2 \mu^3 \nu-\mu^2 \nu (\nu+3)-\nu^3\right)\no\\
   &\quad+\la^5 \Bigl(3 \mu^6 \nu (4 \nu-1)+3 \mu^5 \left(4 \nu^3-3 \nu^2+2 \nu-1\right)+\mu^4 \nu \left(-9 \nu^2+12 \nu-10\right)\no\\
   &\quad-\mu^3 \nu \left(3 \nu^3-14 \nu^2+13 \nu+6\right)+\mu^2 \nu^2 \left(4 \nu^2-9 \nu-12\right)-2 \mu \nu^3 \left(2 \nu^2+2 \nu+3\right)-\nu^5\Bigr)\no\\
   &\quad+\la^4 \Bigl(\mu^6 \left(-6 \nu^3+3 \nu^2+15 \nu-1\right)+2 \mu^5 \nu \left(3 \nu^2-3 \nu+2\right)+\mu^4 \left(3 \nu^4-39 \nu^3+16 \nu^2+6 \nu+3\right)\no\\
   &\quad+4 \mu^3 \nu \left(-6 \nu^3+5 \nu^2+3 \nu+2\right)+\mu^2 \nu \left(6 \nu^4+13 \nu^3-6 \nu^2+9 \nu+3\right)\no\\
   &\quad+2 \mu \nu^2 \left(2 \nu^3-6 \nu^2+3 \nu+3\right)+\nu^3 \left(-6 \nu^2+2 \nu+3\right)\Bigr)\no\\
   &\quad-\la^3 \Bigl(\mu^6 \nu \left(\nu^2+12 \nu-2\right)+\mu^5 \left(\nu^4-20 \nu^3+2 \nu^2+30 \nu-3\right)+\mu^4 \nu \left(-28 \nu^3+13 \nu^2+24 \nu-2\right)\no\\
   &\quad+\mu^3 \left(4 \nu^5+15 \nu^4-18 \nu^3+8 \nu^2+6 \nu+1\right)+\mu^2 \nu \left(6 \nu^4-36 \nu^3+13 \nu^2+12 \nu+2\right)\no\\
   &\quad+2 \mu \nu^2 \left(-12 \nu^3+3 \nu^2+3 \nu+1\right)+\nu^3\Bigr)\no\\
   &\quad+\la^2 \mu \Bigl(\mu^5 \nu \left(\nu^2-\nu-6\right)+2 \mu^4 \nu \left(-5 \nu^3+\nu^2+12 \nu-2\right)+\mu^3 \left(\nu^5+7 \nu^4+6 \nu^3-2 \nu^2+15 \nu-3\right)\no\\
   &\quad+2 \mu^2 \nu \left(2 \nu^4-18 \nu^3+4 \nu^2+9 \nu-2\right)+\mu \left(-36 \nu^5+6 \nu^4+15 \nu^3+\nu^2\right)+2 \nu^3\Bigr)\no\\
   &\quad-\la \mu^2 \left(\nu^2-1\right) \Bigl(\mu^3 \nu (\nu+12)+\mu^2 \nu \left(\nu^2-12 \nu+2\right)+\mu \left(-24 \nu^3+2 \nu^2+1\right)+\nu\Bigr)-6 \mu^4 \nu \left(\nu^2-1\right)^2\Bigr)\,,\\
    h^{3,4}&=-\la \nu (\mu+\nu) \Bigl(\la^5 \Bigl(3 \mu^5+\mu^4 (1-14 \nu)+\mu^3 \nu (3-11 \nu)+\mu^2 \nu \left(6 \nu^2+3 \nu+1\right)+\mu \nu^3-\nu^3\Bigr)\no\\
   &\quad+\la^4 \Bigl(\mu^5 (14 \nu-1)+\mu^4 \left(22 \nu^2-5 \nu-10\right)-\mu^3 \left(18 \nu^3+7 \nu^2-24 \nu+3\right)\no\\
   &\quad+\mu^2 \nu \left(-3 \nu^2+26 \nu-7\right)+\mu \nu \left(14 \nu^2-5 \nu-2\right)-\nu^3 (6 \nu+1)\Bigr)\no\\
   &\quad+\la^3 \Bigl(\mu^5 \left(-11 \nu^2+2 \nu-2\right)+\mu^4 \left(18 \nu^3+5 \nu^2-9 \nu+3\right)+\mu^3 \left(3 \nu^3-58 \nu^2+11 \nu+12\right)\no\\
   &\quad+\mu^2 \left(-36 \nu^3+11 \nu^2-6 \nu+3\right)+\mu \nu \left(24 \nu^3+3 \nu^2-13 \nu+5\right)-16 \nu^3+2 \nu^2+\nu\Bigr)\no\\
   &\quad-\la^2 \Bigl(\mu^5 \nu \left(6 \nu^2+\nu+16\right)+\mu^4 \left(\nu^3-38 \nu^2+4 \nu-6\right)+\mu^3 \left(-34 \nu^3+7 \nu^2+24 \nu+3\right)\no\\
   &\quad+\mu^2 \left(36 \nu^4+3 \nu^3-32 \nu^2+7 \nu+6\right)+\mu \left(-48 \nu^3+4 \nu^2+4 \nu+1\right)+\nu (2 \nu+1)\Bigr)\no\\
   &\quad+\la \mu \Bigl(-6 \mu^4 \nu^2+\mu^3 \nu \left(-11 \nu^2+\nu+32\right)+\mu^2 \left(24 \nu^4+\nu^3-25 \nu^2+2 \nu-6\right)\no\\
   &\quad+\mu \left(-48 \nu^3+2 \nu^2+19 \nu+1\right)+4 \nu^2+\nu+1\Bigr)-2 \mu^2 \left(\nu^2-1\right) \left(3 \mu^2 \nu^2-8 \mu \nu+1\right)\Bigr)\,,\\
    h^{2,4}&=-\nu \Bigl(\la^6 (\mu+\nu)^2 \left(\mu^4 (2 \nu-1)-2 \mu^3 (\nu+1)-\mu^2 \nu (\nu+9)+6 \mu \nu^2+3 \nu^3\right)\no\\
   &\quad+\la^5 \Bigl(\mu^6 \nu (3-4 \nu)+\mu^5 \left(-4 \nu^3+9 \nu^2+18 \nu+3\right)+\mu^4 \left(9 \nu^3+14 \nu^2+10 \nu+6\right)\no\\
   &\quad+\mu^3 \nu \left(3 \nu^3-38 \nu^2+13 \nu+30\right)+\mu^2 \nu^2 \left(-46 \nu^2+9 \nu+38\right)+2 \mu \nu^3 \left(-6 \nu^2+2 \nu+3\right)+(\nu-8) \nu^4\Bigr)\no\\
   &\quad+\la^4 \Bigl(\mu^6 \left(2 \nu^3-3 \nu^2-13 \nu+1\right)+2 \mu^5 \nu \left(-3 \nu^2+\nu-2\right)-\mu^4 \left(3 \nu^4-61 \nu^3+16 \nu^2+46 \nu+3\right)\no\\
   &\quad+\mu^3 \left(66 \nu^4-20 \nu^3-64 \nu^2-8 \nu-6\right)+\mu^2 \nu \left(18 \nu^4-13 \nu^3+6 \nu^2-9 \nu-21\right)\no\\
   &\quad-2 \mu \nu^2 \left(2 \nu^3-14 \nu^2+3 \nu+13\right)+\nu^3 \left(2 \nu^2-2 \nu-9\right)\Bigr)\no\\
   &\quad+\la^3 \Bigl(\mu^6 \left(\nu^3-2 \nu^2-2 \nu+2\right)+\mu^5 \left(\nu^4-32 \nu^3+2 \nu^2+26 \nu-3\right)+\mu^4 \nu \left(-42 \nu^3+13 \nu^2+26 \nu-2\right)\no\\
   &\quad+\mu^3 \left(-12 \nu^5+15 \nu^4-26 \nu^3+8 \nu^2+30 \nu+1\right)+\mu^2 \left(6 \nu^5-36 \nu^4+13 \nu^3+46 \nu^2+2 \nu+2\right)\no\\
   &\quad+2 \mu \nu \left(-4 \nu^4+3 \nu^3+9 \nu^2+\nu+2\right)+\nu^2 \left(2 \nu^2+\nu+2\right)\Bigr)\no\\
   &\quad+\la^2 \mu \Bigl(\mu^5 \nu \left(3 \nu^2+\nu+2\right)+2 \mu^4 \left(5 \nu^4-\nu^3+2 \nu-3\right)+\mu^3 \left(3 \nu^5-7 \nu^4+10 \nu^3+2 \nu^2-13 \nu+3\right)\no\\
   &\quad-2 \mu^2 \nu \left(2 \nu^4-10 \nu^3+4 \nu^2+9 \nu-2\right)+\mu \nu \left(12 \nu^4-6 \nu^3-13 \nu^2-\nu-4\right)-2 \nu^3 (3 \nu+1)\Bigr)\no\\
   &\quad+\la \mu^2 \left(\nu^2-1\right) \left(\mu^3 \nu (\nu+4)+\mu^2 \left(\nu^3-4 \nu^2+2 \nu-6\right)+\mu \left(-8 \nu^3+2 \nu^2+1\right)+\nu (6 \nu+1)\right)\no\\
   &\quad+2 \mu^3 \left(\nu^2-1\right)^2 (\mu \nu-1)\Bigr)\,,\\
    h^{1,4}&=-\la (\mu+\nu) \Bigl(\la \mu^5 \left(\la^4 (\nu-1)-\la^3 (\nu-1)^2-\la^2 \left(\nu^3+\nu\right)+\la (\nu-1)^2 \nu^2+\nu^3 (\nu+1)\right)\no\\
   &\quad+\la \nu \left(\la^4 (\nu-1) \nu^3+\la^3 (\nu-1)^2 \nu^2+\la^2 \left(\nu^3+\nu\right)-\la (\nu-1)^2-\nu-1\right)\no\\
   &\quad+\mu^4 \Bigl(\la^5 (\nu-3) \nu+2 \la^4 \left(\nu^3+\nu^2-2 \nu+2\right)+\la^3 \left(-3 \nu^4+6 \nu^3+\nu^2-6 \nu+4\right)\no\\
   &\quad-\la^2 \nu \left(6 \nu^3+\nu^2-3\right)+\la \nu^2 \left(\nu^3-\nu^2+2 \nu-2\right)+\nu^3 \left(\nu^2-1\right)\Bigr)\no\\
   &\quad-\mu^3 \Bigl(\la^5 \nu \left(\nu^2+2 \nu-1\right)+\la^4 \nu \left(-3 \nu^3+6 \nu^2+3 \nu-10\right)+2 \la^3 \left(-6 \nu^4+\nu^3+3 \nu^2-3 \nu+3\right)\no\\
   &\quad+\la^2 \left(4 \nu^5-4 \nu^4+6 \nu^3-2 \nu^2-6 \nu+6\right)+\la \nu \left(4 \nu^4-2 \nu^3-3 \nu^2+3\right)+\nu^2 \left(\nu^2-1\right)\Bigr)\no\\
   &\quad+\mu^2 \Bigl(\la^5 \left(-\nu^4+2 \nu^3+\nu^2\right)+\la^4 \nu \left(-10 \nu^3+3 \nu^2+6 \nu-3\right)+2 \la^3 \nu \left(3 \nu^4-3 \nu^3+3 \nu^2+\nu-6\right)\no\\
   &\quad+\la^2 \left(6 \nu^5-6 \nu^4-2 \nu^3+6 \nu^2-4 \nu+4\right)+\la \left(3 \nu^4-3 \nu^2-2 \nu+4\right)-\nu^3+\nu\Bigr)\no\\
   &\quad+\mu \Bigl(\la^5 \nu^3 (3 \nu-1)-2 \la^4 \nu^2 \left(2 \nu^3-2 \nu^2+\nu+1\right)-\la^3 \nu \left(4 \nu^4-6 \nu^3+\nu^2+6 \nu-3\right)\no\\
   &\quad+\la^2 \nu \left(-3 \nu^3+\nu+6\right)+\la \left(2 \nu^3-2 \nu^2+\nu-1\right)+\nu^2-1\Bigr)\Bigr)\,,\\
    h^{0,4}&=-2 \la^2 \nu (\la-\mu) (\la \mu-1)^2 (\mu+\nu)^2 (\la (\mu-\nu)+\mu \nu-1)\,,
\end{align}
\begin{align}
    h^{3,3}&=2 \nu \Bigl(\la^6 \left(-(\mu+\nu)^2\right) \left(\mu^4+2 \mu^3 (\nu+3)+\mu^2 (\nu-11) \nu+4 \mu \nu^2-3 \nu^3\right)\no\\
   &\quad+\la^5 \Bigl(\mu^6 (3 \nu-2)+\mu^5 \left(9 \nu^2-10 \nu+3\right)+\mu^4 \left(9 \nu^3-18 \nu^2+10 \nu+18\right)+\mu^3 \nu \left(3 \nu^3-10 \nu^2+13 \nu+14\right)\no\\
   &\quad+\mu^2 \nu^2 \left(-12 \nu^2+9 \nu-14\right)-2 \mu \nu^3 \left(6 \nu^2-2 \nu+7\right)+(\nu-4) \nu^4\Bigr)\no\\
   &\quad-\la^4 \Bigl(\mu^6 \left(3 \nu^2-11 \nu-1\right)+\mu^5 \left(6 \nu^3-2 \nu^2+4 \nu-6\right)+\mu^4 \left(3 \nu^4-5 \nu^3+16 \nu^2-8 \nu+3\right)\no\\
   &\quad+\mu^3 \left(-24 \nu^4+20 \nu^3-26 \nu^2+8 \nu+18\right)+\mu^2 \nu \left(-18 \nu^4+13 \nu^3-20 \nu^2+9 \nu+25\right)\no\\
   &\quad+2 \mu \nu^2 \left(2 \nu^3-6 \nu^2+3 \nu+2\right)+\nu^3 (2 \nu-11)\Bigr)\no\\
   &\quad+\la^3 \Bigl(\mu^6 \left(\nu^3+4 \nu^2-2 \nu+8\right)+\mu^5 \left(\nu^4-4 \nu^3+2 \nu^2-22 \nu-3\right)-\mu^4 \left(20 \nu^4-13 \nu^3+16 \nu^2+2 \nu+6\right)\no\\
   &\quad+\mu^3 \left(-12 \nu^5+15 \nu^4+2 \nu^3+8 \nu^2+14 \nu+1\right)+\mu^2 \left(6 \nu^5-12 \nu^4+13 \nu^3-10 \nu^2+2 \nu+6\right)\no\\
   &\quad+2 \mu \nu \left(3 \nu^3-11 \nu^2+\nu+6\right)+\nu^2 \left(8 \nu^2+\nu+6\right)\Bigr)\no\\
   &\quad+\la^2 \mu \Bigl(\mu^5 \nu^2 (3 \nu+1)+\mu^4 \left(6 \nu^4-2 \nu^3+4 \nu^2+4 \nu-24\right)+\mu^3 \left(3 \nu^5-7 \nu^4-8 \nu^3+2 \nu^2+11 \nu+3\right)\no\\
   &\quad+\mu^2 \left(-4 \nu^5+4 \nu^4-8 \nu^3+38 \nu^2+4 \nu+2\right)-\mu \nu \left(6 \nu^3-11 \nu^2+\nu+12\right)-2 \nu^2 \left(12 \nu^2+\nu-1\right)\Bigr)\no\\
   &\quad+\la \mu^2 \left(\nu^2-1\right) \left(\mu^3 \nu^2+\mu^2 \left(\nu^3+2 \nu-24\right)+2 \mu \nu^2+\mu+\nu (24 \nu+1)\right)-8 \mu^3 \left(\nu^2-1\right)^2\Bigr)\,,\\
    h^{2,3}&=\la (\mu+\nu) \Bigl(\la \mu^5 \left(\la^4 (5 \nu-1)+\la^3 \left(5 \nu^2+2 \nu+1\right)+\la^2 \left(5 \nu-9 \nu^3\right)-\la \nu^2 \left(\nu^2+2 \nu+5\right)+(\nu-1) \nu^3\right)\no\\
   &\quad+\la \nu \left(\la^4 (\nu-5) \nu^3-\la^3 \nu^2 \left(\nu^2+2 \nu+5\right)+\la^2 \left(9 \nu-5 \nu^3\right)+\la \left(5 \nu^2+2 \nu+1\right)+\nu-1\right)\no\\
   &\quad+\mu^4 \Bigl(\la^5 (-\nu) (5 \nu+3)+2 \la^4 \left(9 \nu^3+\nu^2-12 \nu+2\right)+\la^3 \left(3 \nu^4+6 \nu^3+5 \nu^2-6 \nu-4\right)\no\\
   &\quad-3 \la^2 \nu \left(2 \nu^3-7 \nu^2+5\right)+\la \nu^2 \left(\nu^3-9 \nu^2+2 \nu+10\right)-\nu^5+\nu^3\Bigr)\no\\
   &\quad-\mu^3 \Bigl(\la^5 \nu \left(9 \nu^2+2 \nu+1\right)+\la^4 \nu \left(3 \nu^3+6 \nu^2+9 \nu-10\right)+\la^3 \left(-12 \nu^4+34 \nu^3+6 \nu^2-42 \nu+6\right)\no\\
   &\quad+\la^2 \left(4 \nu^5-32 \nu^4+6 \nu^3+26 \nu^2-6 \nu-6\right)-\la \nu \left(4 \nu^4+2 \nu^3-17 \nu^2+15\right)-5 \nu^2 \left(\nu^2-1\right)\Bigr)\no\\
   &\quad+\mu^2 \Bigl(\la^5 \nu^2 \left(\nu^2+2 \nu+9\right)+\la^4 \nu \left(-10 \nu^3+9 \nu^2+6 \nu+3\right)+2 \la^3 \nu \left(3 \nu^4-21 \nu^3+3 \nu^2+17 \nu-6\right)\no\\
   &\quad+\la^2 \left(-6 \nu^5-6 \nu^4+26 \nu^3+6 \nu^2-32 \nu+4\right)-\la \left(15 \nu^4-17 \nu^2+2 \nu+4\right)+5 \nu \left(\nu^2-1\right)\Bigr)\no\\
   &\quad+\mu \Bigl(\la^5 \nu^3 (3 \nu+5)-2 \la^4 \nu^2 \left(2 \nu^3-12 \nu^2+\nu+9\right)+\la^3 \nu \left(4 \nu^4+6 \nu^3-5 \nu^2-6 \nu-3\right)\no\\
   &\quad+3 \la^2 \nu \left(5 \nu^3-7 \nu+2\right)-\la \left(10 \nu^3+2 \nu^2-9 \nu+1\right)-\nu^2+1\Bigr)\Bigr)\,,\\
    h^{1,3}&=\la^3 \left(3 \la^3+(1-8 \nu) \la^2+\left(-9 \nu^2-2 \nu+2\right) \la+\nu \left(2 \nu^2+\nu+2\right)\right) \mu^6\no\\
   &\quad+2 \la^2 \Bigl(6 \nu \la^4+\left(3 \nu^2+2 \nu-6\right) \la^3-\left(13 \nu^3+3 \nu^2-14 \nu+2\right) \la^2+\left(2 \nu^4+\nu^3+9 \nu^2+3 \nu-4\right) \la-\nu (\nu+3)\Bigr) \mu^5\no\\
   &\quad+\la \Bigl(\nu (6 \nu-1) \la^5+\nu \left(38 \nu^2+9 \nu-46\right) \la^4+\left(-21 \nu^4-9 \nu^3+6 \nu^2-13 \nu+18\right) \la^3\no\\
   &\quad+\left(2 \nu^5+2 \nu^4+46 \nu^3+13 \nu^2-36 \nu+6\right) \la^2-\left(4 \nu^4+\nu^3+13 \nu^2+6 \nu-12\right) \la+\nu \left(-\nu^3-6 \nu^2+\nu+6\right)\Bigr) \mu^4\no\\
   &\quad-\Bigl(2 \nu^2 (7 \nu+2) \la^6-\nu \left(30 \nu^3+13 \nu^2-38 \nu+3\right) \la^5+2 \nu \left(3 \nu^4+4 \nu^3+32 \nu^2+10 \nu-33\right) \la^4\no\\
   &\quad-\left(\nu^5+30 \nu^4+8 \nu^3-26 \nu^2+15 \nu-12\right) \la^3+\left(-4 \nu^4+18 \nu^3+8 \nu^2-20 \nu+4\right) \la^2\no\\
   &\quad+\left(\nu^5+\nu^3-8 \nu^2-2 \nu+8\right) \la+2 \nu \left(\nu^2-1\right)^2\Bigr) \mu^3\no\\
   &\quad+\Bigl(\nu^2 \left(-13 \nu^2-6 \nu+2\right) \la^6+\nu^2 \left(6 \nu^3+10 \nu^2+14 \nu+9\right) \la^5-\nu \left(3 \nu^4+46 \nu^3+16 \nu^2-61 \nu+3\right) \la^4\no\\
   &\quad+\nu \left(-2 \nu^3+26 \nu^2+13 \nu-42\right) \la^3+\left(3 \nu^5-13 \nu^4+2 \nu^3+10 \nu^2-7 \nu+3\right) \la^2\no\\
   &\quad+\left(6 \nu^5-2 \nu^4-2 \nu^3+\nu^2-4 \nu+1\right) \la+2 \left(\nu^2-1\right)^2\Bigr) \mu^2\no\\
   &\quad+\la \nu \Bigl(-2 \nu^2 \left(\nu^2+2 \nu-2\right) \la^5+\nu \left(3 \nu^3+18 \nu^2+9 \nu-4\right) \la^4+2 \nu \left(-2 \nu^2+\nu-3\right) \la^3\no\\
   &\quad+\left(-3 \nu^4+26 \nu^3+2 \nu^2-32 \nu+1\right) \la^2-2 \left(3 \nu^4-2 \nu^3+\nu-5\right) \la-4 \nu^3-\nu^2+4 \nu+1\Bigr) \mu\no\\
   &\quad+\la^2 \nu^2 \left(-\left((\nu-2) \nu^2 \la^4\right)+\nu (3 \nu-4) \la^3+\left(\nu^3-13 \nu^2-3 \nu+2\right) \la^2+\left(2 \nu^3-2 \nu^2-2 \nu+1\right) \la+2 \nu^2+\nu+3\right)\,,\\
    h^{0,3}&=\la (\mu+\nu) \Bigl(\la^2 \mu^5 \left(-\la^3 \nu+\la^2 (\nu+2)+\la \nu \left(\nu^2-2 \nu+4\right)+\nu^2 (\nu+2)\right)\no\\
   &\quad-\la \mu^4 \left(\la^4 \nu+\la^3 \nu \left(2 \nu^2-5 \nu-2\right)+\la^2 \left(5 \nu^3+\nu^2+3 \nu+8\right)-\la \nu \left(\nu^3+4 \nu-12\right)+\nu^2 \left(-\nu^2+\nu+4\right)\right)\no\\
   &\quad+\la \nu \left(\la^4 \left(-\nu^3\right)+\la^3 (\nu-2) \nu^2+\la^2 \nu \left(2 \nu^2-2 \nu+1\right)+\la \left(4 \nu^2+\nu+2\right)+2 \nu\right)\no\\
   &\quad+\mu^3 \Bigl(\la^5 \nu \left(\nu^2-3 \nu-2\right)+\la^4 \nu \left(7 \nu^2-2 \nu+3\right)+\la^3 \nu^2 \left(-3 \nu^2+2 \nu-11\right)\no\\
   &\quad+\la^2 \left(-2 \nu^4+7 \nu^3+3 \nu+12\right)-\la \nu \left(\nu^3+5 \nu^2+2 \nu-12\right)-2 \nu^2 \left(\nu^2-1\right)\Bigr)\no\\
   &\quad+\mu^2 \Bigl(\la^5 \left(\nu^2-3 \nu^3\right)+\la^4 \nu \left(3 \nu^3-4 \nu^2+7 \nu+6\right)+\la^3 \nu \left(-11 \nu^2+6 \nu-3\right)\no\\
   &\quad+\la^2 \nu \left(3 \nu^3+8 \nu^2+7 \nu-2\right)+\la \left(6 \nu^4-2 \nu^3+3 \nu^2-\nu-8\right)+4 \nu \left(\nu^2-1\right)\Bigr)\no\\
   &\quad-\mu \Bigl(\la^5 (\nu-2) \nu^3+\la^4 \nu^2 \left(-2 \nu^2-5 \nu+2\right)+\la^3 \nu \left(3 \nu^3+\nu^2+5 \nu+6\right)+\la^2 \nu \left(6 \nu^3-4 \nu^2+6 \nu-1\right)\no\\
   &\quad+\la \nu \left(8 \nu^2+\nu-1\right)+2 \left(\nu^2-1\right)\Bigr)\Bigr)\,,
\end{align}
\begin{align}
    h^{2,2}&=-2 \Bigl(\la^6 (\mu+\nu)^2 \left(3 \mu^4-4 \mu^3 \nu+\mu^2 \nu (11 \nu-1)-2 \mu \nu^2 (3 \nu+1)-\nu^3\right)\no\\
   &\quad+\la^5 \Bigl(\mu^6 (1-4 \nu)-2 \mu^5 \left(7 \nu^2-2 \nu+6\right)+\mu^4 \nu \left(-14 \nu^2+9 \nu-12\right)+\mu^3 \nu \left(14 \nu^3+13 \nu^2-10 \nu+3\right)\no\\
   &\quad+\mu^2 \nu^2 \left(18 \nu^3+10 \nu^2-18 \nu+9\right)+\mu \nu^3 \left(3 \nu^2-10 \nu+9\right)+\nu^4 (3-2 \nu)\Bigr)\no\\
   &\quad+\la^4 \Bigl(\mu^6 \nu (11 \nu-2)-2 \mu^5 \left(2 \nu^3+3 \nu^2-6 \nu+2\right)+\mu^4 \left(-25 \nu^4-9 \nu^3+20 \nu^2-13 \nu+18\right)\no\\
   &\quad-2 \mu^3 \nu \left(9 \nu^4+4 \nu^3-13 \nu^2+10 \nu-12\right)+\mu^2 \nu \left(-3 \nu^4+8 \nu^3-16 \nu^2+5 \nu-3\right)\no\\
   &\quad+2 \mu \nu^2 \left(3 \nu^3-2 \nu^2+\nu-3\right)+\nu^3 \left(\nu^2+11 \nu-3\right)\bigr)\no\\
   &\quad+\la^3 \Bigl(\mu^6 \nu \left(6 \nu^2+\nu+8\right)+2 \mu^5 \nu \left(6 \nu^3+\nu^2-11 \nu+3\right)+\mu^4 \left(6 \nu^5+2 \nu^4-10 \nu^3+13 \nu^2-12 \nu+6\right)\no\\
   &\quad+\mu^3 \left(\nu^5+14 \nu^4+8 \nu^3+2 \nu^2+15 \nu-12\right)-\mu^2 \nu \left(6 \nu^4+2 \nu^3+16 \nu^2-13 \nu+20\right)\no\\
   &\quad+\mu \left(-3 \nu^5-22 \nu^4+2 \nu^3-4 \nu^2+\nu\right)+\nu^2 \left(8 \nu^3-2 \nu^2+4 \nu+1\right)\Bigr)\no\\
   &\quad+\la^2 \Bigl(2 \mu^5 \nu \left(\nu^2-\nu-12\right)-\mu^4 \nu \left(12 \nu^3+\nu^2-11 \nu+6\right)+2 \mu^3 \left(\nu^5+2 \nu^4+19 \nu^3-4 \nu^2+2 \nu-2\right)\no\\
   &\quad+\mu^2 \left(3 \nu^5+11 \nu^4+2 \nu^3-8 \nu^2-7 \nu+3\right)-2 \mu \nu \left(12 \nu^4-2 \nu^3-2 \nu^2+\nu-3\right)+\nu^2 (\nu+3)\Bigr)\no\\
   &\quad-\la \mu \left(\nu^2-1\right) \left(\mu^3 \nu (\nu+24)+\mu^2 \nu \left(\nu^2+2\right)+\mu \left(-24 \nu^3+2 \nu^2+1\right)+\nu\right)-8 \mu^3 \nu \left(\nu^2-1\right)^2\Bigr)\,,\\
    h^{1,2}&=\la (\mu+\nu) \Bigl(-\la^2 \mu^5 \left(\la^3 \nu+\la^2 (\nu+6)-\la \nu \left(\nu^2+2 \nu-16\right)+\nu^2 (\nu+2)\right)\no\\
   &\quad+\la \mu^4 \Bigl(\la^4 \nu+\la^3 \nu \left(-2 \nu^2-5 \nu+14\right)+\la^2 \left(5 \nu^3-13 \nu^2+3 \nu+24\right)\no\\
   &\quad-\la \nu \left(\nu^3+4 \nu^2+4 \nu-48\right)+\nu^2 \left(\nu^2+\nu+4\right)\Bigr)\no\\
   &\quad+\la \nu \left(3 \la^4 \nu^3-\la^3 (\nu-14) \nu^2+\la^2 \nu \left(-2 \nu^2+2 \nu-11\right)-\la \left(16 \nu^2+\nu+6\right)-6 \nu\right)\no\\
   &\quad+\mu^3 \Bigl(\la^5 \nu \left(\nu^2+3 \nu+6\right)+\la^4 \nu \left(-7 \nu^2+26 \nu-3\right)+\la^3 \nu \left(3 \nu^3-6 \nu^2+11 \nu-36\right)\no\\
   &\quad-\la^2 \left(6 \nu^4+7 \nu^3-32 \nu^2+3 \nu+36\right)+\la \nu \left(\nu^3+19 \nu^2+2 \nu-48\right)+2 \nu^2 \left(\nu^2-1\right)\Bigr)\no\\
   &\quad+\mu^2 \Bigl(\la^5 \nu^2 (3 \nu-11)-\la^4 \nu \left(3 \nu^3-24 \nu^2+7 \nu+18\right)+\la^3 \nu \left(12 \nu^3+11 \nu^2-58 \nu+3\right)\no\\
   &\quad-\la^2 \nu \left(3 \nu^3+24 \nu^2+7 \nu-34\right)+\la \left(-6 \nu^4+2 \nu^3-25 \nu^2+\nu+24\right)-16 \nu \left(\nu^2-1\right)\Bigr)\no\\
   &\quad+\mu \Bigl(\la^5 (\nu-14) \nu^3+\la^4 \nu^2 \left(-10 \nu^2-5 \nu+22\right)+\la^3 \nu \left(3 \nu^3-9 \nu^2+5 \nu+18\right)\no\\
   &\quad+\la^2 \nu \left(6 \nu^3-4 \nu^2+38 \nu-1\right)+\la \nu \left(32 \nu^2+\nu-11\right)+6 \left(\nu^2-1\right)\Bigr)\Bigr)\,,\\
    h^{0,2}&=\la^3 \mu^6 \Bigl(-\la^3+\la^2 (2 \nu+1)+\la \left(3 \nu^2-2 \nu+2\right)+\nu (\nu+2)\Bigr)\no\\
   &\quad-2 \la^2 \mu^5 \Bigl(3 \la^4 \nu-\la^3 \left(\nu^2+2 \nu+2\right)+\la^2 \left(-2 \nu^3+3 \nu^2+\nu+2\right)-\la \left(\nu^3-3 \nu^2+3 \nu-4\right)+\nu \left(-\nu^2+\nu+3\right)\Bigr)\no\\
   &\quad+\la^2 \nu^2 \Bigl(-\left(\la^4 (\nu-2) \nu^2\right)+\la^3 \nu \left(-2 \nu^2+3 \nu-4\right)+\la^2 \left(\nu^3-\nu^2-3 \nu+2\right)\no\\
   &\quad+\la \left(2 \nu^3-2 \nu^2+8 \nu+1\right)+2 \nu^2+\nu-1\Bigr)\no\\
   &\quad-\la \mu^4 \Bigl(\la^5 \nu (6 \nu+1)+\la^4 \nu \left(6 \nu^2-9 \nu-16\right)+\la^3 \left(-3 \nu^4+9 \nu^3+6 \nu^2+13 \nu+6\right)\no\\
   &\quad+\la^2 \left(-2 \nu^4+8 \nu^3-13 \nu^2+6 \nu-6\right)+\la \left(\nu^3+\nu^2+6 \nu-12\right)+\nu \left(\nu^3+6 \nu^2-\nu-6\right)\Bigr)\no\\
   &\quad+\la \mu \nu \Bigl(-4 \la^5 (\nu-1) \nu^2+\la^4 \nu \left(3 \nu^3-10 \nu^2+9 \nu-4\right)+2 \la^3 \nu \left(3 \nu^3-2 \nu^2-2 \nu-3\right)\no\\
   &\quad+\la^2 \left(-3 \nu^4+2 \nu^3+2 \nu^2+4 \nu+1\right)-2 \la \left(3 \nu^4-2 \nu^3+5 \nu^2+\nu-1\right)-4 \nu^3-\nu^2+4 \nu+1\Bigr)\no\\
   &\quad+\mu^3 \Bigl(2 \la^6 (\nu-2) \nu^2+\la^5 \nu \left(-6 \nu^3+13 \nu^2+14 \nu+3\right)-4 \la^4 \nu \left(2 \nu^3-2 \nu^2+5 \nu+3\right)\no\\
   &\quad+\la^3 \left(\nu^5-6 \nu^4+8 \nu^3+10 \nu^2+15 \nu+4\right)+2 \la^2 \left(\nu^5+2 \nu^4+3 \nu^3-4 \nu^2+5 \nu-2\right)\no\\
   &\quad-\la \left(\nu^5+\nu^3-8 \nu^2-2 \nu+8\right)-2 \nu \left(\nu^2-1\right)^2\Bigr)\no\\
   &\quad+\mu^2 \Bigl(\la^6 \nu^2 \left(3 \nu^2-6 \nu+2\right)+\la^5 \nu^2 \left(10 \nu^2-6 \nu+9\right)\no\\
   &\quad-\la^4 \nu \left(3 \nu^4-14 \nu^3+16 \nu^2+11 \nu+3\right)+\la^3 \nu^2 \left(-6 \nu^3-2 \nu^2+8 \nu+13\right)\no\\
   &\quad+\la^2 \left(3 \nu^5-\nu^4+2 \nu^3-10 \nu^2-7 \nu-1\right)+\la \left(6 \nu^5-2 \nu^4-2 \nu^3+\nu^2-4 \nu+1\right)+2 \left(\nu^2-1\right)^2\Bigr)\,,
\end{align}
\begin{align}
    h^{1,1}&=2 \Bigl(\la^3 \mu^6 \left(\la^3-\la^2+\la \left(3 \nu^2+2 \nu-6\right)-\nu^2\right)\no\\
   &\quad+2 \la^2 \mu^5 \left(\la^4 \nu-\la^3 \left(3 \nu^2+2 \nu+2\right)+\la^2 \left(3 \nu^3+3 \nu^2-6 \nu+2\right)-\la \left(\nu^3+3 \nu^2+3 \nu-12\right)+\nu^2\right)\no\\
   &\quad+\la^2 \nu^2 \left(\la^4 (\nu-6) \nu^2-3 \la^3 (\nu-4) \nu-\la^2 \left(\nu^3-15 \nu^2-3 \nu+6\right)+\la \left(2 \nu^2-12 \nu-1\right)-6 \nu^2-\nu+1\right)\no\\
   &\quad+\la \mu^4 \Bigl(\la^5 \nu (4 \nu+1)+\la^4 \nu \left(-12 \nu^2-9 \nu+4\right)+\la^3 \left(3 \nu^4+9 \nu^3-6 \nu^2+13 \nu+6\right)\no\\
   &\quad-\la^2 \left(2 \nu^4+12 \nu^3+13 \nu^2-36 \nu+6\right)+\la \left(\nu^3+15 \nu^2+6 \nu-36\right)+\nu^2 \left(\nu^2-1\right)\Bigr)\no\\
   &\quad+\la \mu^3 \Bigl(2 \la^5 \nu^2 (3 \nu+2)-\la^4 \nu \left(6 \nu^3+13 \nu^2-14 \nu+3\right)+4 \la^3 \nu \left(2 \nu^3+3 \nu^2+5 \nu-6\right)\no\\
   &\quad-\la^2 \left(\nu^5+6 \nu^4+8 \nu^3-18 \nu^2+15 \nu+4\right)+\la \left(-4 \nu^4+18 \nu^3+8 \nu^2-36 \nu+4\right)+\nu^5+\nu^3-24 \nu^2-2 \nu+24\Bigr)\no\\
   &\quad+\la \mu \nu \Bigl(4 \la^5 (\nu-3) \nu^2-3 \la^4 \nu \left(\nu^3-2 \nu^2+3 \nu-4\right)+2 \la^3 \nu \left(2 \nu^2-3 \nu+3\right)\no\\
   &\quad+\la^2 \left(3 \nu^4-30 \nu^3-2 \nu^2+20 \nu-1\right)+2 \la \left(-2 \nu^3+12 \nu^2+\nu-5\right)+12 \nu^3+\nu^2-12 \nu-1\Bigr)\no\\
   &\quad+\mu^2 \Bigl(3 \la^6 \nu^2 \left(\nu^2+2 \nu-2\right)+\la^5 \nu^2 \left(-10 \nu^2+12 \nu-9\right)+\la^4 \nu \left(3 \nu^4+6 \nu^3+16 \nu^2-39 \nu+3\right)\no\\
   &\quad+\la^3 \nu \left(2 \nu^3-24 \nu^2-13 \nu+28\right)+\la^2 \left(-3 \nu^5+15 \nu^4-2 \nu^3+6 \nu^2+7 \nu+1\right)\no\\
   &\quad+\la \left(2 \nu^4-12 \nu^3-\nu^2+12 \nu-1\right)-6 \left(\nu^2-1\right)^2\Bigr)\Bigr)\,,\\
    h^{0,1}&=2 \la (\la \mu-1) (\mu+\nu) \Bigl(\la^2 \mu^4 (\la+3 \nu)-\la \mu^3 \left(2 \la^2 \nu-3 \la \left(\nu^2-1\right)+6 \nu\right)+\la \nu \left(3 \la^3 \nu^2-2 \la^2 \nu-\la \left(3 \nu^2+1\right)-\nu\right)\no\\
   &\quad+\mu^2 \left(-\la^4 \nu-5 \la^3 \nu^2+\la^2 \nu \left(3 \nu^2+4\right)+\la \left(3-4 \nu^2\right)-3 \nu \left(\nu^2-1\right)\right)\no\\
   &\quad+\mu \left(2 \la^4 \nu^2+\la^3 \left(2 \nu-6 \nu^3\right)+6 \la^2 \nu^2+\la \left(6 \nu^3-2 \nu\right)+\nu^2-1\right)\Bigr)\,,\\
    h^{0,0}&=4 \left(\la^2 \mu^3+\la \nu \left(\la^2 (-\nu)+\la+\nu\right)-\mu \left(\la^3 \nu-\la^2 \nu^2+\la \nu+\nu^2-1\right)+\la \mu^2 (\la \nu-2)\right)^2\,.
\end{align}

\end{document}